\documentclass[amsmath,amssymb,12pt,nofootinbib]{article}
\usepackage{jheppub}

\usepackage{hyperref}
\usepackage{amsmath}
\usepackage{graphicx}
\usepackage{xspace}
\usepackage{color}
\usepackage{overpic}
\usepackage{url}
\usepackage[utf8]{inputenc}
\usepackage{epstopdf}
\usepackage{tikz}
\usepackage{caption}
\usepackage{makecell}
\usepackage{bm}
\usepackage{revsymb}
\usepackage{slashed}
\usepackage{float}
\usepackage{booktabs}

\usepackage{fixltx2e} 

\makeatletter
\DeclareRobustCommand*\textsubsuperscript[2]{%
	\@textsubsuperscript{\selectfont#1}{\selectfont#2}}
\def\@textsubsuperscript#1#2{%
	{\m@th\ensuremath{_{\mbox{\fontsize\sf@size\z@#1}}
			^{\mbox{\fontsize\sf@size\z@#2}}}}}
\makeatother

\newcommand{\bea}{\begin{eqnarray}}
\newcommand{\eea}{\end{eqnarray}}

\newcommand{\eq}[1]{Eq.~\eqref{#1}}

\newcommand{\bctaunu}{$b\to c\tau\nu$ }

\definecolor{point1}{rgb}{1,0,0}
\definecolor{point2}{rgb}{0.5,0,0.5}
\definecolor{point3}{rgb}{1,0.75,0}
\definecolor{point4}{rgb}{0,0.66,0}
\definecolor{point5}{rgb}{0.2,0.6,0.8}
\definecolor{point6}{rgb}{1,0.5,0}
\definecolor{point7}{rgb}{0,0,0}
\definecolor{point8}{rgb}{0.8,0.1,0.5}

\begin{document}
\preprint{PSI-PR-19-25 , ZU-TH 52/19}

\title{Flavor Phenomenology of the Leptoquark Singlet-Triplet Model}

\author[a]{Andreas Crivellin}
\affiliation[a]{Paul Scherrer Institut, CH--5232 Villigen PSI, Switzerland}

\author[a,b]{Dario M\"uller}
\affiliation[b]{Physik-Institut, Universit\"at Z\"urich,
	Winterthurerstrasse 190, CH-8057 Z\"urich, Switzerland}

\author[c]{Francesco Saturnino}
\affiliation[c]{Albert Einstein Center for Fundamental Physics, Institute
	for Theoretical Physics,\\ University of Bern, CH-3012 Bern,
	Switzerland}
	
\emailAdd{andreas.crivellin@cern.ch}
\emailAdd{dario.mueller@psi.ch}
\emailAdd{saturnino@itp.unibe.ch}

\abstract{
In recent years, experiments revealed intriguing hints for new physics (NP) in semi-leptonic $B$ decays. Both in charged current processes, involving \bctaunu transitions, and in the neutral currents $b\to s\ell^+\ell^-$, a preference for NP compared to the standard model (SM) of more than $3\sigma$ and $5\sigma$ was found, respectively. In addition, there is the long-standing tension between the theory prediction and the measurement of the anomalous magnetic moment (AMM) of the muon ($a_\mu$) of more than $3\sigma$. Since all these observables are related to the violation of lepton flavor universality (LFU), a common NP explanation seems not only plausible but is even desirable. In this context, leptoquarks (LQs) are especially promising since they give tree-level effects in semi-leptonic $B$ decays, but only loop-suppressed effects in other flavor observables that agree well with their SM predictions. Furthermore, LQs can lead to a $m_t/m_\mu$ enhanced effect in $a_{\mu}$, allowing for an explanation even with (multi) TeV particles. However, a single scalar LQ representation cannot provide a common solution to all three anomalies. In this article we therefore consider a model in which we combine two scalar LQs: the $SU(2)_L$ singlet and the $SU(2)_L$ triplet. Within this model we compute all relevant 1-loop effects and perform a comprehensive phenomenological analysis, pointing out various interesting correlations among the observables. Furthermore, we identify benchmark points which are in fact able to explain all three anomalies ($b \to c \tau \nu$, $b\to s\ell^+\ell^-$ and $a_\mu$), without violating bounds from other observables, and study their predictions for future measurements. 
}


\maketitle

\section{Introduction}

While the Large Hadron Collider (LHC) at CERN has not directly observed any particles beyond the ones of the SM (see e.g. Refs.~\cite{Butler:2017afk,Masetti:2018btj} for an overview) intriguing indirect hints for NP have been acquired in flavor observables. In particular, measurements of semi-leptonic $B$ meson decays, involving the charged current \bctaunu or the flavor changing neutral current $b\to s\ell^+\ell^-$, point towards the violation of LFU. Furthermore, also the AMM of the muon, which measures LFU violation as it vanishes in the massless limit, points convincingly towards physics beyond the SM. In order to explain these deviations from the SM predictions -- also called anomalies -- one thus needs NP that couples differently to tau leptons, muons and electrons. As we will see, LQs are prime candidates for such an explanation in terms of physics beyond the SM.
\smallskip

Let us now review these anomalies in more detail. The first anomaly arose in the AMM of the muon $a_\mu=(g-2)_{\mu}/2$ in 2006. Here, the E821 experiment at Brookhaven  
discovered a tantalizing tension between their measurement~\cite{Bennett:2006fi,Mohr:2015ccw}
\begin{equation}
a_\mu^\text{exp}=116,\!592,\!089(63)\times 10^{-11}
\end{equation}
and the SM prediction\footnote{The SM prediction of $a_\mu$ is currently re-evaluated in a community-wide effort prompted by upcoming improved measurements at Fermilab \cite{Grange:2015fou} and J-PARC~\cite{Saito:2012zz}, see also Ref.~\cite{Gorringe:2015cma}. With electroweak~\cite{Czarnecki:1995wq,Czarnecki:1995sz,Gnendiger:2013pva} and QED~\cite{Aoyama:2017uqe} contributions under good control, recent advances in the evaluation of the hadronic part include: hadronic vacuum polarization~\cite{DellaMorte:2017dyu,Davier:2017zfy,Borsanyi:2017zdw,Blum:2018mom,Keshavarzi:2018mgv,Giusti:2018mdh,Colangelo:2018mtw}, hadronic light-by-light scattering~\cite{Gerardin:2016cqj,Blum:2016lnc,Colangelo:2017qdm,Colangelo:2017fiz,Blum:2017cer,Hoferichter:2018dmo,Colangelo:2019uex}, and higher-order hadronic corrections~\cite{Kurz:2014wya,Colangelo:2014qya}.}
\begin{equation}
\label{Delta_amu}
\delta a_\mu=a_\mu^{\rm{exp}} - a_\mu^{\rm{SM}} = (278 \pm 88) \times 10^{-11}  \ 
\end{equation}
of around $3$--$4\sigma$\footnote{During the publication process of this article, the Budapest-Marseilles-Wuppertal collaboration (BMWc) released a lattice QCD calculation from hadronic vacuum polarization (HVP) \cite{Borsanyi:2020mff}. These results would render the SM prediction for $a_\mu$ compatible with the experiment. However, the BMWc results are in tension with the HVP determined from $e^+e^-\to$ hadrons data~\cite{Davier:2017zfy, Keshavarzi:2018mgv, Davier:2019can, Keshavarzi:2019abf}, combined with analyticity and unitarity constraints for the leading $2\pi$~\cite{Davier:2019can, Colangelo:2018mtw, Ananthanarayan:2018nyx} and $3\pi$~\cite{Hoferichter:2019gzf} channels, covering almost $80\%$ of the HVP contribution. Furthermore, the HVP also enters the global EW fit \cite{Passera:2008jk}, whose (indirect) determination disagrees with the BMWc result. Therefore, the BMWc determination of the HVP would lead to a significant tension in EW fit~\cite{Haller:2018nnx} and we therefore use the (conservative) estimate of~\eq{Delta_amu}. }. This discrepancy is of the same order as the electroweak contribution of the SM. Therefore, TeV scale NP needs an enhancement mechanism, called chiral enhancement, to be able to account for the deviation~\cite{Crivellin:2018qmi}. For LQs this factor can be {$m_t/m_\mu \approx 10^3$} which provides the required enhancement, making LQs prime candidates for an explanation in terms of NP~\cite{Bauer:2015knc,Djouadi:1989md, Chakraverty:2001yg,Cheung:2001ip,Popov:2016fzr,Chen:2016dip,Biggio:2016wyy,Davidson:1993qk,Couture:1995he,Mahanta:2001yc,Queiroz:2014pra,Chen:2017hir,Das:2016vkr,Crivellin:2017zlb, Cai:2017wry, Crivellin:2018qmi,Kowalska:2018ulj,Mandal:2019gff,Dorsner:2019itg}. In fact, there are only two LQ representations (under the SM gauge group), out of the 10 possible ones~\cite{Buchmuller:1986zs}, that can have this enhancement: the scalar LQ $SU(2)_L$ singlet and the scalar LQ $SU(2)_L$ doublet with hypercharge $-2/3$ and $-7/3$, respectively.
\smallskip

In tauonic $B$ decays, BaBar measured in 2012 the ratios
\begin{align} 
R(D^{(*)})=\frac{{\rm Br}[B\to D^{(*)}\tau\nu]}{{\rm Br}[B\to D^{(*)} \ell\nu]} ~~~~ \text{with} ~~~~ \ell=\{e,\mu\}\,
\end{align}
significantly above the SM predictions~\cite{Lees:2012xj}. This is in agreement with the later LHCb measurements~\cite{Aaij:2015yra,Aaij:2017uff,Aaij:2017deq} of $R(D^*)$, while BELLE found values closer to the SM in its latest analysis~\cite{Abdesselam:2019dgh}. In combination, these deviations from the SM amount to $3.1\sigma$~\cite{Amhis:2019ckw}\footnote{This tension would even slightly increase by around $0.3\sigma$ if the new theory prediction of $R(D^*)$ of Ref.~\cite{Gambino:2019sif} was taken into account.}. Interestingly, also the ratio 
\begin{align}
R(J/\psi)=\frac{{\rm Br}[B_c\to J/\psi\tau\nu]}{{\rm Br}[B_c\to J/\psi\mu\nu]}
\end{align}
lies above its SM prediction~\cite{Aaij:2017tyk}, supporting the assumption of NP in  $b \to c \tau \nu$~\cite{Watanabe:2017mip,Chauhan:2017uil}. This picture is confirmed by different independent global fits~\cite{Murgui:2019czp,Shi:2019gxi,Blanke:2019qrx,Kumbhakar:2019avh} which include in addition polarization observables. Interestingly, these hints for NP are accompanied by data on $b\to u\tau\nu$ transitions. 

Once more, LQs are prime candidates for an explanation. Despite the $U_1$ vector LQ $SU(2)_L$ singlet~\cite{Alonso:2015sja, Calibbi:2015kma, Fajfer:2015ycq, Hiller:2016kry, Bhattacharya:2016mcc, Buttazzo:2017ixm, Barbieri:2015yvd, Barbieri:2016las, Calibbi:2017qbu, Bordone:2017bld, Bordone:2018nbg, Kumar:2018kmr, Biswas:2018snp, Crivellin:2018yvo, Blanke:2018sro, deMedeirosVarzielas:2019lgb, Cornella:2019hct, Bordone:2019uzc} and scalar LQ $S_2$ option~\cite{Tanaka:2012nw, Dorsner:2013tla, Sakaki:2013bfa, Sahoo:2015wya, Chen:2016dip, Dey:2017ede, Becirevic:2017jtw, Chauhan:2017ndd, Becirevic:2018afm, Popov:2019tyc}, the scalar LQ $\Phi_1$~\cite{Fajfer:2012jt, Deshpande:2012rr, Sakaki:2013bfa, Freytsis:2015qca, Bauer:2015knc, Li:2016vvp, Zhu:2016xdg, Popov:2016fzr, Deshpand:2016cpw, Becirevic:2016oho, Cai:2017wry, Buttazzo:2017ixm, Altmannshofer:2017poe, Kamali:2018fhr, Azatov:2018knx, Wei:2018vmk, Angelescu:2018tyl, Kim:2018oih, Crivellin:2019qnh, Yan:2019hpm} or the combination of $\Phi_1$ and $\Phi_3$\footnote{$\Phi_1$ and $\Phi_3$ are also called $S_1$ and $S_3$, respectively, in the literature.} can explain these data~\cite{Crivellin:2017zlb, Buttazzo:2017ixm, Marzocca:2018wcf, Bigaran:2019bqv}.
\smallskip

Finally, the statistically most significant deviations from the SM predictions were observed in observables involving $b\to s\ell^+\ell^-$ transitions. Here, the LHCb measurements~\cite{Aaij:2017vbb,Aaij:2019wad} of
\begin{align}
R(K^{(*)})=\frac{{\rm Br}[B\to K^{(*)}\mu^+\mu^-]}{{\rm Br}[B\to K^{(*)} e^+e^-]}
\end{align}
indicate LFU violation with a combined significance of $\approx4\sigma$~\cite{Capdevila:2017bsm, Altmannshofer:2017yso, DAmico:2017mtc, Ciuchini:2017mik, Hiller:2017bzc, Geng:2017svp, Hurth:2017hxg,Alguero:2019ptt,Aebischer:2019mlg,Ciuchini:2019usw,Arbey:2019duh}. Taking in addition into account all other $b\to s\mu^+\mu^-$ observables, e.g. the angular observable $P_5^\prime$~\cite{Aaij:2015oid} in the decay $B\to K^*\mu^{+}\mu^{-}$, the global fit of the Wilson coefficients even prefers several NP scenarios above the $5\sigma$ level~\cite{Alguero:2019ptt,Aebischer:2019mlg,Ciuchini:2019usw}. Furthermore, $b\to d\ell^+\ell^-$ transitions measured in $B\to \pi\mu^+\mu^-$~\cite{Hambrock:2015wka} deviate slightly from the LHCb measurement~\cite{Aaij:2015nea}. While this is not significant on its own, the central value is very well in agreement with the expectation from $b\to s\ell^+\ell^-$ assuming a $V_{td}/V_{ts}$-like scaling~\cite{Rusov:2019ixr} of the NP effect as obtained in models possessing an $U(2)$ flavor symmetry in the quark sector (see e.g. Refs.~\cite{Crivellin:2015lwa,Barbieri:2016las,Fuentes-Martin:2019mun,Calibbi:2019lvs} for accounts in the context of the flavor anomalies). This means that an effect of the same order and sign as in $b\to s\ell^+\ell^-$, relative to the SM, is preferred. Once more, LQs are prime candidates for an explanation. In particular the $U_1$ vector LQ $SU(2)_L$ singlet~\cite{Alonso:2015sja, Calibbi:2015kma, Hiller:2016kry, Bhattacharya:2016mcc, Buttazzo:2017ixm, Barbieri:2015yvd, Barbieri:2016las, Calibbi:2017qbu, Bordone:2018nbg, Kumar:2018kmr, Crivellin:2018yvo, Crivellin:2019szf, Cornella:2019hct, deMedeirosVarzielas:2019lgb, Bernigaud:2019bfy, Bordone:2019uzc}, the $U_3$ vector LQ $SU(2)_L$ triplet~\cite{Fajfer:2015ycq, Calibbi:2015kma, Hiller:2016kry, Bhattacharya:2016mcc, Barbieri:2015yvd, Kumar:2018kmr, Blanke:2018sro, deMedeirosVarzielas:2019lgb, Bernigaud:2019bfy,Fuentes-Martin:2019ign} and the $\Phi_3$ scalar LQ  $SU(2)_L$ triplet~\cite{Fajfer:2015ycq, Varzielas:2015iva, Bhattacharya:2016mcc, Buttazzo:2017ixm, Barbieri:2015yvd, Kumar:2018kmr, deMedeirosVarzielas:2019lgb, Bernigaud:2019bfy} can explain data very well via a purely left-handed current.
\smallskip

From the discussion above it is clear that there are several options for a combined explanation of the flavor anomalies with LQs. Here we will consider the singlet-triplet model introduced in Refs.~\cite{Crivellin:2017zlb,Marzocca:2018wcf} which was also studied in the context of Dark Matter~\cite{Choi:2018stw}. Within this model, a combined explanation can be possible since $\Phi_1$ can account for the anomaly in $a_\mu$ and affects \bctaunu~transitions while $\Phi_3$ can explain $b\to s\ell^+\ell^-$ data and enters $b\to c\tau\nu$ processes. Furthermore, their combined effects in $b\to s\nu\bar{\nu}$ processes can be destructive, relieving the bounds. However, in order to perform a complete phenomenological analysis, an inclusion of all relevant loop effects is necessary. We will compute these effects and extend the analysis of Ref.~\cite{Crivellin:2017zlb}, allowing for couplings of $\Phi_1$ to right-handed fermions. 
\smallskip

The outline of the article is as follows: In the next section we will define our setup. The conventions for the various observables as well as the results of the matching, taking into account the relevant loop effects, are given in Sec.~\ref{observables} before we perform our phenomenological analysis in Sec.~\ref{pheno} and conclude in Sec.~\ref{conclusions}.
\smallskip

\section{Setup}
\label{setup}

\begin{table}
	\centering
	\begin{tabular}{c|c|c|c|c|c|c|c} 
		& $\Phi_1$ & $\Phi_3$ & $Q$ & $L$ & $\ell$ & $u$&$d$ \\ \hline \hline
		$Y$ & $-2/3$ &$-2/3$ & $1/3$& $-1$ & $-2$ & $4/3$ & $-2/3$ \\ \hline
	\end{tabular}
	\caption{Values of the hypercharges for the LQ and fermion fields.}
	\label{tab:hypercharge}
\end{table}

The scalar LQ singlet-triplet model is obtained by adding a scalar LQ $SU(2)_L$ singlet ($\Phi_1$) and an $SU(2)_L$ triplet ($\Phi_3$), each carrying hypercharge $-2/3$, to the SM particle content. While the couplings to gauge bosons are completely determined by the representations of the LQs under the SM gauge symmetry, their couplings to the SM fermions and the SM Higgs\footnote{Couplings to the Higgs lead to mixing among different LQ representations. Via this mixing LQs are able to generate Majorana masses for neutrinos~\cite{Hirsch:1996qy, Popov:2016fzr, Deppisch:2016qqd, Dorsner:2017wwn, Heeck:2018ntp, Cata:2019wbu, Bigaran:2019bqv, Babu:2019mfe}.} are free parameters of the Lagrangian
\begin{align}
\begin{split}
\mathcal{L}_{\text{LQ}} = \left( \lambda _{fi}^{I}\overline {Q_f^c} i{\tau _2}{L_i}  +\hat{\lambda}_{fi}^I\overline{u^c_f}\ell_i\right)\Phi_1^{I\dagger} 
+ \kappa ^{J}_{fi}\overline {Q_f^c} i{\tau _2}{\left( {\tau  \cdot \Phi _3^{J}} \right)^\dag }{L_i}  + \rho_{IJ} \Phi _1^{I\dag}\left( {{H^\dag }\left( {\tau  \cdot \Phi _3^{J}} \right)H} \right)  \\
 -\!\! \sum_{\{I,I'\}=1}^{N}\! \left(\left(M_{\Phi_{1}}^2\right)_{II'} -\xi^{\Phi_1}_{II'}H^{\dagger}H\right)\Phi^{I\dagger}_{1}\Phi_{1}^{I'} - \!\sum_{\{J,J'\}=1}^{M}\! \left(\left(M_{\Phi_{3}}^2\right)_{JJ'}-\xi^{\Phi_3}_{JJ'}H^{\dagger}H\right)\Phi^{J\dagger}_{3}\Phi_{3}^{J'}+ {\rm{h.c.}}\,.\label{LLQ}
\end{split}
\end{align}
Here, $Q$ ($L$) is the quark (lepton) $SU(2)_L$ doublet and $u$ ($\ell$) the quark (charged lepton) singlet. The superscript $c$ denotes charge conjugation, $f,i$ are flavor indices and $I^{(\prime)},J^{(\prime)}$ denote the number of LQs in a given representation (i.e. $\{I,I'\}=1,...,N$ for $\Phi_1$ and $\{J,J'\}=1,...,M$ for $\Phi_3$)\footnote{In the R-parity violating MSSM this would correspond to the number of generations for the singlet. However, in general $N$ and $M$ do not need to be equal.}. For the hypercharge $Y$ we use the convention \mbox{$Q_{em}=T_3 + Y/2$}, where $Q_{em}$ is the electric charge and $T_3$ the third component of weak isospin (see Tab.~\ref{tab:hypercharge}).
\smallskip

After electroweak symmetry breaking the Higgs acquires its vacuum expectation value $v\approx 174\,$GeV. The last term in \eq{LLQ} then leads to a shift in the bi-linear mass terms of the LQs. However, this shift can be absorbed by defining
\begin{align}
\left(M_{\Phi_{1,3}}^2\right)_{KK'}-v^2\xi^{\Phi_{1,3}}_{KK'}\equiv\left(\tilde{M}_{\Phi_{1,3}}^{2}\right)_{KK'}\,.
\end{align}
Thus, the terms $\xi_{KK'}^{\Phi_{1,3}}$ have (at leading order in perturbation theory) no impact on the low energy flavor phenomenology of the singlet-triplet model but would only enter processes with an external Higgs (or at higher loop level). Furthermore, by unitary rotations of the LQ fields, we can now diagonalize their bi-linear mass terms via unitary rotations $U_{1,2}$:
\begin{align}
\begin{aligned}
U_{1}^{\dagger}\tilde{M}_{\Phi_{1}}^2 U_{1}={\rm{diag}}\Big(\hat{m}_{1}^2,...\,,\hat{m}_{N}^2\Big)&\equiv  m_{\Phi_1}^2\,,\\
U_{3}^{\dagger}\tilde{M}_{\Phi_{3}}^2 U_{3}={\rm{diag}}\Big(\bar{m}_{1}^2,...\,,\bar{m}_{M}^2\Big)&\equiv m_{\Phi_3}^2\,.\label{ULQ}
\end{aligned}
\end{align}
In turn, these rotations lead to an effect in the couplings to the Higgs which can however be absorbed by the definition
\begin{equation}
U_{1}^{\dagger}\rho U_{3}\equiv \hat{\rho}\,.
\end{equation}
The LQ field rotations in \eq{ULQ} have to be applied to their fermionic interactions as well. Here, they can again be absorbed by a redefinition of the couplings 
\begin{align}
\begin{aligned}
\lambda_{fi}^{I}U_{1,KI}^{*}\equiv\lambda_{fi}^{K}\,, &&
\hat{\lambda}_{fi}^{I}U_{1,KI}^{*}\equiv\hat{\lambda}_{fi}^{K}\,, &&\kappa_{fi}^{J}U_{3,KJ}^{*}\equiv\kappa^{K}_{fi}\,.
\end{aligned}
\end{align}
\sloppy{Hence, we are left with diagonal bi-linear mass terms with entries $\left(m_{\Phi_1}^2\right)_{II}$ and $\left(m_{\Phi_3}^2\right)_{JJ}$ and off-diagonal $\Phi_{1}-\Phi_{3}$ mixing governed by $\hat{\rho}_{IJ}$. While the LQs with $Q_{em}=\{2/3,-4/3\}$ are already in their mass eigenstates, we have to diagonalize the resulting full matrix of the $\Phi_{1}-\Phi_{3}$ system with $Q_{em}=-1/3$}
\begin{align}
W^{\dagger}
\begin{pmatrix}
m_{\Phi_{1}}^2 & v^2\hat{\rho}\\
v^2\hat{\rho}^{\dagger}& m_{\Phi_{3}}^2
\end{pmatrix}
W={\rm{diag}}\left(m^2_{1},...\,\,,m^2_{M+N}\right)\,,
\end{align}
with a unitary matrix $W$. Working in the down basis, i.e. in the basis where no CKM elements appear in flavor changing neutral currents of down-type quarks, this leads to the following interaction terms with fermions
\begin{align}
\begin{split}
\mathcal{L}_{\text{LQ}}=&\Gamma _{{u_f}{\ell _i}}^{L,K}\bar u_f^c{P_L}{\ell _i}\Phi _K^{ - 1/3*}+ \Gamma _{{u_f}{\ell _i}}^{R,K}\bar u_f^c{P_R}{\ell _i}\Phi _K^{ - 1/3*} + \Gamma _{{d_f}{\nu _i}}^{L,K}\bar d_f^c{P_L}{\nu _i}\Phi _K^{ - 1/3*}  \,\\
&+ \Gamma _{{u_f}{\nu _i}}^J\bar u_f^c{P_L}{\nu _i}\Phi _J^{2/3*} + \Gamma _{{d_f}{\ell _i}}^J\bar d_f^c{P_L}{\ell _i}\Phi _J^{ - 4/3*}\,,
\end{split}
\end{align}
where the superscripts of the LQ fields refer to their electric charge and
\begin{align}
\begin{split}
\Gamma_{u_{f}\ell_{i}}^{L,K}&=V_{fj}^{*}\left(\lambda_{ji}^{I}W^{*}_{IK}-\kappa^{J}_{ji}W^{*}_{J+N,K}\right)\,,\\
\Gamma_{u_{f}\ell_{i}}^{R,K}&=\hat{\lambda}_{fi}^{I}W_{IK}^{*}\,,\\
\Gamma_{d_{f}\nu_{i}}^{L,K}&=-\lambda_{fi}^{I}W_{IK}^{*}-\kappa_{fi}^{J}W_{J+N,K}^{*}\,,\\
\Gamma_{u_{f}\nu_{i}}^{J}&=\sqrt{2}V_{fj}^{*}\kappa_{ji}^{J}\,,\\
\Gamma_{d_{f}\ell_{i}}^{J}&=-\sqrt{2}\kappa_{fi}^{J}\,.
\end{split}
\end{align}
Recall that the indices take the numbers $I=\{1,...,N\}$, $J=\{1,...,M\}$ and \mbox{$K=\{1,...,M+N\}$}. In the limit with only one generation of each LQ and without mixing we have
\begin{align}
\begin{aligned}
&\Gamma_{u_{f}\ell_{i}}^{L,K}=V_{fj}^{*}\left(\lambda_{ji}\delta_{1K}-\kappa_{ji}\delta_{2K}\right)\,,&& \Gamma_{u_{f}\ell_{i}}^{R,K}=\hat{\lambda}_{fi}\delta_{1K}\,,&&\\
&\Gamma_{d_{f}\nu_{i}}^{L,K}=-\lambda_{fi}\delta_{1K}-\kappa_{fi}\delta_{2K}\,, &&
\Gamma_{u_{f}\nu_{i}}=\sqrt{2}V_{fj}^{*}\kappa_{ji}\,,&&
\Gamma_{d_{f}\ell_{i}}=-\sqrt{2}\kappa_{fi}\,,
\end{aligned}
\end{align}
where the indices 1 and 2 correspond to $\Phi_1$ and $\Phi_3$, respectively.
\smallskip

\section{Processes and Observables}
\label{observables}

In order to illustrate the phenomenology of our model, we will limit ourselves to the case of one LQ singlet $\Phi_1$ and one LQ triplet $\Phi_3$ without mixing among them. Therefore, we will derive the corresponding expressions for the relevant processes in this simplified limit in this section and denote by $M_1$ and $M_3$ the singlet and triplet mass, respectively. In the appendix we will provide the most general expressions for the Wilson coefficients allowing for an arbitrary number of LQs and include mixing among them.
\smallskip

Let us now study the various classes of processes. For each class, we will first define the effective Hamiltonians governing these processes and perform the matching of the model on them. Then we discuss the relation of the Wilson coefficients to observables and review the related available experimental information.
\smallskip

\begin{boldmath}
\subsection{$d d\ell\ell$ and $d d\gamma$ Processes}
\end{boldmath}

\begin{figure}
	\centering
	\begin{overpic}[scale=.5,,tics=10]
		{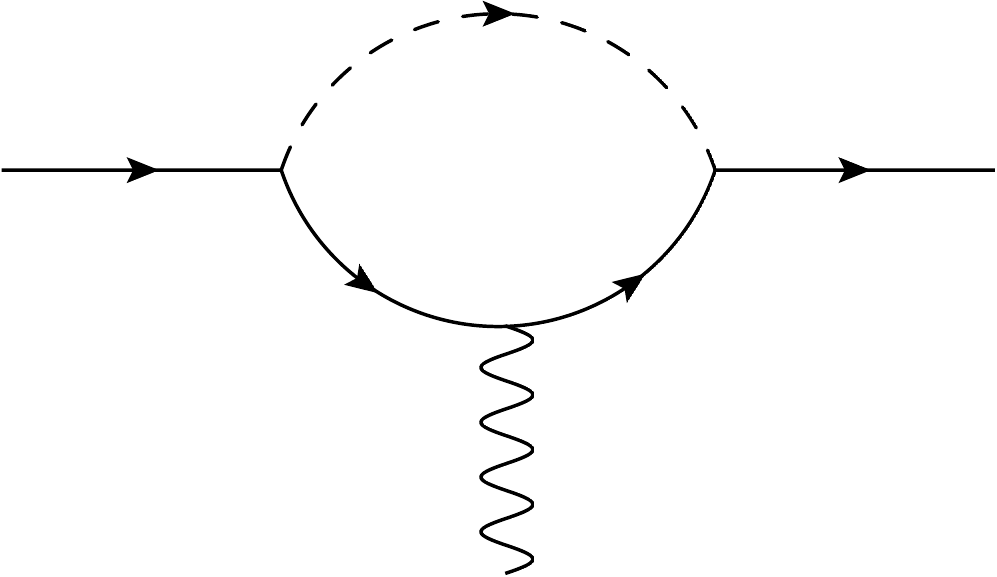}
		\put(2,44){$b$}
		\put(87,44){$s$}
		\put(57,5){$\gamma$}
		\put(47,47){$\Phi$}
		\put(29,25){$\ell$}
		\put(67,25){$\ell$}
	\end{overpic}
	\hfill
	\begin{overpic}[scale=.5,,tics=10]
		{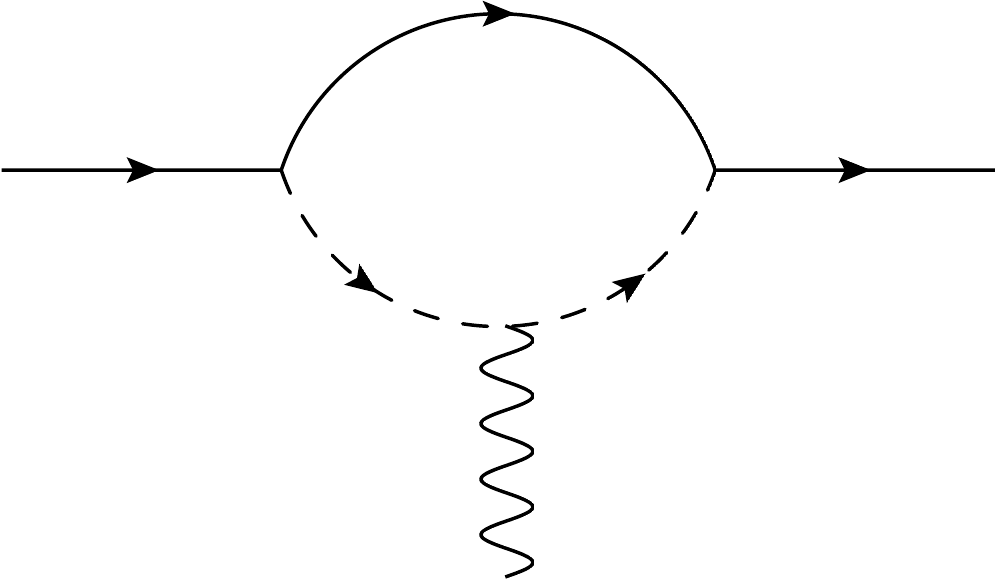}
		\put(2,44){$b$}
		\put(87,44){$s$}
		\put(57,5){$\gamma,g$}
		\put(46,47){$\ell,\nu$}
		\put(28,22){$\Phi$}
		\put(66,22){$\Phi$}
	\end{overpic}
	\hfill
	\begin{overpic}[scale=.46,,tics=10]
		{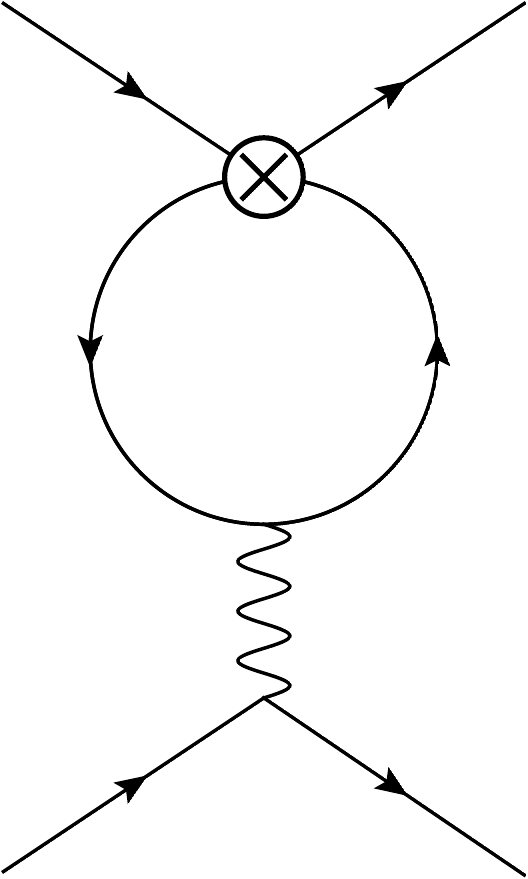}
		\put(8,97){$b$}
		\put(48,97){$s$}
		\put(37,28){$\gamma^*$}
		\put(41,58){$\tau$}
		\put(14,58){$\tau$}
		\put(52,8){$\ell$}
		\put(4,8){$\ell$}
	\end{overpic}
	\caption{Feynman diagrams in our LQ singlet-triplet model generating contributions to $b\to s\gamma$ and $b\to s\ell^{+}\ell^{-}$ at the 1-loop level. The left two diagrams show the matching contribution to the (chromo) magnetic operator. The diagram on the right, with an off-shell photon, is generating the mixing of ${\mathcal O}_9^{\tau\tau}$ into ${\mathcal O}_9^{\ell\ell}$.}
	\label{fig:diagramm_bsll_EFT}
\end{figure}
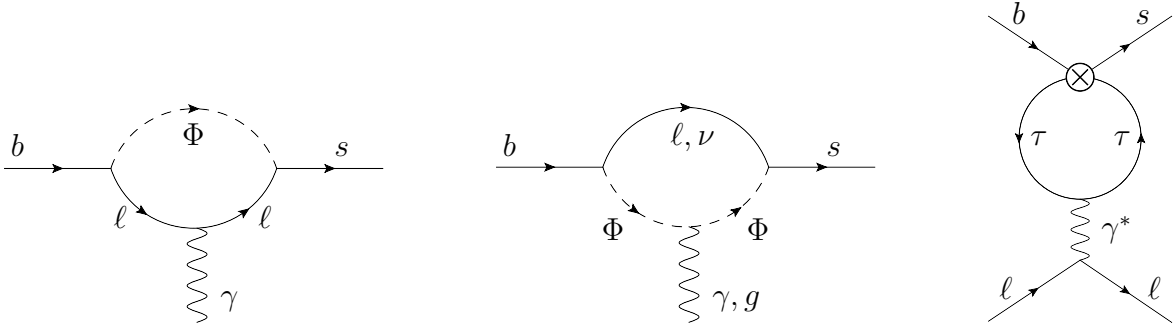

To describe $d_k\to d_j\ell_f^-\ell_i^+$ transitions, we use the effective Hamiltonian
\begin{align}
\begin{split}
\mathcal{H}_{{\rm eff}}^{dd\ell\ell}&=- \dfrac{ 4 G_F }{\sqrt 2}V_{td_k}V_{td_j}^{*} \left[ \sum_{A=7,8}  C^{jk}_A\,\mathcal{O}^{jk}_A + \sum_{A =9,10} C_{A,jk}^{fi} \mathcal{O}_{A,jk}^{fi}\, \right],\\
{\mathcal{O}_{7(8)}^{jk}} &=\dfrac{e (g_s)}{16\pi^2}m_k[\bar d_j{\sigma^{\mu\nu} } (T^a) P_R d_k]F_{\mu\nu}(G^a_{\mu\nu})\,,\\
{\mathcal{O}_{9,jk}^{fi}} &=\dfrac{\alpha }{4\pi}[\bar d_j{\gamma ^\mu } P_L d_k]\,[\bar\ell_f{\gamma _\mu }\ell_i] \,,\\
{\mathcal{O}_{10,jk}^{fi}} &=\dfrac{\alpha }{4\pi}[\bar d_j{\gamma ^\mu } P_L d_k]\,[\bar\ell_f{\gamma _\mu }\gamma_5\ell_i] \,,
\end{split}
\label{eq:effHam}
\end{align}
and define the covariant derivate as 
\begin{equation}
D_\mu=\partial_\mu+ieQ A_\mu +i g_s G^a_\mu T^a\,.
\end{equation}
At tree level, the only matching contribution to ${C}_{9,jk}^{fi}$ and ${C}_{10,jk}^{fi}$ stems from $\Phi_3$
\begin{align}
C_{9,jk}^{fi}=-C_{10,jk}^{fi}=&\frac{{\sqrt 2}} {{2{G_F}{V_{td_k}}V_{td_j}^*}}\frac{\pi }{\alpha }\frac{\kappa_{ki}\kappa_{jf}^{*}}{{{M_{3}^2}}}\,.
\label{eq:C_9C_10}
\end{align}
As in any model, the Wilson coefficients of the (chromo) magnetic operator can only be generated at the loop level. The left two diagrams in Fig.~\ref{fig:diagramm_bsll_EFT} (given for concreteness for $b\to s$ transitions) with on-shell photon and gluons result in
\begin{align}
\begin{split}
C_{7}^{jk}(\mu_{\text{LQ}})&=\frac{-\sqrt{2}}{4G_F V_{td_k}V_{td_j}^{*}}\frac{1}{24}\Bigg(\frac{1}{3}\frac{\lambda_{ki}\lambda_{ji}^{*}}{M_{1}^2} +7\frac{\kappa_{ki}\kappa_{ji}^{*}}{M_{3}^2}\Bigg)\,,\\
C_{8}^{jk}(\mu_{\text{LQ}})&=\frac{\sqrt{2}}{4G_F V_{td_k}V_{td_j}^{*}}\frac{1}{24}\Bigg(\frac{\lambda_{ki}\lambda_{ji}^{*}}{M_{1}^2} +3\frac{\kappa_{ki}\kappa_{ji}^{*}}{M_{3}^2}\Bigg)\,,
\end{split}
\end{align}
at the matching scale $\mu_{\rm LQ}$.
\smallskip

Concerning the QCD evolution of these coefficients, $O_{8}$ mixes into $O_{7}$ at $\mathcal{O}(\alpha_{s})$, yielding the relation~\cite{Borzumati:1998tg, Borzumati:1999qt}
\begin{align}
\begin{pmatrix} C_{7}(\mu_l)\\C_{8}(\mu_l) \end{pmatrix} =\hat{U}^{f}(\mu_l,\mu_h)
\begin{pmatrix} C_{7}(\mu_h)\\ C_{8}(\mu_h) \end{pmatrix}\,,
\end{align}
with 
\begin{align}
\hat{U}^{f}(\mu_l,\mu_h)=
\begin{pmatrix}
\eta^{\frac{16}{33-2f}} & \frac{8}{3}\Big(\eta^{\frac{14}{33-2f}}-\eta^{\frac{16}{33-2f}}\Big)\\
0 & \eta^{\frac{14}{33-2f}}
\end{pmatrix} \ .
\end{align}
Here, $f$ denotes the number of active quark flavors, $\mu_{h(l)}$ refers to the high (low) energy scale and
\begin{align}
\eta=\frac{\alpha_{s}(\mu_h)}{\alpha_{s}(\mu_l)}\,,
\end{align}
where $\alpha_{s}$ needs to be evaluated with the number of active flavors at a given scale as well. 
\smallskip

Even though $b\to s\ell^+\ell^-$ can be induced at tree level in our model, there are still scenarios in which loop effects are phenomenologically important. As pointed out in Ref.~\cite{Crivellin:2018yvo}, the large couplings to tau leptons, needed to explain $b\to c\tau\nu$ data, also lead to huge Wilson coefficients $C_{9,sb}^{\tau\tau}=-C_{10,sb}^{\tau\tau}$. In turn, $\mathcal{O}_{9,sb}^{\tau\tau}$ mixes into $\mathcal{O}_{9,sb}^{\ell\ell}$ via the off-shell photon penguin~\cite{Bobeth:2014rda}, shown in the right diagram of Fig.~\ref{fig:diagramm_bsll_EFT}. In our UV complete model, we cannot only calculate this mixing, but also the finite part of the effect, contained in the matching contribution
\begin{align}
C_{9,jk}^{\ell\ell}(\mu_\text{LQ})=\frac{\sqrt{2}}{216 G_{F}V_{td_{k}}V_{td_{j}}^{*}} \Bigg[\frac{\lambda_{kl}\lambda_{jl}^{*}}{M_1^2}  +3\frac{\kappa_{kl}\kappa_{jl}^{*}}{M_{3}^2} \left(19+12\log\left(\frac{\mu_\text{LQ}^2}{M_{3}^2}\right)\right)\Bigg]\,.
\label{eq:C9_LFU}
\end{align}
This means that even if couplings to light leptons are absent at tree level, they are generated via loop effects in the presence of tau couplings. Since we will mainly focus on $b\to s$ transitions, we shorten our notation in the following and write $C_{7(8)}^{sb}\equiv C_{7(8)}$, $C_{9(10),sb}^{fi}\equiv C_{9(10)}^{fi}$. The logarithm involving $\mu_{\rm LQ}$ in \eq{eq:C9_LFU} originates from the fact that the right-diagram in Fig.~\ref{fig:diagramm_bsll_EFT} is divergent. To get rid of this dependence one has to solve the RGE governing the mixing between $\mathcal{O}_9^{ii}$ with different lepton flavors:
\begin{align}
\mu \frac{\partial \, C_{9}^{ii}(\mu)}{\partial \mu} = \gamma \, C_{9}^{ff}(\mu) \qquad (f \neq i)
\end{align}
with $\gamma=\frac{2\alpha}{3 \pi}$. Here, we do not take into account the running of $\alpha$ and do not consider the running of $C_{9}^{ii}$ (i.e. just the mixing of $O_{9}^{ii}$ into $C_{9}^{jj}$ with $i\ne j$). This then has the solution
\begin{align}
C_{9}^{ii}(\mu)=C_9^{ii}(\mu_{\text{LQ}})+\gamma \log \left( \frac{\mu}{\mu_{\text{LQ}}}\right) C_9^{ff} \qquad (f \neq i)\,.
\end{align}
For $B$ meson decays, this amounts to replacing the high scale $\mu_\text{LQ}$ in \eq{eq:C9_LFU} by the low scale of the processes $\mu_b$. In addition, at the $B$ meson scale, $\mathcal{O}_{9}^{\tau\tau}$ gives a $q^2$ dependent contribution to $C_{9,\rm eff}^{\ell\ell}$, which however is numerically  small~\cite{Bobeth:2014rda} and currently not accessible with the SM independent fit. However, there are intriguing prospects that with improved future data this effect could be distinguished from the $q^2$-independent $C_9$ effect~\cite{Cornella:2020aoq}.
\smallskip

QCD corrections to the matching of scalar LQs for semi-leptonic processes (both charged and neutral current) can be taken into account by applying the following shifts to the Wilson coefficients of vector ($V$), scalar ($S$) and tensor ($T$) operators~\cite{Aebischer:2018acj}
\begin{align}
\begin{split}
C_V &\to C_V \left(1 + \frac{\alpha_s}{4 \pi} C_F \left( 3 l_\mu +\frac{17}{2} \right) \right) \ , \\
C_S & \to C_S \left(1 + \frac{3 \alpha_s}{2 \pi} C_F \right) \ , \\
C_T & \to C_T \left( 1 + \frac{\alpha_s}{\pi} C_F \left(l_\mu + 2  \right) \right) \ ,
\end{split}
\label{eq:matching_corrections}
\end{align} 
with $l_\mu = \log\left(\mu^2/M^2 \right)$ (where $M$ can be either $M_1$ or $M_3$) and $C_F = 4/3$ as the color factor. Since QCD is insensitive to flavor, electric charge and chirality, these corrections can be applied in a straightforward way to all other semi-leptonic processes, particularly to $b\to s\nu\bar{\nu}$ and $b\to c\tau\nu$.
\smallskip\\

\subsubsection*{Observables}
As mentioned in the introduction, a main motivation for this anlysis is the explanation of the hints for NP in $b\to s\ell^+\ell^-$ data. In order to resolve this discrepancy between SM and experiment, an $\mathcal{O}(20 \%)$ effect to $C_{9,10}$ is required compared to the SM contribution which is given by~\cite{Bobeth:1999mk, Huber:2005ig}
\begin{align}
C_{9}^{\mathrm{SM}}(4.8~{\mathrm{GeV}})= 4.07 \,,&&
C_{10}^{\mathrm{SM}}(4.8~{\mathrm{GeV}})=-4.31\,.
\end{align}
In a global fit one finds preference for scenarios like $C_9^{\mu \mu}=-C_{10}^{\mu \mu}$, as generated in our model at tree level. However, a $C_9^{\mu \mu}=-C_{10}^{\mu \mu}$ effect complemented by a LFU one in $C_9^{\ell\ell}$ gives an even better fit to data~\cite{Alguero:2018nvb,Alguero:2019ptt}. As we will see, this is exactly the pattern that arises in our model, taking into account the loop effects discussed above.
\smallskip

For $b\to s\tau^+\tau^-$ transitions we have on the experimental side~\cite{Aaij:2017xqt}
\begin{eqnarray}
{\rm{Br}}{\left[{B_s \to {\tau ^ + }{\tau ^ - }} \right]_{{\rm{exp}}}} \le 6.8 \times {10^{ - 3}}\quad(95\%\,\mathrm{C.L.})\,.
\end{eqnarray}
For $B_d\to\tau^+\tau^-$ there is a (unpublished) measurement of BELLE~\cite{Ziegler:2016mtm} and an upper limit of LHCb~\cite{Aaij:2017xqt}
\begin{align}
\begin{split}
{\rm{Br}}\left[{B_d \to \tau^+\tau^-}\right]^{\rm{BELLE}}_{ \rm{exp} } &=  \left( 4.39{}^{+ 0.80}_{-0.83} \pm 0.45\right) \times 10^{-3}\,,\\
{\rm{Br}}\left[B_d \to \tau^+\tau^-\right]^{\rm{LHCb}}_{\rm{exp}} &\le  2.1\times {10^{ - 3}} \quad(95\%\,\mathrm{C.L.})\,.
\end{split}
\end{align}
These measurements are compatible at the $2\sigma$ level. The SM predictions read~\cite{Bobeth:2013uxa,Bobeth:2014tza}
\begin{eqnarray}
\begin{aligned}
{\rm{Br}}{\left[{B_{s} \to {\tau ^ + }{\tau ^ - }} \right]_{{\rm{SM}}}} = \left( {7.73 \pm 0.49} \right) \times {10^{ - 7}}\,,\\
{\rm{Br}}{\left[{B_{d} \to {\tau ^ + }{\tau ^ - }} \right]_{{\rm{SM}}}} = \left( 2.22\pm 0.19 \right)  \times {10^{ - 8}}\,.
\end{aligned}
\end{eqnarray}
In our model we find
\begin{equation}
\dfrac{{\rm{Br}}\left[{{B_{s}} \to {\tau ^ + }{\tau ^ - }} \right]}{{\rm{Br}}{\left[{{B_{s}} \to {\tau ^ + }{\tau ^ - }} \right]_{\rm
		SM}}}=\left|{ {1 + \frac{{C_{10}^{\tau\tau}}}{{C_{10}^{\rm SM}}}}}\right|^2\,,
	\label{Bstautau}
\end{equation}
and the analogous expression for $b\to d$ transitions.  Also the branching ratios of semi-leptonic $b\to s\tau^{+}\tau^{-}$ processes can be expressed in terms of NP Wilson coefficients~\cite{Capdevila:2017iqn}
\begin{align}
\begin{aligned}
{\mathrm{Br}}\big[B_{(s)}\to X&\tau^{+}\tau^{-}\big]\times 10^{7}= A_{0}^{X}+A^{X}_{1}C_{9}^{\tau\tau}+A^{X}_{2}C_{10}^{\tau\tau}+A^{X}_{3}C_{9}^{\prime\tau\tau}+A^{X}_{4}C_{10}^{\prime\tau\tau}+A^{X}_{5}(C_{9}^{\tau\tau})^2\\
&+A^{X}_{6}(C_{10}^{\tau\tau})^2+A^{X}_{7}(C_{9}^{\prime\tau\tau})^2+A^{X}_{8}(C_{10}^{\prime\tau\tau})^2+A_{9}^{X}C_{9}^{\tau\tau}C_{9}^{\prime\tau\tau}+A_{10}^{X}C_{10}^{\tau\tau}C_{10}^{\prime\tau\tau}\,.
\end{aligned}
\label{eq:BXtautau_prediction}
\end{align}
These branching ratios together with the corresponding coefficients are shown in Tab.~\ref{tab:BXtautau_prediction}.

\begin{table}
	\centering
	\renewcommand{\arraystretch}{1.15}
	\resizebox{\textwidth}{!}
	{
		\begin{tabular}{|c|c|c|ccccc|}
			\hline
			$X$ & $q^2[\text{GeV}^2]$ & $A_{0}$ & $A_{1}$ & $A_{2}$ & $A_{3}$ & $A_{4}$ & $A_{5}$\\
			\hline
			$K$ & $[15,22]$ & $1.20\pm 0.12$& $0.15\pm 0.02$& $-0.42\pm 0.04$ & $0.15\pm 0.01$ & $0.15\pm 0.04$ & $0.02$\\
			$K^{*}$ & $[15,19]$ & $0.98\pm0.09$ & $0.38\pm 0.03$ & $-0.14\pm 0.01$ & $-0.30\pm 0.03$ & $0.12$ & $0.05$ \\
			$\phi$ & $[15,18.8]$ & $0.86\pm 0.06$ & $0.34\pm 0.02$ & $-0.11$ & $-0.28\pm 0.02$ & $0.10$ & $0.05$\\
			\hline
			\multicolumn{3}{c|}{} & $A_{6}$ & $A_{7}$ & $A_{8}$ & $A_{9}$ & $A_{10}$ \\
			\cline{4-8}
			\multicolumn{3}{c|}{} & $0.05\pm 0.01$ & $0.02$ & $0.05\pm 0.01$ & $0.04$& $0.10\pm 0.01$ \\
			\multicolumn{3}{c|}{} & $0.02$& $0.05\pm 0.01$ & $0.02\pm 0.01$ & $-0.08\pm 0.01$ & $-0.03$ \\
			\multicolumn{3}{c|}{} & $0.01$ & $0.05$ & $0.01\pm 0.02$ & $-0.08$ & $-0.02$\\
			\cline{4-8}
		\end{tabular}
	}
	\caption{Numerical values for the coefficients given in \eq{eq:BXtautau_prediction} for the different decay modes involving  $b\to s\tau^{+}\tau^{-}$ transitions together with the corresponding $q^2$ ranges.}
	\label{tab:BXtautau_prediction}
\end{table}

Turning to $b\to s\tau\mu$ transitions, we have~\cite{Crivellin:2015era}
\begin{align}
{\rm{Br}}[B \to K \tau^\pm \mu^\mp] =  10^{-9} \left[9.6 \left(\left|C_{9}^{\mu\tau}\right|^2 + \left|C_{9}^{\tau\mu}\right|^2 \right) + 10 \left(\left|C_{10}^{\mu\tau}\right|^2 + \left|C_{10}^{\tau\mu}\right|^2 \right) \right] \,,
\end{align}
and
\begin{align}
\begin{aligned}
{\mathrm{Br}}\big[&\bar{B}_{s}\to\ell_{f}^{-}\ell_{i}^{+}\big]=\frac{G_F^2 \alpha ^2}{64\pi^3} \big|V_{tb}V_{ts}^{*}\big|^2f_{B_s}^2\tau_{B_s}m_{B_s}(m_{\ell_i}+m_{\ell_f})^2\eta(x_i,x_f)\\
&\times\Bigg[\Big|C_{10}^{fi}-C_{10}^{\prime fi}\Big|^2\big(1-(x_{i}-x_{f})^2\big)+\bigg|\frac{m_{\ell_i}-m_{\ell_f}}{m_{\ell_i}+m_{\ell_f}}\left(C_{9}^{fi}-C_{9}^{\prime fi} \right) \bigg|^2\big(1-(x_{i}+x_{f})^2\big)\Bigg]\,,
\end{aligned}
\label{eq:letptonic_Bmeson_decay}
\end{align} 
with $x_k=m_{\ell_k}/m_{B_s}$ and
\begin{align}
\eta(x,y)=\sqrt{1-2(x+y)+(x-y)^2}\,.
\end{align}
We neglected the contributions of (pseudo-)scalar operators, since they do not appear in our model. The relevant experimental limits are~\cite{Lees:2012zz,Aaij:2019okb}
\begin{align}
\begin{aligned}
{\rm{Br}}[B \to K \tau^\pm \mu^\mp]_{\rm{exp}} \leq 4.8 \times 10^{-5} \ ,\\
{\rm{Br}}[B_s \to \tau^\pm \mu^\mp]_{\rm{exp}} \leq 4.2 \times 10^{-5}\,.
\end{aligned}
\end{align}

$\bar d d\bar \ell\ell$ operators contribute to $\tau \to \phi \mu$ as well. This gives relevant constraints on the parameter space of our model. We use the result of Ref.~\cite{Bhattacharya:2016mcc} and obtain
\begin{align}
{\rm{Br}}\left[\tau\to\phi\mu\right]=\frac{f_{\phi}^{2}m_{\tau}^{3}\tau_{\tau}}{128\pi}\frac{\left|\kappa_{22}\kappa_{23}^{*}\right|^2}{M_{3}^4}\left(\!1-\frac{m_{\phi}^{2}}{m_{\tau}^{2}}\right)^{\!2}\!\!\left(\!1+2\frac{m_{\phi}^{2}}{m_{\tau}^{2}}\right)\,,
\end{align}
which has to be compared to the current experimental limit of~\cite{Miyazaki:2011xe}
\begin{equation}
{\rm{Br}}\left[\tau\to\phi\mu\right]<8.4\times10^{-8} \quad(90\%\,\mathrm{C.L.})\,.
\end{equation}
\smallskip

\begin{boldmath}
\subsection{$dd\nu\nu$ Processes}
\end{boldmath}

To describe $d_k\to d_j \nu_{f}\bar{\nu}_i$ processes we use the Hamiltonian
\begin{align}
\begin{split}
\mathcal{H}_{{\rm eff}}^{dd\nu\nu}&=-\dfrac{4G_F}{\sqrt{2}}V_{td_k}V_{td_j}^{*}\left(C^{fi}_{L,jk}\mathcal{O}_{L,jk}^{fi}+C^{fi}_{R,jk}\mathcal{O}_{R,jk}^{fi}\right)\,,\\
\mathcal{O}_{L(R),jk}^{fi}&=\frac{\alpha}{4\pi}\left[\bar{d}_{j}\gamma^{\mu}P_{L(R)}d_{k}\right] \left[\bar{\nu}_{f}\gamma_{\mu}\left(1-\gamma_5\right)\nu_i\right]\,.
\end{split}
\label{eq:Heff_ddnunu}
\end{align}
At tree level we find contributions from $\Phi_1$ and $\Phi_3$ resulting in
\begin{align}
C_{L,jk}^{fi}&=\frac{\sqrt{2}}{4G_{F}V_{td_{k}}V_{td_{j}}^{*}}\frac{\pi}{\alpha}\left[\frac{\lambda_{ki}\lambda_{jf}^{*}}{M_1^2}+\frac{\kappa_{ki}^{}\kappa_{jf}^{*}}{M_3^2}\right]\,.
\label{eq:tree-level_bsvv}
\end{align}
Since these processes are generated at tree level, we do not need to calculate loop effects, which would only amount to numerically small corrections. Again, we simplify the notation for $b\to s$ transitions, writing $C_{L,sb}^{fi}\equiv C_{L}^{fi}$. The QCD matching corrections are given in \eq{eq:matching_corrections} and there is no QCD evolution of these operators.
\smallskip\\

\subsubsection*{Observables}

For $B\to K^{(*)}\nu\bar{\nu}$ we follow Ref.~\cite{Buras:2014fpa} and use $C_{L}^{{\rm SM}}\approx-1.47/s_w^2$. The branching ratios normalized to the SM read
\begin{equation}
{R_{K^{(*)}}^{\nu\bar{\nu}}} = 
\frac{1}{3}\sum\limits_{f,i=1}^3 \dfrac{ \big|C_{L}^{{\rm SM}}\delta_{fi}+{C_{L}^{fi}}\big|^2}{\big|{C_{L}^{{\rm SM}}}\big|^2} \,.
\end{equation}
This has to be compared to the current experimental limits ${R_K^{\nu\bar{\nu}}} < 3.9$ and ${R_{{K^*}}^{\nu\bar{\nu}}} < 2.7$~\cite{Grygier:2017tzo} (both at $90\%\,\mathrm{C.L.}$). The expected BELLE II sensitivity for $B\to K^{(*)}\nu\bar{\nu}$ is 30\% of the SM branching ratio~\cite{Abe:2010gxa}.
\smallskip\\

\begin{boldmath}
\subsection{$du\ell\nu$ Processes}
\end{boldmath}

For the charged current semi-leptonic processes we define the effective Hamiltonian as
\begin{align}
\begin{split}
{\mathcal{H}_{{\rm{eff}}}^{du\ell\nu}} = \dfrac{{4{G_F}}}{{\sqrt 2 }}{V_{jk}}\big(&C_{VL,jk}^{fi}\left[ {\bar u_{j}{\gamma ^\mu }{P_L}d_k} \right]\left[ {{{\bar \ell }_f}{\gamma _\mu }{P_L}{\nu _i}} \right]+C_{SL,jk}^{fi}\left[ {\bar u_{j}{P_L}d_k} \right]\left[ {{{\bar \ell }_f}{P_L}{\nu _i}} \right]\\
&+C_{TL,jk}^{fi}\left[ {\bar u_{j}\sigma^{\mu\nu}{P_L}d_k} \right]\left[ {{{\bar \ell }_f}\sigma_{\mu\nu}{\nu _i}} \right]\big)\,,
\end{split}
\label{eq:Heff_dulnu}
\end{align}
where in the SM $C_{VL}^{\rm SM}=1$. The contribution of our model to the SM Wilson coefficient from $\Phi_1$ and $\Phi_3$ is given by
\begin{align}
C_{VL,jk}^{fi}&=\frac{-\sqrt{2}}{8G_{F}V_{jk}}\left[-\frac{V_{jl}\lambda_{lf}^{*}\lambda_{ki}}{M_1^2}+\frac{V_{jl}\kappa_{lf}^{*}\kappa_{ki}}{M_3^2}\right]\,,
\end{align}
while scalar and tensor operators are generated by $\Phi_1$ only
\begin{align}
C_{SL,jk}^{fi}=-4C_{TL,jk}^{fi}&=\frac{-\sqrt{2}}{8G_{F}V_{jk}}\frac{\lambda_{ki}\hat{\lambda}^{*}_{jf}}{M_{1}^2}\,.
\end{align}
Since we are mainly interested in $b\to c$ transitions, we abbreviate
\begin{align}
C_{VL,cb}^{fi}\equiv C_{VL}^{fi}\,,&& C_{SL,cb}^{fi}\equiv C_{SL}^{fi}\,, && C_{TL,cb}^{fi}\equiv C_{TL}^{fi}\,.
\end{align}
Again, the QCD matching corrections are given in \eq{eq:matching_corrections}. 
We also include the 2-loop QCD and the 1-loop EW RGE. Using the results of Ref.~\cite{Gonzalez-Alonso:2017iyc}, we have
\begin{align}
\begin{split}
C_{VL}^{fi}(\mu_{b})&=C_{VL}^{fi}(1\,{\mathrm{TeV}})\,,\\
C_{SR}^{fi}(\mu_{b})&=1.737~C_{SR}^{fi}(1\,{\mathrm{TeV}})\,,\\
\begin{pmatrix}C_{SL}^{fi}(\mu_b)\\ C_{TL}^{fi}(\mu_b)\end{pmatrix}&=
\begin{pmatrix}1.752 & -0.287\\ -0.004 & 0.842\end{pmatrix}
\begin{pmatrix}C_{SL}^{fi}(1\,{\mathrm{TeV}})\\ C_{TL}^{fi}(1\,{\mathrm{TeV}})\end{pmatrix}\,.
\end{split}
\end{align}
\smallskip\\

\subsubsection*{Observables}
With these conventions, the ratios $R(D^{(*)})$ are given by~\cite{Blanke:2018yud}
\begin{align}
\begin{aligned}
\frac{{R(D)}}{{{R(D)_{{\rm{SM}}}}}}& \simeq \big|1 + C_{VL}^{\tau\tau}\big|^2 + 1.54{\mkern 1mu} \text{Re} \Big[\big(1 + C_{VL}^{\tau\tau}\big)C_{SL}^{\tau\tau*}\Big]   + 1.09\big|C_{SL}^{\tau\tau}\big|^2\\
&+ 1.04{\mkern 1mu} \text{Re} \Big[\big(1 + C_{VL}^{\tau\tau}\big)C_{TL}^{\tau\tau*}\Big] + 0.75|C_{TL}^{\tau\tau}{|^2}{\mkern 1mu} ,\\ 
\frac{{R({D^*})}}{{{R({D^*})_{{\rm{SM}}}}}} &\simeq \big|1 + C_{VL}^{\tau\tau}\big| - 0.13{\mkern 1mu} \text{Re} \Big[\big(1 + C_{VL}^{\tau\tau}\big)C_{SL}^{\tau\tau*}\Big] + 0.05\big|C_{SL}^{\tau\tau}\big|^2
\\ & - 5.0{\mkern 1mu} \text{Re} \Big[\big(1 + C_{VL}^{\tau\tau}\big)C_{TL}^{\tau\tau*}\Big]+16.27\big|C_{TL}^{\tau\tau}\big|^2\,,
\end{aligned}
\end{align}
in terms of the Wilson coefficients given at the $B$ meson scale. Furthermore, the branching ratio of $B_c \to \tau \nu$ reads~\cite{Blanke:2018yud, Iguro:2018vqb}
\begin{align}
\text{Br}[B_c \to \tau \nu] =0.02\bigg(\frac{f_{B_c}}{0.43\,\mathrm{GeV}}\bigg)^{\!2}\Big|1+C_{VL}^{\tau\tau}+4.3\big(C_{SR}^{\tau\tau}-C_{SL}^{\tau\tau}\big)\Big|^2\,.
\end{align}
In this work we use the most stringent limit of Ref.~\cite{Akeroyd:2017mhr}
\begin{align}
\text{Br}[B_c \to \tau \nu] \leq 0.1 \ ,
\end{align}
even though this bound might be too restrictive (see Refs.~\cite{Akeroyd:2017mhr, Blanke:2019qrx} for theoretical discussions). However, we will see that even this limit does not constrain our model significantly.
\smallskip

A further constraint comes from the determination of the CKM element $V_{cb}$ when comparing electron and muon final states. Here Ref.~\cite{Jung:2018lfu} finds that
\begin{align}
\frac{\tilde{V}_{cb}^e}{\tilde{V}_{cb}^\mu} = 1.011 \pm 0.012 \ ,
\end{align}
where
\begin{align}
\tilde{V}_{cb}^\ell = V_{cb} \bigg[ \big|1+C_{VL}^{\ell \ell}\big|^2 + \sum_{\ell\neq\ell^{\prime}} \big|C_{VL}^{\ell\ell^{\prime}}\big|^2 \bigg]^{1/2} \ .
\end{align}
For observables including first and second generation quarks such as \mbox{$\tau\to \pi\nu$}, \mbox{$K\to\mu\nu/K\to e\nu$} or $D$ decays, the Wilson coefficients can be applied using appropriate indices. The corresponding formulas and analyses can be found e.g. in Refs.~\cite{deBoer:2015boa, Mandal:2019gff}. 
\smallskip

\begin{boldmath}
\subsection{$\Delta F=2$ processes}
\end{boldmath}

Dealing with $\Delta F=2$ processes, concretely $B_s-\bar B_s$ mixing, we use the effective Hamiltonian
\begin{align}
\mathcal{H}^{B\bar{B}}_{\rm{eff}}=C_{1}\left[\bar{s}_{\alpha}\gamma_{\mu}P_{L}b_{\alpha}\right]\left[\bar{s}_{\beta}\gamma^{\mu}P_{L}b_{\beta}\right]\,.
\label{eq:Heff_Bs-mixing}
\end{align}
In our model we obtain
\begin{align}
\begin{aligned}
{C_1} = \frac{ - 1}{128{\pi ^2}}\bigg(& \lambda _{2i}^{*}\lambda_{3j}\lambda_{2j}^{*}\lambda_{3i}C_0\left( {0,{M_1^2},{M_1^2}} \right)  + 5 \kappa _{2i}^{*}\kappa_{3j}\kappa _{2j}^{*}\kappa _{3i}C_0\left( {0,{M_{3}^2},{M_{3}^2}} \right)\\
&+ 2\lambda _{2j}^{*}\lambda_{3i}\kappa _{2i}^{*}\kappa _{3j}C_0\left( {0,{M_1^2},{M_{3}^2}} \right) \bigg)\,
\end{aligned}
\label{BsMixing}
\end{align}
at the high scale $\mu_{\text{LQ}}$. Here the first term originates only from $\Phi_1$ and the second one only from $\Phi_3$. The last term originates from a box diagram where both LQ representations contribute. One of the corresponding Feynman diagram is shown in Fig. \ref{fig:Bsmixing}. The formula for $B_d$ and Kaon mixing follow trivially. We can write the mass difference $\Delta m_{B_s}$ (including NP) normalized to the SM one as 
\begin{align}
\dfrac{\Delta m_{B_s}}{{\Delta m_{B_s}^{\rm{SM}}}} =  \left|1 + \frac{C_1}{C_{1}^{\rm{SM}}} \right|\,,
\label{eq:bs_mixing_observable}
\end{align}
with~\cite{Lenz:2010gu}
\begin{align}
C_{1}^{\text{SM}}=2.35\frac{\big(V_{tb}V_{ts}^{*}G_{F}m_{W}\big)^2}{4\pi^2}\,
\end{align}
given at the high scale. Since both the SM and LQ contribute to $C_{1}$, the QCD running down to $\mu_{b}$ is the same for both and therefore cancels in \eq{eq:bs_mixing_observable}, neglecting the evolution from $\mu_\text{LQ}$  to the EW scale.
\smallskip\\

\subsubsection*{Observables}

$B_{s}-\bar{B}_{s}$ mixing has been measured to very good precision~\cite{Bona:2008jn} and the current world average reads~\cite{Tanabashi:2018oca}
\begin{align}
\Delta m_{B_s}^{\exp}=(17.757\pm 0.021)\times 10^{12}\,\text{s}^{-1}\,.
\end{align}
The theoretical prediction suffers strongly from the uncertainties in QCD effects. While Ref.~\cite{Jubb:2016mvq} and Ref.~\cite{Bona:2006sa} fit well to the measurement (with rather large errors)
\begin{align}
\Delta m_{B_s}^{\text{SM}}=(18.3\pm 2.7)\times 10^{12}\,\text{s}^{-1}\,,
\end{align}
Ref.~\cite{DiLuzio:2017fdq} obtains a larger SM value
\begin{figure}[t]
	\centering
	\begin{overpic}[scale=.47,,tics=10]
		{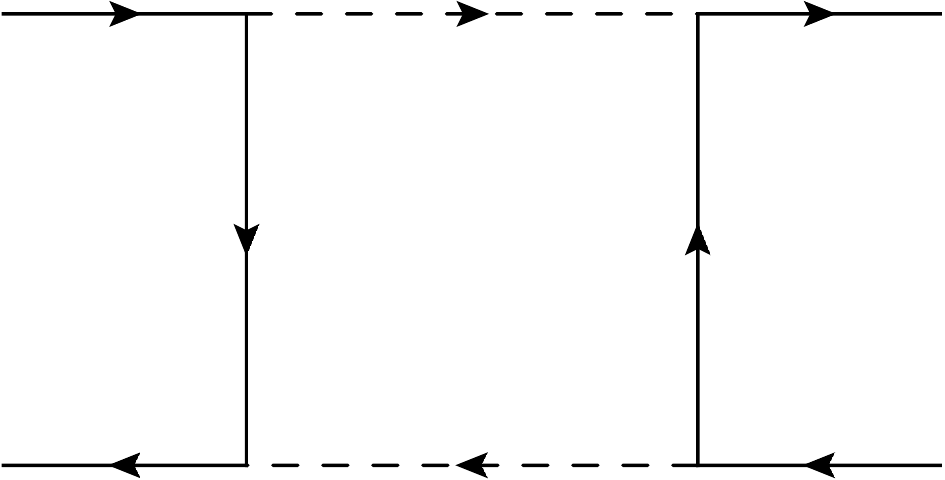}
		\put(5,51){$b$}
		\put(5,4){$s$}
		\put(11,24){$\ell,\nu$}
		\put(78,24){$\ell,\nu$}
		\put(47,40){$\Phi$}
		\put(47,4){$\Phi$}
		\put(92,4){$b$}
		\put(92,51){$s$}
	\end{overpic}
	\hfill
	\begin{overpic}[scale=.47,,tics=10]
		{LQ_on_shell_photon}
		\put(5,44){$\ell_{i}$}
		\put(90,44){$\ell_{f}$}
		\put(57,5){$\gamma$}
		\put(47,48){$\Phi_{1}$}
		\put(27,25){$t^{c}$}
		\put(67,25){$t^{c}$}
	\end{overpic}
	\hfill
	\begin{overpic}[scale=.47,,tics=10]
		{LQ_on_shell_photon2}
		\put(5,44){$\ell_{i}$}
		\put(90,44){$\ell_{f}$}
		\put(57,5){$\gamma$}
		\put(48,48){$t^{c}$}
		\put(27,24){$\Phi_{1}$}
		\put(65,24){$\Phi_{1}$}
	\end{overpic}
	\caption{Left: LQ boxes contributing to $B_{s}-\bar{B}_{s}$ mixing. Middle and right: Loop diagrams induced by $\Phi_{1}$, generating effects in $\ell_{i}\to\ell_{f}\gamma$. In case of a top quark, as depicted, a chirally enhanced term can arise.}
	\label{fig:Bsmixing}
\end{figure}
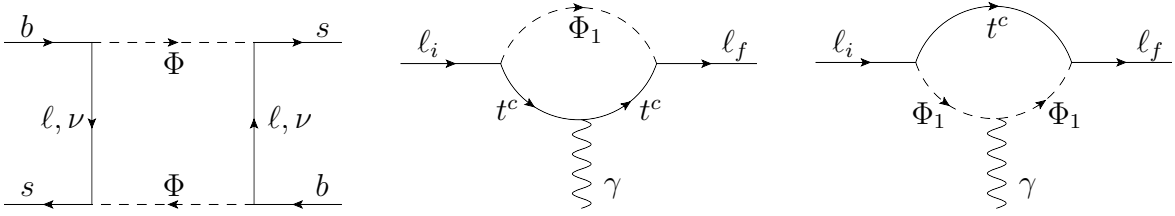
\begin{align}
\Delta m_{B_s}^{\text{SM}}=(20.01\pm 1.25)\times 10^{12}\,\text{s}^{-1}\,.
\end{align}
The bounds on the imaginary part of the Wilson coefficient is even more stringent. In our phenomenological anlysis we will assume real couplings and allow for NP effects of up to 20\% with respect to the SM prediction.
\smallskip

\begin{boldmath}
\subsection{$\ell\ell\gamma$ Processes}
\end{boldmath}

In case of charged lepton interactions with on-shell photons we define
\begin{align}
\mathcal{H}^{\ell\ell\gamma}_{\rm{eff}}=C^{L}_{\ell_{f}\ell_{i}}O^{L}_{\ell_{f}\ell_{i}}+C^{R}_{\ell_{f}\ell_{i}}O^{R}_{\ell_{f}\ell_{i}}\,,
\label{eq:Heff_llgamma}
\end{align}
with
\begin{align}
O_{\ell_{f}\ell_{i}}^{L(R)}=\frac{e}{16\pi^2}\big[\bar{\ell}_{f}\sigma^{\mu\nu}P_{L(R)}\ell_{i}\big]F_{\mu\nu}\,.
\end{align}
We obtain the following matching contribution in case of a top quark in the loop
\begin{align}
\begin{split}
C^{L}_{\ell_{f}\ell_{i}}=-\frac{m_{\ell_{f}}\lambda_{3f}^{*} \lambda_{3i}+m_{\ell_{i}}\hat{\lambda}_{3f}^{*}\hat{\lambda}_{3i}}{8 M_1^2} +\frac{m_{t} \hat{\lambda}_{3f}^{*} V_{3k}^{*}\lambda_{ki}}{4M_1^2}\left(7+4\log\left(\frac{m_t^2}{M_{1}^2}\right)\right)+\frac{3m_{\ell_f}\kappa_{3f}^{*}\kappa_{3i}}{8M_{3}^2}\,
\end{split}
\label{eq:llgamma_CL}
\end{align}
from the Feynman diagram given in Fig.~\ref{fig:Bsmixing} with $N_{c}=3$ already included. Note that we have $C^{R}_{\ell_{f}\ell_{i}}=C^{L*}_{\ell_{i}\ell_{f}}$ due to the hermiticity of the Hamiltonian. Here we quoted explicitly the formula for the top quark, which we integrated out together with the LQ at the scale $M\approx M_1\approx M_3$. In case of light quarks, some comments concerning the use of \eq{eq:llgamma_CL} are in order: in principle, one has to integrate out only the LQ at the scale $M$ but keep the quark as a dynamical degree of freedom. In this way, the matching contribution to $C^L_{\ell_{f}\ell_{i}}$ acquires an infrared divergence, which is cancelled by the corresponding UV divergence of the contribution of the tensor operator\footnote{See Sec.~\ref{sec:app_EDM} for the matching to the $uu\gamma$ and $uu\ell\ell$ operators.}, obtained by integrating out the LQ at tree level. This amounts to a replacement of $m_t$ by $\mu_{\rm LQ}$ in the logarithm in \eq{eq:llgamma_CL}. Now, at the low scale, the solution to the RGE (disregarding QED effects) leads to a replacement of $\mu_{\rm LQ}$ by the scale of the processes, or by the quark mass in case this mass is bigger than the scale. Therefore, in the case of light quarks, \eq{eq:llgamma_CL} can be considered as an effective Wilson coefficient at the low scale, which includes the effect of 4-fermion operators (up to QED corrections) and can therefore be used for the numerical evaluation.
\smallskip

Considering $\ell_i\to\ell_{f}$ transition with an off-shell photon, we define the amplitude
\begin{equation}
\mathcal{A} (\ell _i \to \ell _f \gamma ^* ) = -e{q^2} \,\bar{\ell}_f(p_f) \, \slashed{\varepsilon}^*(q^2) \left( {\widehat{\Xi} _{fi}^L}P_L + {\widehat{\Xi} _{fi}^R}P_R + \delta_{fi}\right) \ell _i(p_i)
\label{eq:offshell_photon_amp}
\end{equation}
with
\begin{align}
\begin{aligned}
\widehat{\Xi}_{fi}^{L}&=\frac{-N_{c}}{576\pi^2 }\Bigg(\frac{V_{jk}\lambda_{kf}^{*}V_{jl}^{*}\lambda_{li}}{M_{1}^2}F\bigg(\!\frac{m_{u_j}^{2}}{M_{1}^2}\!\bigg)+\frac{V_{jk}\kappa_{kf}^{*}V_{jl}^{*}\kappa_{li}}{M_{3}^2}F\bigg(\!\frac{m_{u_j}^{2}}{M_{3}^2}\!\bigg)+\frac{2\kappa_{jf}^{*}\kappa_{ji}}{M_{3}^2}G\bigg(\!\frac{m_{d_j}^{2}}{M_{3}^2}\!\bigg)\Bigg)\,,\\
\widehat{\Xi}_{fi}^{R}&=\frac{-N_{c}}{576\pi^2 }\frac{\hat{\lambda}_{jf}^{*}\hat{\lambda}_{ji}}{M_{1}^2}F\bigg(\!\frac{m_{u_j}^{2}}{M_{1}^2}\!\bigg)\,,
\end{aligned}
\label{eq:l3l_off_photon}
\end{align}
where
\begin{align}
\begin{split}
F(y)&=\frac{y^3-18y^2+27y-10+2\left(y^3+6y-4\right)\log(y)}{(y-1)^4}
\,,\\
G(y)&=\frac{-17y^3+36y^2-27y+8+\left(8y^3-6y+4\right)\log(y)}{(y-1)^4} \,.
\end{split}
\label{eq:loop_functions_photon}
\end{align}
\smallskip\\

\subsubsection*{Observables}
We can now express the branching ratios of flavor changing radiative lepton decays in terms of the Wilson coefficients as
\begin{align}
{\mathrm{Br}}\left[\ell_{i}\rightarrow\ell_{f}\gamma\right]=\frac{\alpha m_{\ell_{i}}^3}{256\pi^4} \tau_{\ell_i}\Big(\big|C_{\ell_{f}\ell_{i}}^{L}\big|^2+\big|C_{\ell_{f}\ell_{i}}^{R}\big|^2\Big)~,
\label{eq:Br_ell_ellgamma}
\end{align}
where $\tau_{\ell_i}$ is the life time of the initial state lepton. The AMM of a charged lepton $\ell_i$ is given by
\begin{align}
a_{\ell_{i}}= -\frac{m_{\ell_{i}}}{4\pi^2}\text{Re} \Big[C^{R}_{\ell_{i}\ell_{i}}\Big]\,.
\end{align}
The expression for the electric dipole moment of the lepton is quite similar to the one for the AMM, namely
\begin{align}
d_{\ell_i}=-\frac{e}{8\pi^2}\text{Im}\Big[C^{R}_{\ell_{i}\ell_{i}}\Big]\,.
\end{align}
In case of the AMM of the muon we already discussed the experimental situation in the introduction. In summary, the difference between the experiment and the SM prediction is
\begin{align*}
\delta a_\mu = (278 \pm 88) \times 10^{-11} \ ,
\end{align*}
corresponding to a $3.5\sigma$ deviation. Note that in our case the Wilson coefficient is in general complex and could therefore lead to sizable EDMs~\cite{Crivellin:2018qmi}. 
\smallskip
The current limits for radiative LFV decays are~\cite{TheMEG:2016wtm, Aubert:2009ag}
\begin{align}
\begin{aligned}
{\rm{Br}}[\mu \to e \gamma] <& 4.2 \times 10^{-13}\,,\\
{\rm{Br}}[\tau \to e \gamma] <&  3.3 \times 10^{-8}\,,\\
\rm{Br}[\tau \to \mu \gamma] <&  4.4 \times 10^{-8}\,,
\end{aligned}
\end{align}
representing relevant constraints for our analysis.
\smallskip
The off-shell photon penguins contribute to processes like $\tau\to 3\mu$ which we will consider later.
\smallskip

\begin{boldmath}
\subsection{$Z\ell\ell$ and $Z\nu\nu$ Processes}
\end{boldmath}

In this subsection we compute the amplitudes for $Z\to\ell_{i}^{-}\ell_{f}^{+}$ and $Z\to\nu_{f}\bar{\nu}_{i}$ processes for massless leptons. At zero momentum transfer (or equivalently vanishing $Z$ mass), these amplitudes are directly related to effective $Z\ell\ell$ and $Z{\nu}\nu$ couplings, which will enter flavor observables like for example in $\tau\to 3\mu$. We write the amplitude in an analogous way to the case with the off-shell photon
\begin{align}
\begin{aligned}
\mathcal{A} (Z\to \ell_{f}^{-} \ell_{i}^{+})&=\frac{g}{c_{w}}\bar{u}(p_f,m_{\ell_f})\gamma_{\mu}\left(\Lambda^{L}_{\ell_f\ell_i}\big(q^2\big)P_L+\Lambda_{\ell_f\ell_i}^{R}\big(q^2\big)\right)v(p_i,m_{\ell_i}) \varepsilon^{\mu}(q)\,,\\
\mathcal{A} (Z\to \nu_{f}\bar{\nu}_{i})&= \frac{g_2}{c_w}\Sigma_{\nu_{f}\nu_{i}}\big(q^2\big) \bar{u}(p_f) \gamma_{\mu} P_L v(p_i) \varepsilon^{\mu}(q)\,,
\label{eq:def_Zll_and_Zvv}
\end{aligned}
\end{align}
where $\varepsilon^\mu$ is the polarization vector of the $Z$ and
\begin{align}
\Lambda_{\ell_f\ell_i}^{L(R)}\big(q^2\big)=\Lambda_{\text{SM}}^{L(R)}(q^2)\delta_{fi}+\Delta_{fi}^{L(R)}\big(q^2\big)\,,&&
\Sigma_{\nu_{f}\nu_{i}}\big(q^2\big)=\Sigma_{\text{SM}}(q^2)\delta_{fi} + \Sigma_{fi}^{\text{LQ}}\big(q^2\big) \,.
\label{eq:effectice_Z_couplings}
\end{align}
At tree-level the SM couplings read
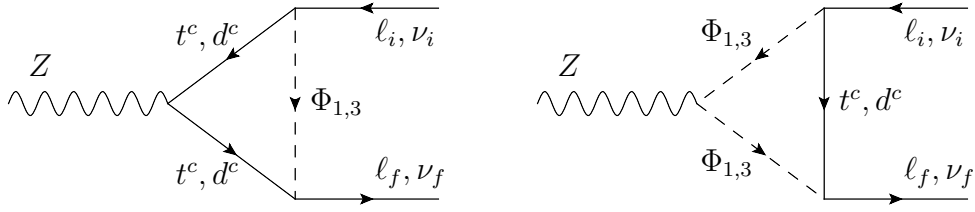
\begin{figure}[t]
	\centering
	\begin{overpic}[scale=.60,,tics=10]
		{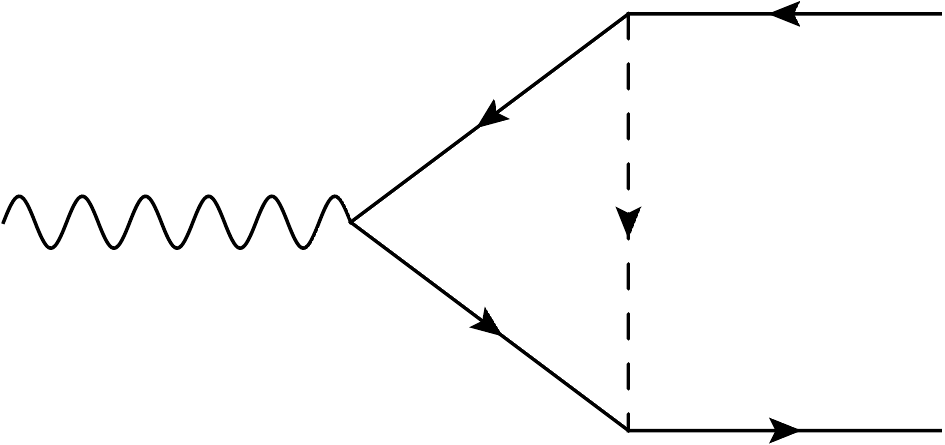}
		\put(5,30){$Z$}
		\put(85,38){$\ell_{i},\nu_{i}$}
		\put(85,6){$\ell_{f},\nu_{f}$}
		\put(39,37){$t^{c},d^{c}$}
		\put(39,5){$t^{c},d^{c}$}
		\put(70,22){$\Phi_{1,3}$}
	\end{overpic}
	\hspace{1cm}
	\begin{overpic}[scale=.60,,tics=10]
		{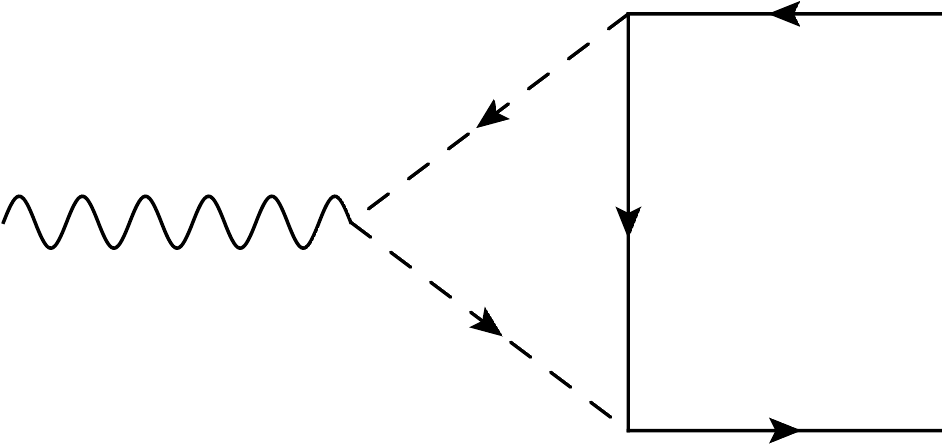}
		\put(5,30){$Z$}
		\put(85,38){$\ell_{i},\nu_{i}$}
		\put(85,6){$\ell_{f},\nu_{f}$}
		\put(38,38){$\Phi_{1,3}$}
		\put(38,8){$\Phi_{1,3}$}
		\put(70,22){$t^{c},d^{c}$} 
	\end{overpic}
	\caption{Feynman diagrams in our singlet-triplet model contributing to $Z\to\ell_{f}^{-}\ell_{i}^{+}$ and $Z\to\nu_{f}\bar{\nu}_{i}$ processes. }
	\label{fig:Zll_diagrams}
\end{figure}
\begin{align}
\Lambda_{\text{SM}}^{L}=\left(\frac{1}{2}-s_w^2\right)\,, &&\Lambda_{\text{SM}}^{R}=-s_w^2\,,&&\Sigma_{\text{SM}}=-\frac{1}{2}\,,
\end{align}
with $s_w$ being the Weinberg angle. Beyond tree-level, the SM coefficients receive momentum dependent corrections which are included in the predictions for EW observables.  The corresponding Feynman diagrams, generating these amplitudes in our model, are depicted in Fig.~\ref{fig:Zll_diagrams}. For the calculation we include the up-type quark masses (which become relevant in case of the top) and the $Z$ mass up to the order $m_u^2/M_{\text{LQ}}^2$ and $m_Z^2/M_{\text{LQ}}^2$, respectively. In this setup we obtain
\begin{align}
\begin{aligned}
\Delta^{L}_{fi}\big(q^2\big)&=V_{jk}\lambda_{kf}^{*}V_{jl}^{*}\lambda_{li} \mathcal{F}_{L}\big(m_{u_j}^2,q^2,M_{1}^2\big) +V_{jk}\kappa_{kf}^{*}V_{jl}^{*}\kappa_{li} \mathcal{F}_{L}\big(m_{u_j}^2,q^2,M_{3}^2\big)\\
&+2\kappa_{jf}^{*}\kappa_{ji}\mathcal{G}_{L}\big(q^2,M_{3}^2\big)\,,\\
\Delta^{R}_{fi}\big(q^2\big)&=\hat{\lambda}_{jf}^{*}\hat{\lambda}_{ji}\mathcal{F}_{R}\big(m_{u_j}^2,q^2,M_{1}^2\big)\,,\\
\Sigma_{fi}^{\text{LQ}}\big(q^2\big)&=\lambda_{jf}^{*}\lambda_{ji}\mathcal{H}_{1}\big(q^2,M_1^2\big)+\kappa_{jf}^{*}\kappa_{ji}\mathcal{H}_{1}\big(q^2,M_{3}^2\big)+2V_{jk}\kappa_{kf}^{*}V_{jl}^{*}\kappa_{li}\mathcal{H}_{2}\big(m_{u_{j}}^2,q^2,M_{3}^2\big)\,.
\end{aligned}
\end{align}
The corresponding loop functions $\mathcal{F}_{L,R}$, $\mathcal{G}_{L}$ and $\mathcal{H}_{1,2}$ are given in \eq{eq:Zll_loop_function} and \eq{eq:Zvv_loop_functions}. In case of $Z$ decays we have $q^2=m_Z^2$.
\smallskip

For the effective $Z\ell\ell$ and $Z\nu\nu$ couplings (at zero momentum transfer), we define
\begin{align}
\begin{aligned}
\mathcal{L}^{Z\ell\ell}_{\rm{int}}& =\frac{g_2}{c_w}\big[\bar{\ell}_{f} \left(\Lambda_{\ell_f\ell_i}^{L}(0)\gamma_{\mu}P_{L}+\Lambda^{R}_{\ell_f\ell_i}(0)\gamma_{\mu}P_{R}\right)\ell_{i}\big]Z^{\mu}\,,\\
\mathcal{L}^{Z\nu\nu}_{\text{int}}& =\frac{g_2}{c_w}\Sigma_{\nu_{f}\nu_{i}}(0)\left[\bar{\nu}_{f}\gamma_{\mu}P_{L}\nu_{i}\right]Z^{\mu}\,.
\end{aligned}
\label{eq:Zll_eff_Lag}
\end{align}
In this case, only the top contribution is relevant and the effective couplings become
\begin{align}
\begin{split}
\Lambda_{\ell_{f}\ell_{i}}^{L}(0)&=\Lambda_{\text{SM}}^{L}(0)\delta_{fi}\\
&+\frac{N_{c} m_{t}^2}{32\pi^2} \Bigg(\frac{V_{3k}\lambda_{kf}^{*}V_{3l}^{*}\lambda_{li}}{M_{1}^2}\bigg(\!1+\log\!\bigg(\!\frac{m_{t}^2}{M_{1}^2}\!\bigg)\!\bigg)\!+\frac{V_{3k}\kappa_{kf}^{*}V_{3l}^{*}\kappa_{li}}{M_{3}^2}\bigg(\!1+\log\!\bigg(\!\frac{m_{t}^2}{M_{3}^2}\!\bigg)\!\bigg)\!\Bigg)\,,\\
\Lambda_{\ell_{f}\ell_{i}}^{R}(0)&=\Lambda_{\text{SM}}^{R}(0)\delta_{fi}-\frac{N_{c}m_{t}^2}{32\pi^2} \frac{\hat{\lambda}_{3f}^{*}\hat{\lambda}_{3i}}{M_{1}^2}\bigg(\!1+\log\!\bigg(\!\frac{m_{t}^2}{M_{1}^2}\!\bigg)\!\bigg)\,,\\
\Sigma_{\nu_{f}\nu_{i}}(0)&=\Sigma_{\text{SM}}(0)\delta_{fi}+\frac{N_{c}m_{t}^2}{16\pi^2} \frac{V_{3k}\kappa_{kf}^{*}V_{3l}^{*}\kappa_{li}}{M_{3}^2}\bigg(\!1+\log\!\bigg(\!\frac{m_{t}^2}{M_{3}^2}\!\bigg)\bigg)\,.
\end{split}
\label{eq:Zcouplings_eff}
\end{align}
Note that $Z\to\ell_{i}^{-}\ell_{f}^{+}$ has also been considered in Ref.~\cite{Arnan:2019olv}.
\smallskip\\

\subsubsection*{Observables}

The branching ratio of a $Z$ decaying into a charged lepton pair reads
\begin{align}
{\rm Br}\left[Z \to \ell^{-}_{f}\ell^{+}_{i}\right]=\frac{G_F}{\sqrt{2}}\frac{m_Z^3}{3\pi}\frac{1}{\Gamma_{\text{tot}}}\left(\big|\Lambda^{L}_{\ell_{f}\ell_{i}}(m_Z^2)\big|^2+\big|\Lambda^{R}_{\ell_{f}\ell_{i}}(m_Z^2)\big|^2\right)\,.
\end{align}
with $\Gamma_{\text{tot}}\approx 2.5\, \text{GeV}$. The case for a pair of neutrinos in the final state follows trivially. The effective number of active neutrinos, including the corrections in our model, are given by
\begin{align}
N_{\nu}=\sum_{f,i}\bigg|\delta_{fi}+\frac{\Sigma_{fi}^{\rm{LQ}}(m_Z^2)}{\Sigma_{\rm{SM}}(m_Z^2)}\bigg|^2\,.
\end{align}

At LEP~\cite{ALEPH:2005ab} the lepton flavor conserving $Z$ boson couplings were measured precisely. We give the experimental results for each flavor separately
\begin{align}
\begin{aligned}
\Lambda^{Le}_{\mathrm{exp}}(m_Z^2)&=0.26963\pm0.00030\,, && \Lambda^{Re}_{\mathrm{exp}}(m_Z^2) =-0.23148\pm0.00029\,,\\
\Lambda^{L\mu}_{\mathrm{exp}}(m_Z^2)&=0.2689\pm0.0011\,, && \Lambda^{R\mu}_{\mathrm{exp}}(m_Z^2)=-0.2323\pm0.0013\,,\\
\Lambda^{L\tau}_{\mathrm{exp}}(m_Z^2)&=0.26930\pm0.00058 \,, && \Lambda^{R\tau}_{\mathrm{exp}}(m_Z^2) =-0.23274\pm0.00062\,,\\
\Sigma^{L\nu}_{\mathrm{exp}}(m_Z^2)&=-0.5003\pm0.0012\,.
\end{aligned}
\end{align}
The SM predictions at the $Z$ pole are
\begin{align}
\begin{split}
\Lambda^{Le}_{\mathrm{SM}}(m_Z^2)=\Lambda^{L\mu}_{\mathrm{SM}}(m_Z^2)=\Lambda^{L\tau}_{\mathrm{SM}}(m_Z^2)& =0.26919\pm0.00020\,,\\
\Lambda^{Re}_{\mathrm{SM}}(m_Z^2)=\Lambda^{R\mu}_{\mathrm{SM}}(m_Z^2)=\Lambda^{R\tau}_{\mathrm{SM}}(m_Z^2)&=-0.23208\substack{+0.00016\\-0.00018}\,,\\
\Sigma^{L\nu}_{\mathrm{SM}}(m_Z^2)&=-0.50199\substack{+0.00017\\-0.00020}\,.
\end{split}
\end{align}
Concerning lepton flavor violating $Z$ decays the limits from LEP are~\cite{Aad:2014bca,Akers:1995gz,Abreu:1996mj}
\begin{align}
\begin{split}
{\rm Br}\left[Z\to e^{\pm}\mu^{\mp}\right]&\leq 7.5\times 10^{-7}\,,\\
{\rm Br}\left[Z\to e^{\pm}\tau^{\mp}\right]&\leq 9.8\times 10^{-6}\,,\\
{\rm Br}\left[Z\to \mu^{\pm}\tau^{\mp}\right]&\leq 1.2\times 10^{-5}\,.
\end{split}
\end{align}
From $Z\to\nu\bar{\nu}$ one can determine the number of active neutrinos to be~\cite{ALEPH:2005ab}
\begin{align}
N_\nu = 2.9840 \pm 0.0082 \ .
\end{align}
\smallskip
As mentioned before, $Z\ell\ell$ couplings (at zero momentum transfer) contribute to processes like $\tau\to 3\mu$. Furthermore, $Z\ell\ell$ couplings in $Z$ decays can be measured much more precisely at an FCC-ee which could produce more than $10^{11}$ $Z$ bosons~\cite{Abada:2019zxq}. 
\smallskip

\begin{figure}[t]
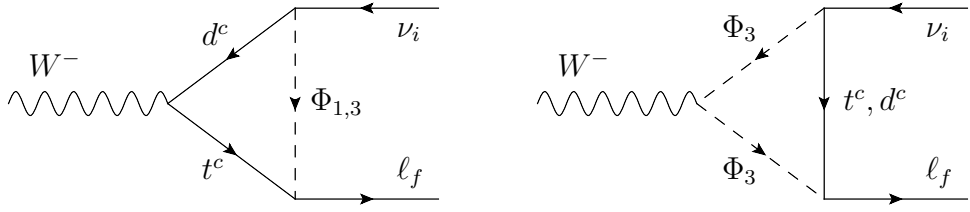

	\centering
	\begin{overpic}[scale=.60,,tics=10]
		{z_decay}
		\put(5,30){$W^{-}$}
		\put(90,39){$\nu_{i}$}
		\put(90,6){$\ell_{f}$}
		\put(45,37){$d^{c}$}
		\put(45,6){$t^{c}$}
		\put(70,22){$\Phi_{1,3}$}
	\end{overpic}
	\hspace{1cm}
	\begin{overpic}[scale=.60,,tics=10]
		{z_decay2}
		\put(5,30){$W^{-}$}
		\put(90,39){$\nu_{i}$}
		\put(90,6){$\ell_{f}$}
		\put(43,39){${\Phi_3}$}
		\put(43,6){${\Phi_3}$}
		\put(71,22){$t^{c},d^{c}$} 
	\end{overpic}
	\caption{Feynman diagrams contributing to $W^{-}\to\ell_{f}^{-}\bar{\nu}_{i}$. The right diagram is only present for the triplet since the singlet does not couple to the $W$ boson (at tree-level).}
	\label{fig:Wlv_decays}
\end{figure}

\begin{boldmath}
\subsection{$W\ell\nu$  Processes}
\end{boldmath}

Computing the amplitude of this process (also considered in Ref.~\cite{Arnan:2019olv}), we obtain
\begin{align}
\mathcal{A} (W^{-}\to \ell_{f}^{-} \bar{\nu}_{i})= -\frac{g_2}{\sqrt{2}}\Lambda_{\ell_{f}\nu_{i}}^{W}\big(q^2\big) \bar{u}(p_{\ell_f},m_{\ell_f}) \gamma_\mu P_L \, u(p_{\nu_i}) \varepsilon^{\mu}(q) \ ,
\label{eq:ampl_wlnu}
\end{align}
where
\begin{align}
\Lambda^W_{\ell_{f}\nu_{i}}\big(q^2\big)&= \Lambda^{W}_{\text{SM}}(q^2)\delta_{fi} + \Lambda^{\text{LQ}}_{fi}\big(q^2\big) \ .
\label{eq:wlnu_eff}
\end{align}
At tree level in the SM we have $\Lambda^{W}_{\text{SM}}(q^2)=1$. The Feynman diagrams shown in Fig. \ref{fig:Wlv_decays} result in
\begin{align}
\begin{split}
\Lambda_{fi}^{\text{LQ}}\big(q^2\big)=\frac{N_{c}}{288\pi^2} \Bigg[&V_{jk}\lambda_{kf}^{*}V_{jl}\lambda_{li} \mathcal{F}_1\left(m_{u_j}^2,q^2,M_{1}^2\right) +V_{jk}\kappa_{kf}^{*}V_{jl}\kappa_{li} \mathcal{F}_2\left(m_{u_j}^2,q^2,M_{3}^2\right)\\ &+ \frac{8 \kappa_{jf}^{*} \kappa_{ji} \, q^2}{9 M_3^2}\Bigg] \,,
\end{split}
\label{eq:wlnu}
\end{align}
with the loop functions $\mathcal{F}_{1,2}$ given in \eq{eq:Wlv_loop_functions}. Again, we set all down-type quark masses to zero but included the up-type quark masses, which are relevant for the top. At the level of effective couplings, we define the Lagraigian
\begin{align}
\mathcal{L}_{{\rm int}}^{W\ell\nu}=-\frac{g}{\sqrt{2}}\Lambda_{\ell_{f}\nu_{i}}^{W}(0) \big[\bar{\ell}_{f}\gamma^{\mu}P_{L}\nu_{i}\big]W_{\mu}^{-}\,.
\end{align}
The LQ contribution then reads
\begin{align}
\begin{split}
\Lambda_{ji}^{\text{LQ}}(0)=\frac{N_c   m_t^2 }{64 \pi ^2}\Bigg[\frac{V_{3l} \lambda _{lj}^{*} V_{3k}^*\lambda _{ki}}{M_{1}^2}\left(\!1+2 \log\! \left(\!\frac{m_t^2}{M_{1}^2}\!\right)\!\right)
- \frac{V_{3l} \kappa _{lj}^{*} V_{3k}^*\kappa _{ki}}{M_{3}^2}\left(\!1+2 \log\! \left(\!\frac{m_t^2}{M_{3}^2}\!\right)\!\right)\Bigg]\,.
\label{Weff}
\end{split}
\end{align}
Out of this formula one deduces a destructive interference between the contribution of the singlet and the triplet in case of lepton flavor conservation.
\smallskip\\

\subsubsection*{Observables}

Experimentally, the modification of the $W\tau \nu$ coupling extracted from $\tau \to \mu \nu \bar\nu$ and $\tau \to e \nu \bar\nu$ decays reads~\cite{Pich:2013lsa,Tanabashi:2018oca}
\begin{align}
	{|\Lambda_{\tau\nu}^{W}(0)|}_{\exp}\approx 1.002 \pm 0.0015
\end{align}
and provides a better constraint than data of $W$ decays. Here we averaged the central values of the muon and tau mode, but did not add the errors in quadrature in order to be conservative. We see that a positive NP effect is preferred which means that the triplet contribution should exceed the one of the singlet.
\smallskip\\

\begin{boldmath}
\subsection{$4\ell$ Processes}
\end{boldmath}

We define the effective Hamiltonian as
\begin{align}
{\cal H}_{{\rm{eff}}}^{4\ell} = {\cal H}_{{\rm{eff}}}^{\ell\ell \gamma } + \sum\limits_{a,b,f,i} {\left( {C_{abfi}^{VLL}O_{abfi}^{VLL} + C_{abfi}^{VLR}O_{abfi}^{VLR} + C_{abfi}^{SLL}O_{abfi}^{SLL}} \right) + L \leftrightarrow R} {\mkern 1mu} \,,
\label{eq:Heff_tau3mu}
\end{align}
with
\begin{align}
\begin{split}
O_{abfi}^{V{\kern 1pt} LL} &= \left[ {{{\bar \ell }_a}{\gamma ^\mu }{P_L}{\ell _b}} \right]\left[ {{{\bar \ell }_f}{\gamma _\mu }{P_L}{\ell _i}} \right]{\mkern 1mu} \,,\\
O_{abfi}^{V{\kern 1pt} LR} &= \left[ {{{\bar \ell }_a}{\gamma ^\mu }{P_L}{\ell _b}} \right]\left[ {{{\bar \ell }_f}{\gamma _\mu }{P_R}{\ell _i}} \right]{\mkern 1mu} \,,\\
O_{abfi}^{S{\kern 1pt} LL} &= \left[ {{{\bar \ell }_a}{P_L}{\ell _b}} \right]\left[ {{{\bar \ell }_f}{P_L}{\ell _i}} \right]\,{\mkern 1mu} .
\end{split}
\end{align}
Here we sum over flavor indices. In this way, no distinction for the cases of equal flavors are necessary in the matching and tensor and scalar $LR$ operators do not need to be included since they follow from Fierz identities.
\smallskip

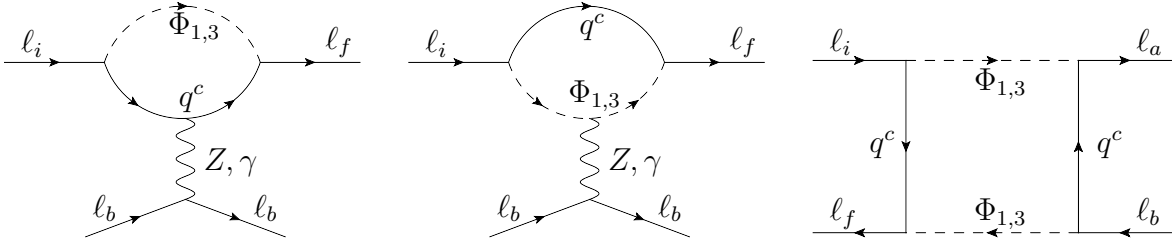
\begin{figure}[t]
	\centering
	\begin{overpic}[scale=.47,,tics=10]
		{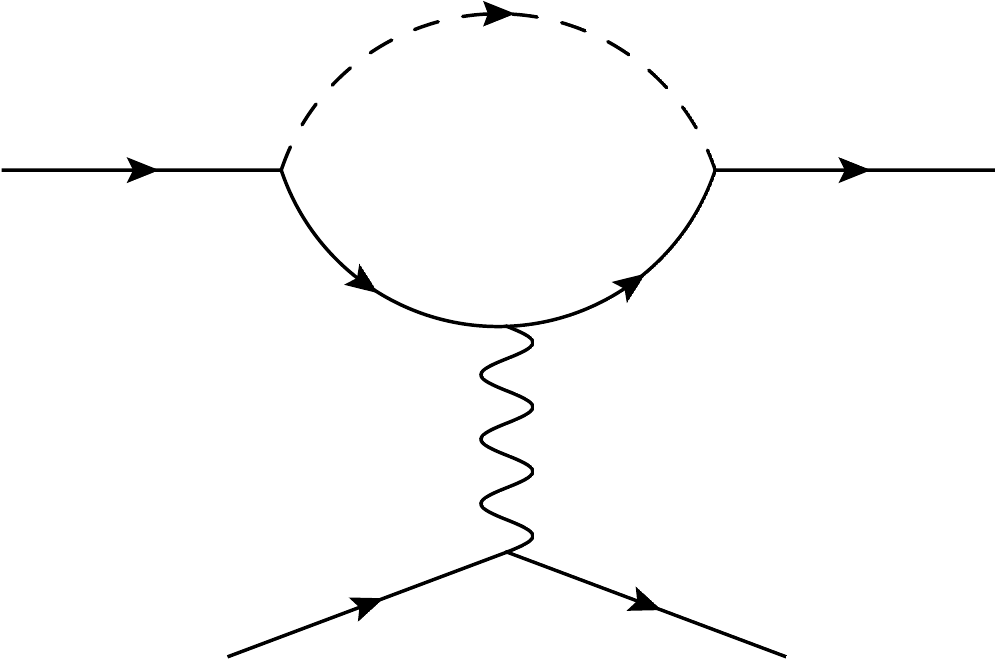}
		\put(5,52){$\ell_{i}$}
		\put(90,53){$\ell_{f}$}
		\put(25,6){$\ell_{b}$}
		\put(70,6){$\ell_{b}$}
		\put(49,37){$q^{c}$}
		\put(56,19){$Z,\gamma$}
		\put(46,57){$\Phi_{1,3}$}
	\end{overpic}
\hfill
	\begin{overpic}[scale=.47,,tics=10]
		{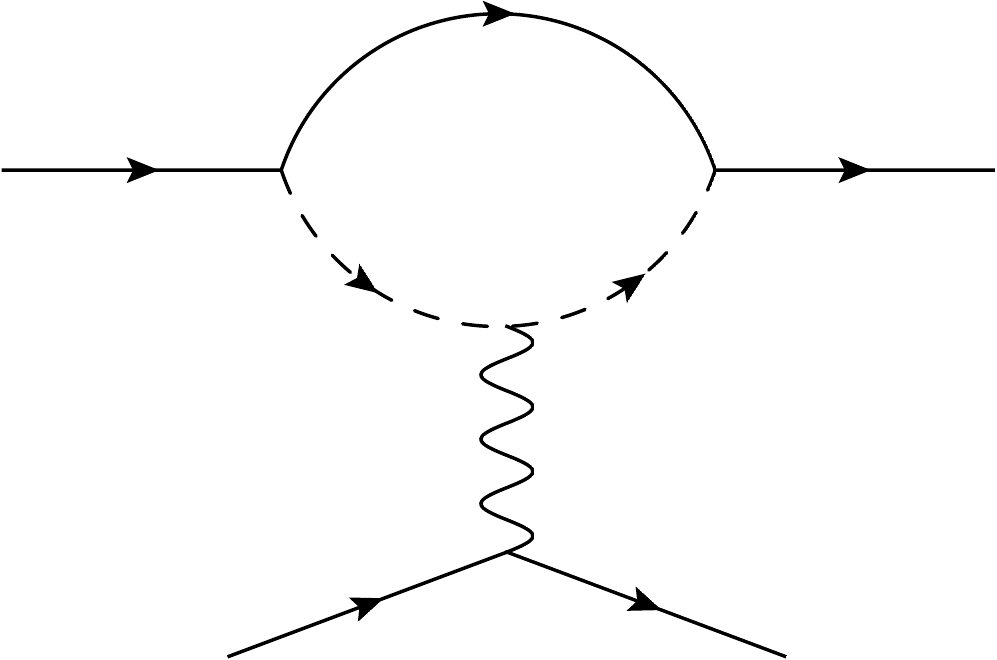}
		\put(5,52){$\ell_{i}$}
		\put(90,53){$\ell_{f}$}
		\put(25,6){$\ell_{b}$}
		\put(70,6){$\ell_{b}$}
		\put(45,38){$\Phi_{1,3}$}
		\put(56,19){$Z,\gamma$}
		\put(48,57){$q^{c}$} 
	\end{overpic}
\hfill
	\begin{overpic}[scale=.50,,tics=10]
		{box_HH}
		\put(5,52){$\ell_{i}$}
		\put(5,5){$\ell_{f}$}
		\put(90,52){$\ell_{a}$}
		\put(90,5){$\ell_{b}$}
		\put(45,41){$\Phi_{1,3}$}
		\put(45,5){$\Phi_{1,3}$}
		\put(79,24){$q^c$}
		\put(16,24){$q^{c}$} 
	\end{overpic}
	\caption{Feynman diagrams contributing to $\ell_{i}\to\ell_{f}\ell_{a}\ell_{b}$. Left and centre: Penguin diagrams with $Z$ boson and photon exchange. Right: Box diagram involving two LQs.}
	\label{fig:l_lll_decays}
\end{figure}

The photon contribution reads
\begin{align}
\begin{split}
C_{abfi}^{V{\kern 1pt} LL} &=  - \pi \alpha \left( {\Xi _{ab}^L\Xi _{fi}^L + \Xi _{ai}^L\Xi _{fb}^L} \right)\,{\mkern 1mu} ,\\
C_{abfi}^{V{\kern 1pt} LR} &=  - 2\pi \alpha {\mkern 1mu} \Xi _{ab}^L\Xi _{fi}^R\,{\mkern 1mu} ,
\end{split}
\label{eq:tau3mu_photon}
\end{align}
where
\begin{align}
\Xi_{fi}^{L(R)}=\delta_{fi}+\widehat{\Xi}_{fi}^{L(R)}\,.
\label{eq:photon_withSM}
\end{align}
The effective photon off-shell couplings $\widehat{\Xi}_{fi}^{L(R)}$ are defined in \eq{eq:l3l_off_photon}.
Using the effective couplings defined in \eq{eq:effectice_Z_couplings}, the $Z$ penguins give
\begin{align}
\begin{split}
C^{V\,LL}_{abfi}&=\frac{2G_{F}}{\sqrt{2}}\Big(\Lambda_{ab}^{L}(0)\Lambda_{fi}^{L}(0)+\Lambda_{fb}^{L}(0)\Lambda_{ai}^{L}(0)\Big)\,,\\
C^{V\,LR}_{{a}{b}{f}{i}}&=\frac{4G_{F}}{\sqrt{2}}\Lambda_{ab}^{L}(0)\Lambda_{fi}^{R}(0)\,.
\end{split}
\label{eq:tau3mu_Z}
\end{align}
Note that $C^{V \, RL(RR)}_{abfi}$ are obtained from $C^{V \, LR(LL)}_{abfi}$ by interchanging $L$ and $R$ for both the photon and the $Z$ contribution. Finally, we have contributions from box diagrams involving two LQs. Since they turn out to be numerically irrelevant in our model, we omit to list them here analytically. However, in \eq{eq:l_lll_boxes} we give the results in full generality, i.e. including LQ mixing with multiple generations. The LQ contributions are depicted in Fig.~\ref{fig:l_lll_decays}.
\smallskip

The expression for the branching ratios, which are in agreement with Ref.~\cite{Crivellin:2018ahj}, read
\begin{align}
\begin{split}
\mathrm{Br}\left[\tau^\mp\to \mu^\mp e^{+}e^{-}\right] =&\frac{m_{\tau}^3}{768\pi^3 \Gamma_{\tau}^{\text{tot}}} \bigg[\frac{\alpha^2}{\pi^2}\big|C^{L}_{\mu\tau}\big|^2\Big(\!\log\!\Big(\frac{m_{\tau}^2}{m_{e}^2}\Big)-3\Big) + \frac{m_{\tau}^2}{2} \bigg(\big|C_{\mu\tau ee}^{SLL}\big|^2+\big|C_{\mu e e\tau}^{SLL}\big|^2\\
&\phantom{1234}-\,\text{Re}\Big[C_{\mu\tau ee}^{SLL}C_{\mu ee\tau}^{SLL*}\Big]+16\big|C_{\mu\tau ee}^{VLL}\big|^2+4\big|C_{\mu\tau ee}^{VLR}\big|^2+4\big|C_{\mu ee\tau}^{VLR}\big|^2\bigg)\\
&\phantom{1234}-\frac{2\alpha}{\pi} m_{\tau}\, \text{Re}\Big[C_{\mu\tau}^{L*}\big(C_{\mu\tau ee}^{VRL}+2C_{\mu\tau ee}^{VRR}\big)\Big]+L\leftrightarrow R\,\bigg]
\label{brtauemumu}
\end{split}
\end{align}
and
\begin{align}
\begin{split}
\mathrm{Br}\left[\tau^\mp\to \mu^\mp\mu^{+}\mu^{-}\right] =&\frac{m_{\tau}^3}{768\pi^3\Gamma_{\tau}^{\text{tot}}}\bigg[ \frac{\alpha^2}{\pi^2} \big|C^{L}_{\mu\tau}\big|^2 \Big(\!\log\!\Big(\frac{m_{\tau}^2}{m_{\mu}^2}\Big)-\frac{11}{4}\Big)\\
&\phantom{1234}+ \frac{m_{\tau}^2}{4}\bigg(\big|C_{\mu\mu\mu\tau}^{SLL}\big|^2+16\big|C_{\mu\tau\mu\mu}^{VLL}\big|^2+4\big|C_{\mu\tau\mu\mu}^{VLR}\big|^2+4\big|C_{\mu\mu\mu\tau}^{VLR}\big|^2\bigg)\\
&\phantom{1234}-\frac{2\alpha}{\pi} m_{\tau}\, \text{Re} \Big[C_{\mu\tau}^{L*} \big(C_{\mu\tau\mu\mu}^{VRL}+2C_{\mu\tau\mu\mu}^{VRR}\big)\Big]+L\leftrightarrow R\,\bigg]
\label{brmu3e}
\end{split}
\end{align}
with $\Gamma_{\tau}^{\text{tot}}$ as the tau lepton's total decay width. The experimental bounds are~\cite{Hayasaka:2010np,BELLGARDT19881}
\begin{align}
\begin{split}
\mathrm{Br}\left[\tau^\mp\to \mu^\mp e^{+}e^{-}\right]&< 1.5\times 10^{-8}\,,\\
\mathrm{Br}\left[\tau^\mp\to \mu^\mp\mu^{+}\mu^{-}\right]&< 2.1\times 10^{-8}\,,\\
\mathrm{Br}\left[\mu^\mp\to e^\mp e^{+}e^{-}\right]&< 1.0\times 10^{-12}\,.
\end{split}
\end{align}
\smallskip

\begin{figure}[t]
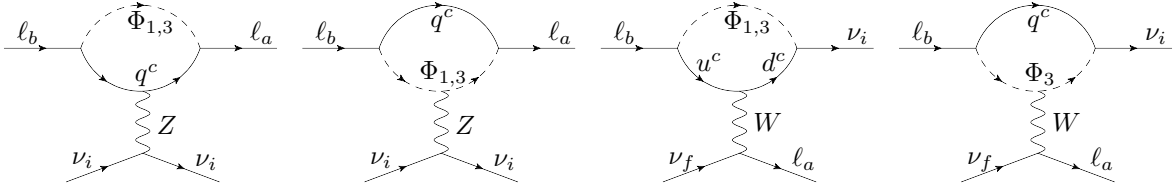

	\centering
	\begin{overpic}[scale=.36,,tics=10]
		{higgs_Z_pengiuin}
		\put(5,52){\footnotesize$\ell_{b}$}
		\put(90,52){\footnotesize$\ell_{a}$}
		\put(25,7){\footnotesize$\nu_{i}$}
		\put(70,6){\footnotesize$\nu_{i}$}
		\put(48,37){\footnotesize$q^{c}$}
		\put(56,18){\footnotesize$Z$}
		\put(45,56){\footnotesize$\Phi_{1,3}$}
	\end{overpic}
\hfill
	\begin{overpic}[scale=.36,,tics=10]
		{higgs_Z_pengiuin2}
		\put(5,52){\footnotesize$\ell_{b}$}
		\put(90,52){\footnotesize$\ell_{a}$}
		\put(25,7){\footnotesize$\nu_{i}$}
		\put(70,6){\footnotesize$\nu_{i}$}
		\put(43,38){\footnotesize$\Phi_{1,3}$}
		\put(56,18){\footnotesize$Z$}
		\put(47,57){\footnotesize$q^{c}$} 
	\end{overpic}
\hfill
	\begin{overpic}[scale=.36,,tics=10]
		{higgs_Z_pengiuin}
		\put(5,52){\footnotesize$\ell_{b}$}
		\put(90,52){\footnotesize$\nu_{i}$}
		\put(25,7){\footnotesize$\nu_{f}$}
		\put(70,6){\footnotesize$\ell_{a}$}
		\put(35,41){\footnotesize$u^{c}$}
		\put(59,41){\footnotesize$d^{c}$}
		\put(56,19){\footnotesize$W$}
		\put(45,56){\footnotesize$\Phi_{1,3}$}
	\end{overpic}
\hfill
	\begin{overpic}[scale=.36,,tics=10]
		{higgs_Z_pengiuin2}
		\put(5,52){\footnotesize$\ell_{b}$}
		\put(90,52){\footnotesize$\nu_{i}$}
		\put(25,7){\footnotesize$\nu_{f}$}
		\put(70,6){\footnotesize$\ell_{a}$}
		\put(46,37){\footnotesize$\Phi_{3}$}
		\put(56,19){\footnotesize$W$}
		\put(47,57){\footnotesize$q^{c}$} 
	\end{overpic}
	\caption{Penguin diagrams that contribute to $\ell_{b}\to\ell_{a}\nu_{i}\bar{\nu}_{f}$ transitions. In case of the $Z$ boson, lepton flavor is conserved at tree-level vertex ($f=i$). For the $W$ penguins we applied Fierz identities in order to match on the effective operators. The box diagrams look similar to the one in Fig.~\ref{fig:Zll_diagrams} but turn out to be numerically insignificant.}
	\label{fig:l_lvv_diagrams}
\end{figure}

\begin{boldmath}
\subsection{$\ell\ell\nu\nu$ Processes}
\end{boldmath}

We define the effective Hamiltonian as
\begin{align}
\mathcal{H}^{2\ell 2\nu}_{\rm{eff}}&= \left(D_{\ell_{a}\ell_{b}}^{L,fi}O_{\ell_{a}\ell_{b}}^{L,fi} +D_{\ell_{a}\ell_{b}}^{R,fi}O_{\ell_{a}\ell_{b}}^{R,fi}\right)\,,
\label{eq:Heff_taumununu}
\intertext{with}
O^{L(R),fi}_{\ell_{a}\ell_{b}}&=\left[\bar{\ell}_{a}\gamma_{\mu}P_{L(R)}\ell_{b}\right]\left[\bar{\nu}_f\gamma^{\mu}P_{L}\nu_{i}\right]\,.
\end{align}
At the 1-loop level, LQs can contribute to these processes through three types of Feynman diagrams: $W$-penguins, $Z$-penguins and pure LQ box diagrams, see Fig.~\ref{fig:l_lvv_diagrams}. Again, the boxes are numerically not relevant due to the small couplings to muons. Therefore, we only present these results with full generality in the appendix.
\smallskip

The $W$ penguin given in terms of the modified $W\ell\nu$ couplings of \eq{Weff} gives
\begin{align}
D^{L,fi}_{\ell_{a}\ell_{b}}=\frac{4G_{F}}{\sqrt{2}}\Lambda^{W*}_{\ell_{b}\nu_{f}}(0)\Lambda^{W}_{\ell_{a}\nu_{i}}(0) \ .
\label{eq:taumununu_W}
\end{align}
Finally we also have the $Z$-penguins, yielding
\begin{align}
D^{L,fi}_{\ell_{a}\ell_{b}}=\frac{8G_{F}}{\sqrt{2}}\Lambda_{\ell_{a}\ell_{b}}^{L}(0)\Sigma_{\nu_{f}\nu_{i}}(0)\,,&&
D^{R,fi}_{\ell_{a}\ell_{b}}=\frac{8G_{F}}{\sqrt{2}}\Lambda_{\ell_{a}\ell_{b}}^{R}(0)\Sigma_{\nu_{f}\nu_{i}}(0)\,,
\label{eq:taumununu_Z}
\end{align}
where we used the effective $Z\ell\ell$ and $Z\nu\nu$ couplings given in \eq{eq:Zcouplings_eff}.
\smallskip\\

\section{Phenomenology}
\label{pheno}

\begin{figure}[t]
	\centering
	\begin{overpic}[scale=.47,,tics=10]
		{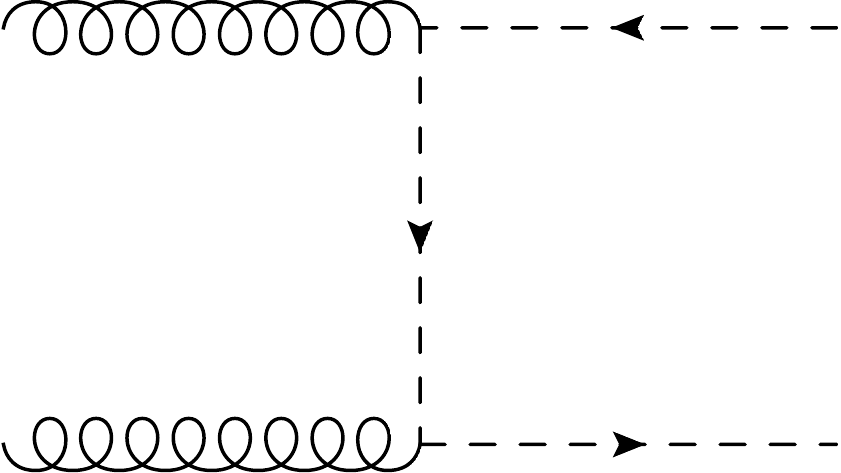}
		\put(90,40){$\bar{\Phi}$}
		\put(90,7){$\Phi$}
		\put(5,42){$g$}
		\put(5,11){$g$}
	\end{overpic}
	\hspace{1cm}
	\begin{overpic}[scale=.47,,tics=10]
		{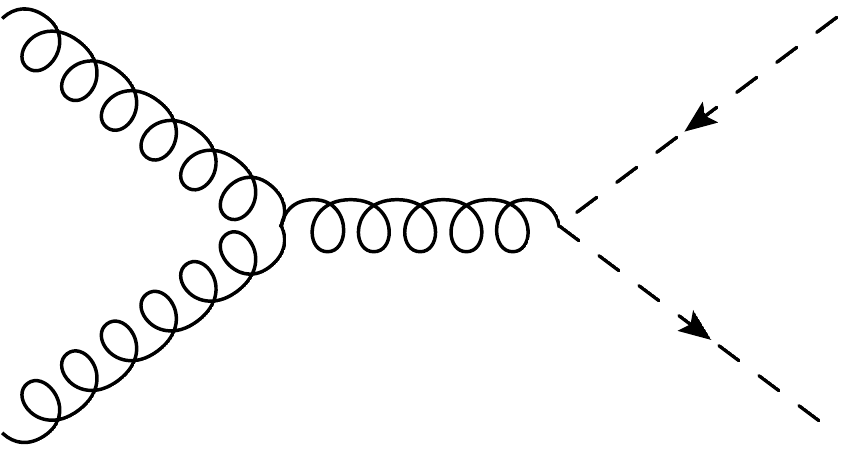}
		\put(91,35){$\bar{\Phi}$}
		\put(91,11){$\Phi$}
		\put(47,34){$g$}
		\put(5,35){$g$}
		\put(5,16){$g$}
	\end{overpic}
	\hspace{1cm}
	\begin{overpic}[scale=.47,,tics=10]
		{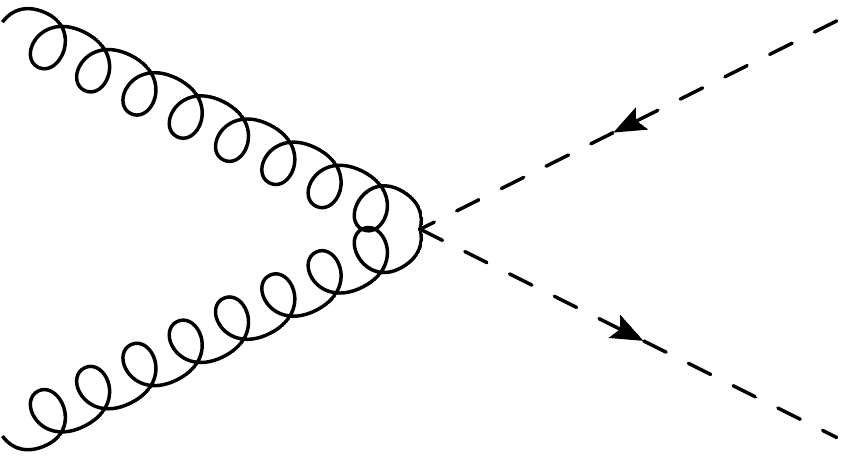}
		\put(91,36){$\bar{\Phi}$}
		\put(91,10){$\Phi$}
		\put(5,37){$g$}
		\put(5,14){$g$}
	\end{overpic}
	\caption{Tree-level Feynman diagrams contributing to $gg\to \Phi\bar{\Phi}$.}
	\label{fig:gg_to_LQLQ}
\end{figure}

\begin{figure}[t]
	\centering
		\begin{overpic}[scale=.43,,tics=10]
			{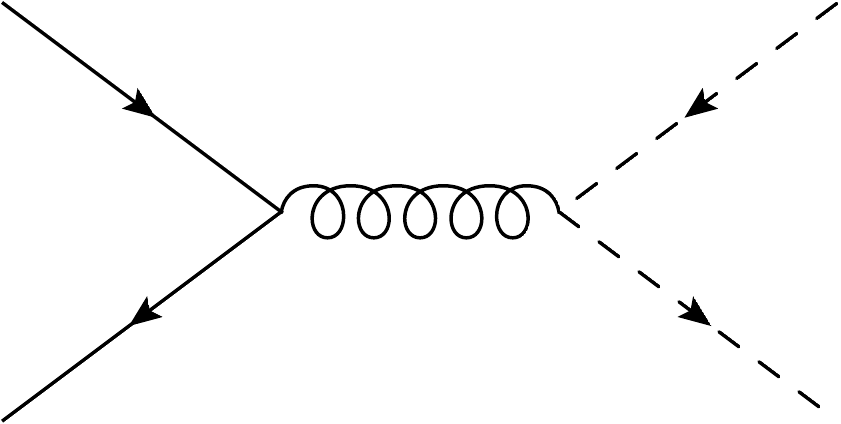}
			\put(91,33){$\bar{\Phi}$}
			\put(91,10){$\Phi$}
			\put(47,33){$g$}
			\put(5,36){$q$}
			\put(5,12){$\bar{q}$}
		\end{overpic}
	\hfill
		\begin{overpic}[scale=.43,,tics=10]
			{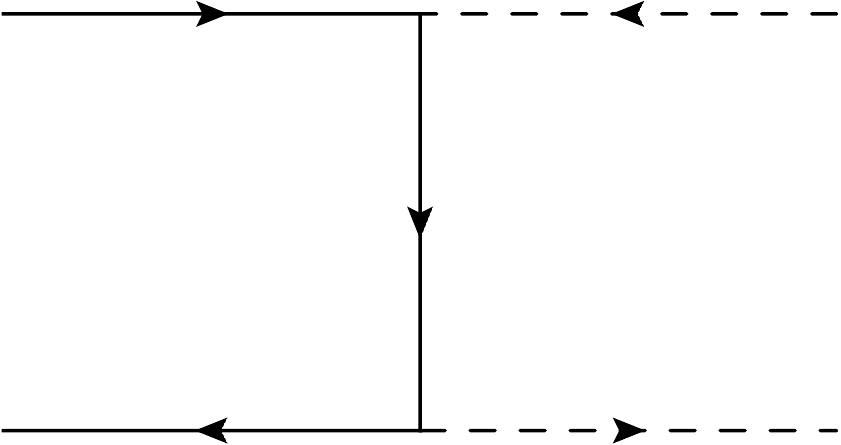}
			\put(90,39){$\bar{\Phi}$}
			\put(90,5){$\Phi$}
			\put(38,23){$L$}
			\put(5,42){$b$}
			\put(5,5){$\bar{b}$}
		\end{overpic}
	\hfill
		\begin{overpic}[scale=.43,,tics=10]
			{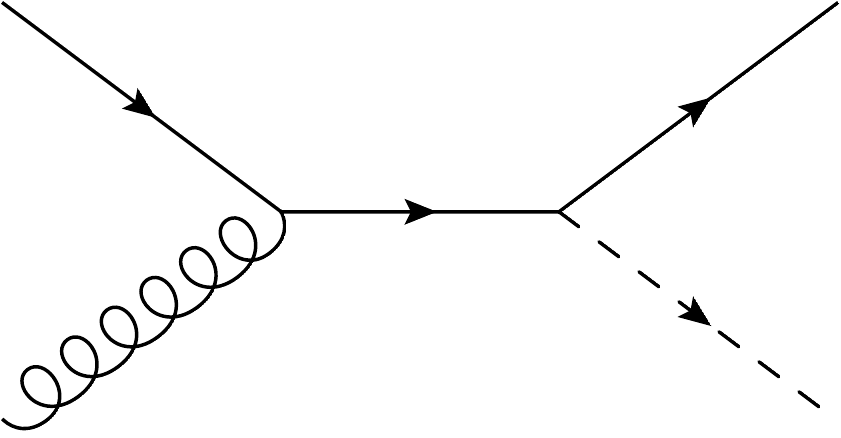}
			\put(91,36){$L$}
			\put(91,11){$\Phi$}
			\put(47,31){$b$}
			\put(5,36){$b$}
			\put(5,15){$g$}
		\end{overpic}
	\hfill
		\begin{overpic}[scale=.43,,tics=10]
			{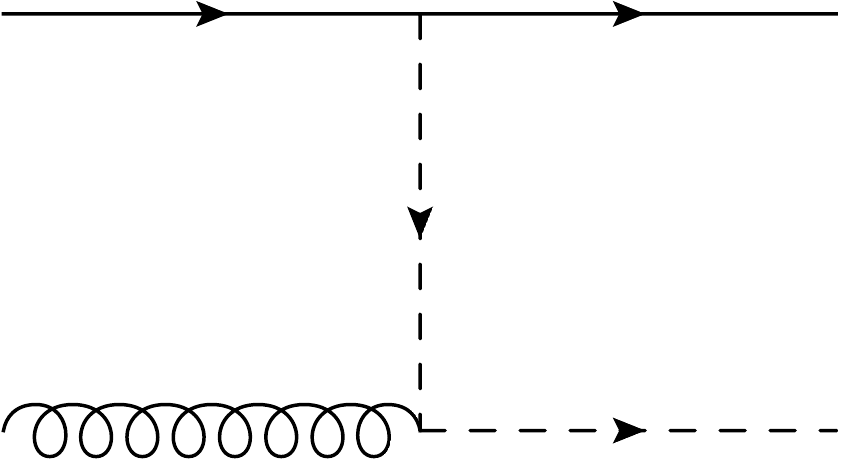}
			\put(90,41){$L$}
			\put(90,6){$\Phi$}
			\put(38,25){$\Phi$}
			\put(5,43){$b$}
			\put(5,11){$g$}
		\end{overpic}
	\caption{Tree-level diagrams contributing to $q\bar{q}\to \Phi\bar{\Phi}$ and $gq\to L\Phi$. Except for the left diagram, the cross-sections depend on the couplings of the LQ to SM fermions. $L$ can be either a neutrino or a charged lepton, depending on the specific LQ representation.}
	\label{fig:qq_to_LQLQ}
\end{figure}
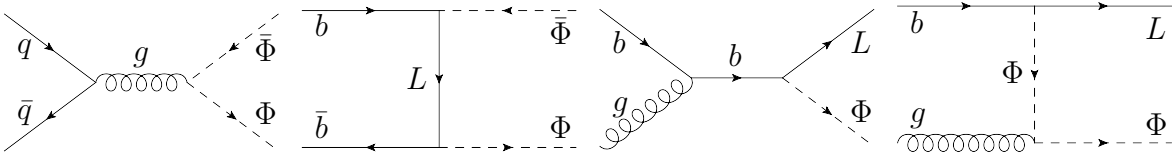

Now we turn to the phenomenological analysis of our singlet-triplet model. We consider the processes discussed above and include the loop effects calculated in the previous section. Our strategy is as follows: First we will discuss the LHC bounds on third-generation LQs. Then we will consider how one can explain $b\to c\tau\nu$ data taking into account these limits and then study the impacts on other observables like $B_s\to\tau^{+}\tau^{-}$ and $W \to \tau \nu$. For this purpose, only couplings to tau leptons (but not to muons or electrons) are necessary. In a next step we will include $b\to s\ell^+\ell^-$ data in our analysis and thus allow for non-zero couplings to left-handed muons, while disregarding couplings to electrons due to the strong constraints from $\mu\to e\gamma$~\cite{Crivellin:2017dsk}. In a final step, we search for benchmark points which can explain $b\to c\tau\nu$, $b\to s\ell^+\ell^-$ and $a_\mu$ simultaneously. For this purpose we also include couplings to right-handed muons in our analysis.
\smallskip

\begin{boldmath}
	\subsection{LHC bounds}
\end{boldmath}

Both $\Phi_1$ and $\Phi_3$ could obviously be produced at the LHC. Since LQs are charged under $SU(3)_c$ they can be pair produced via gluons (depicted in Fig.~\ref{fig:gg_to_LQLQ}), which in general gives the best bound. However, for a third generation LQ, which is the case for our model to a good approximation, also t-channel production from bottom fusion is possible as well as single production via bottom-gluon fusion (see Fig.~\ref{fig:qq_to_LQLQ}). ATLAS and CMS performed searches in these channels. In particular, in Ref.~\cite{Sirunyan:2018ruf} CMS analyzed data taken at a center-of-mass energy of 13 TeV with an integrated luminosity of 35.9 $\mathrm{fb}^{-1}$ for the scalar singlet $\Phi_1$. Assuming $\mathrm{Br}\big[\Phi_{1}\to t\tau\big]=100\%$, LQ masses up to 900 GeV are excluded. ATLAS searched for typical signals of the scalar triplet $\Phi_3$, using 36.1 $\mathrm{fb}^{-1}$ of data at $\sqrt{s}=13$ TeV~\cite{Aaboud:2019bye}. Focusing on NP effects in third generation quarks and leptons, i.e. $\Phi_{3}\to t\nu / b\tau$ and $\Phi_{3}\to t\tau/b\nu$, they find a lower limit on the LQ mass of 800 GeV. This limit can be raised up to 1 TeV if one of the aforementioned decay channels is dominating. Therefore, a third generation scalar LQ with mass above 1 TeV is consistent with LHC searches. We will assume this as a lower limit in the following phenomenological analysis of flavor observables. For more extensive analyses of LQ searches in combination with the flavor anomalies we refer e.g. to Refs.~\cite{Faroughy:2016osc, Greljo:2017vvb, Dorsner:2017ufx, Angelescu:2018tyl, Cerri:2018ypt, Hiller:2018wbv, Schmaltz:2018nls}.
\smallskip

\begin{boldmath}
\subsection{$b\to c\tau\nu$}
\end{boldmath}

\begin{figure}
	\centering
	\includegraphics[scale=0.42]{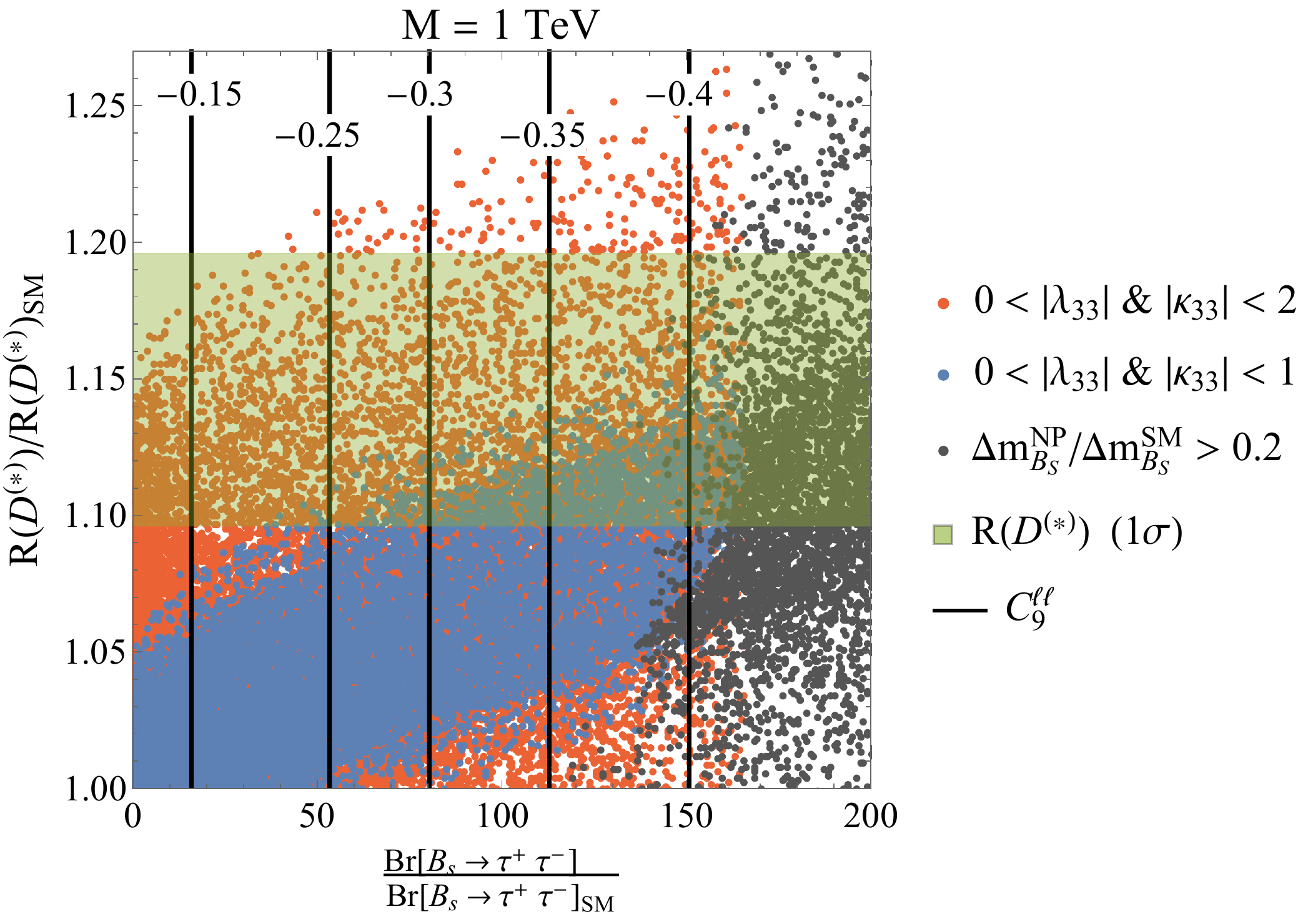}
	\includegraphics[scale=0.42]{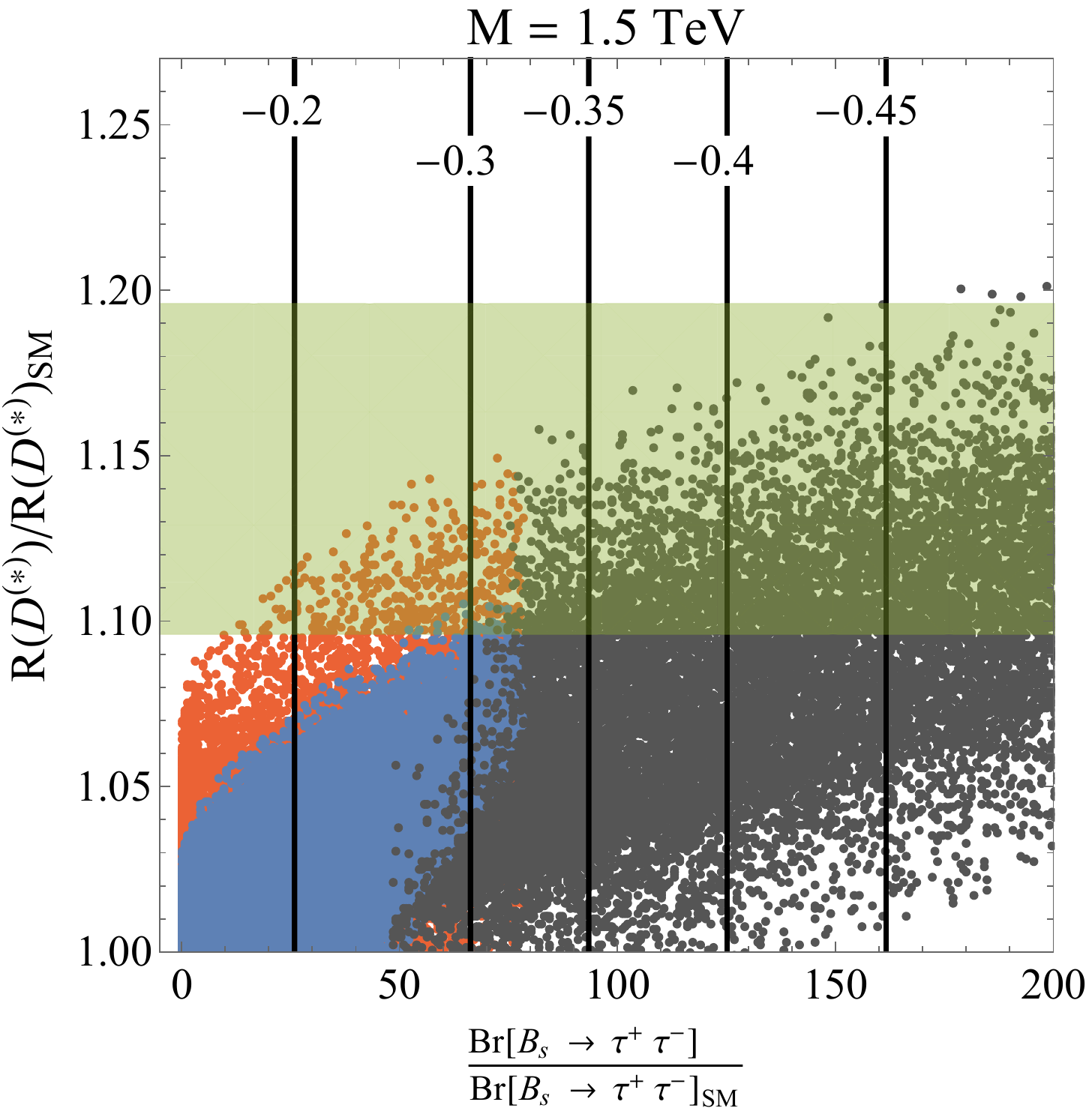}
	\caption{Correlation between $\text{Br}\big[B_{s} \to\tau^{+}\tau^{-}\big]$ and $R(D^{(*)})$, both normalized to their SM values, in the scenario with only left-handed couplings for $M_{1}=M_{3}\equiv M=1\,$TeV (left plot) and $M_{1}=M_{3}\equiv M=1.5\,$TeV (right plot). Here we scanned over $\lambda_{23}, \kappa_{23} \in\big[\!-\!1,1\big]$ for all points and $\lambda_{33}, \kappa_{33} \in\big[\!-\!1,1\big]$ (blue) or $\lambda_{33}, \kappa_{33}\in\big[\!-\!2,2\big]$ (red), respectively. The blue points are displayed on top of the red ones, showing only points that are allowed by $R_{K^{*}}^{\nu\bar\nu}$. The dark gray points are in agreement with $R_{K^{*}}^{\nu\bar\nu}$, but excluded by $B_s-\bar B_s$ mixing. The horizontal contour lines depict the LFU contribution to $C_{9}^{\ell\ell}$ while the green band represents the region for $R(D^{(*)})$ preferred by data at the $1\sigma$ level.}
	\label{fig:Bstautau_RDs}
\end{figure}
	
Concerning $b\to c\tau\nu$ processes one can address the anomalies with couplings to third generation leptons, i.e. the tau lepton and the tau neutrino, while disregarding couplings to muons and electrons. In a first step we consider the simplified case of left-handed couplings only, i.e. $\hat\lambda=0$. Furthermore, we can safely neglect CKM suppressed effects from first-generation quark couplings and are therefore left with the couplings $\lambda_{23,33}$ and $\kappa_{23,33}$, involving second and third generation quarks (i.e. bottom and strange quark in the down-basis). In this case the box contributions to $B_s-\bar B_s$ in \eq{BsMixing}, together with the tree-level effect in $b\to s\nu\bar\nu$ in \eq{eq:tree-level_bsvv} put an upper limit on the possible contribution to $b\to c\tau\nu$ processes (see Fig. \ref{fig:Bstautau_RDs}). While the relative effect in $b\to s\nu\bar\nu$ compared to $b\to c\tau\nu$ is independent of the LQ mass, the relative effect in $B_s-\bar B_s$ mixing compared to $b\to c\tau\nu$ amplitudes turns out to have a quadratic scaling with the mass. In fact, assuming real couplings and an exact cancellation in $R_{K^{(*)}}^{\nu\bar\nu}$, $\Delta m_{B_s}$ can be expressed in terms of the NP effect in $R(D^{(*)})$ as
\begin{align}
\frac{\Delta m_{B_s}}{\Delta m_{B_s}^{\text{SM}}}=1+\frac{1}{4\pi^2}\frac{G_{F}^2 V_{cb}^2M^2}{C_{1}^{\text{SM}}}\Bigg(\sqrt{\frac{R(D^{(*)})}{R(D^{(*)})_{\text{SM}}}}-1\Bigg)^{\!\!2}\,,
\end{align}
with $M_1=M_3=M$. This relation holds once small CKM rotations are neglected which is possible in the case of an anarchic flavor structure, i.e. $V_{cb}\lambda_{33}\ll\lambda_{23}$ and $V_{cb}\kappa_{33}\ll\kappa_{23}$.
\begin{figure}
	\centering
	\includegraphics[width=0.9\textwidth]{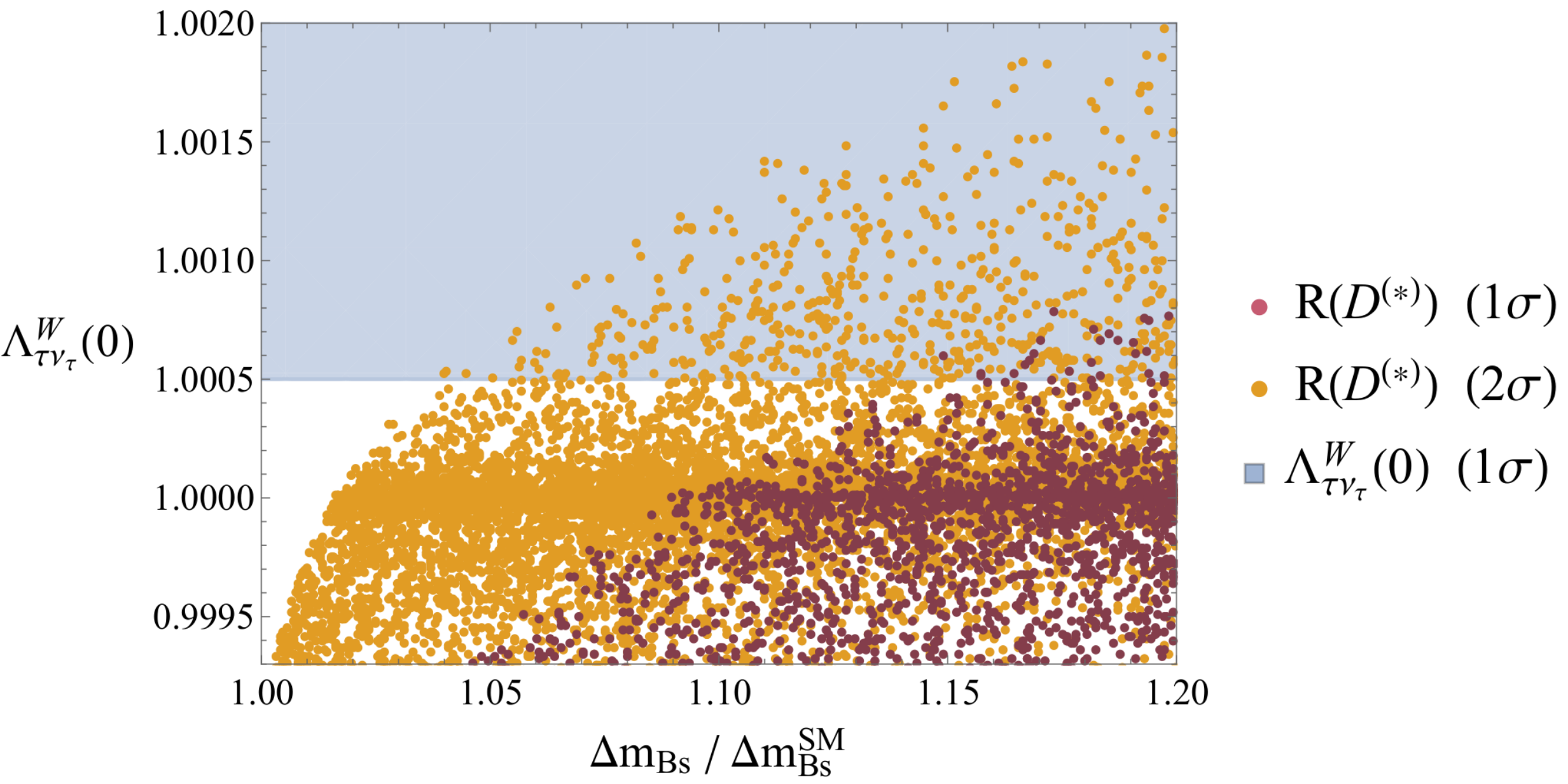}
	\caption{Correlations between the NP effect in $\Delta m_{B_s}$ and the corrections to the effective $W\tau\nu_{\tau}$ coupling $\Lambda^W_{\tau\nu_{\tau}}(0)$, constrained from $\tau\to\mu\nu\bar\nu$ and $\tau\to e\nu\bar\nu$. Like in Fig. \ref{fig:Bstautau_RDs} we only considered the couplings $\lambda_{23,33}$ and $ \kappa_{23,33}$, i.e. only couplings to left-handed taus, scanning over $\lambda_{23}$ and $\kappa_{23}$ ($\lambda_{33}$ and $\kappa_{33}$) between $\pm 1$ ($\pm 2$) and setting $M_1=M_3=M=1$ TeV. The blue region is preferred by $\tau\to\mu\nu\bar\nu$ and $\tau\to e\nu\bar\nu$ data at the $1\sigma$ level.}
	\label{fig:bladibladibla}
\end{figure}
The tau loops also generate an effect in $C_{7}$ as well as a LFU contribution to $C_{9}^{\ell\ell}$. Both these effects are directly correlated to $b\to s\tau^{+}\tau^{-}$ processes, induced by the tree-level coefficients $C_{9}^{\tau\tau}=-C_{10}^{\tau\tau}$. We find
\begin{align}
\begin{aligned}
C_{9}^{\ell\ell}(\mu_b)&=\frac{\alpha}{27\pi}\bigg(14+9\log\!\bigg(\!\frac{\mu_{b}^2}{M^2}\!\bigg)\bigg)C_{9}^{\tau\tau}\,,\\
C_{7}(\mu_{b})&=-\frac{5\alpha}{36\pi}\bigg(\frac{27}{11}\eta^{\frac{16}{23}}-\frac{48}{33}\eta^{\frac{14}{23}}\bigg)C_{9}^{\tau\tau}\,,
\end{aligned}
\end{align}
neglecting the different running of $C_{7}$ from $\mu_{\text{LQ}}$ down to $m_{t}$. One can also relate these two coefficients, yielding
\begin{align}
C_{9}^{\ell\ell}(\mu_b)=-\frac{4}{15}\dfrac{14+9\log\!\Big(\!\frac{\mu_{b}^2}{M^2}\!\Big)}{\frac{27}{11}\eta^{\frac{16}{23}}-\frac{48}{33}\eta^{\frac{14}{23}}}C_{7}(\mu_b)\,.
\end{align}

\begin{figure}
	\centering
	\includegraphics[width=0.9\textwidth]{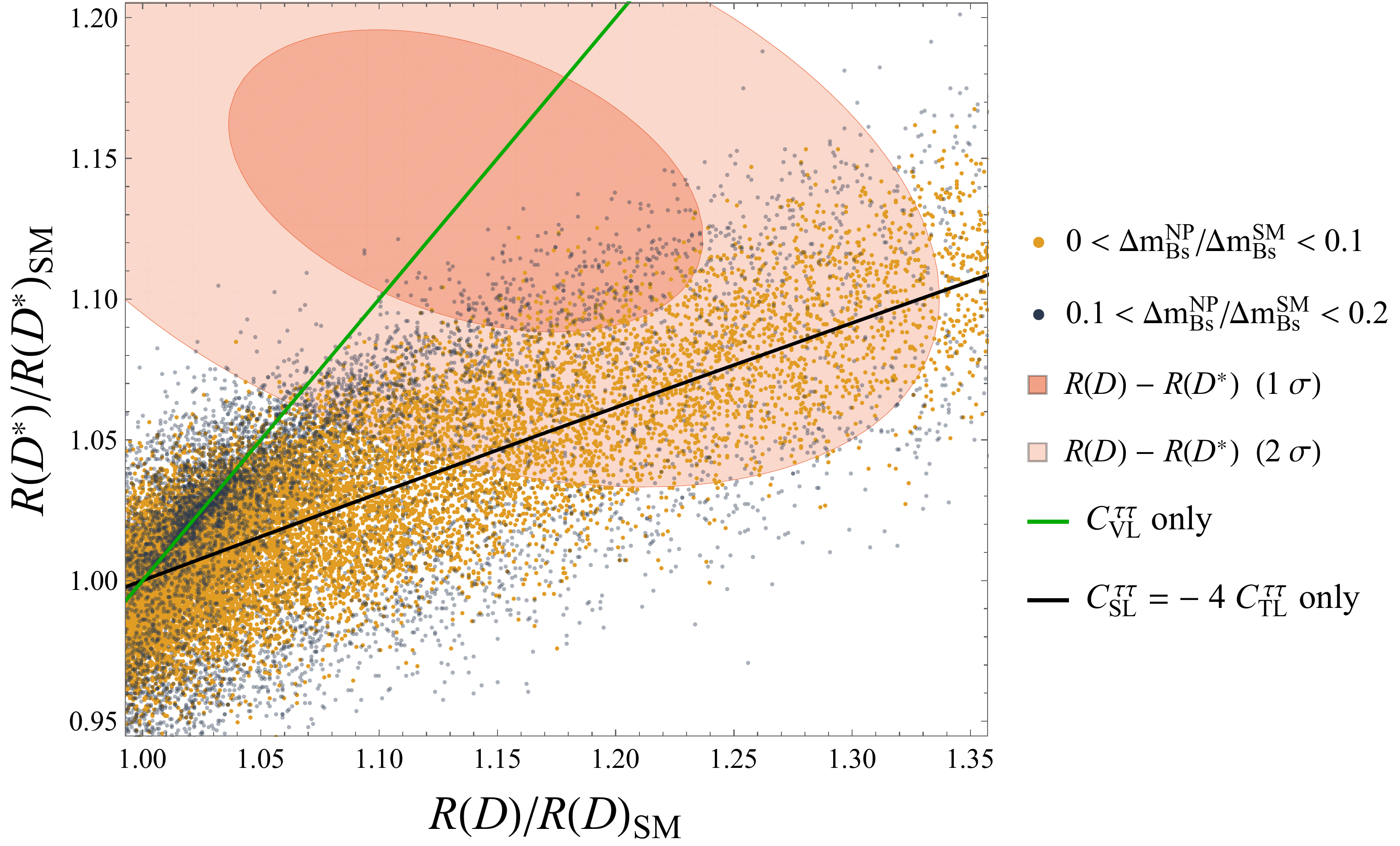}
	\caption{Correlation between $R(D)$ and $R(D^*)$, both normalized to their SM values. The (light) red ellipse shows the preferred region at the $1\sigma$ ($2\sigma$) level. The yellow points yield an effect in $B_s-\bar{B}_{s}$ mixing of $< 10\%$ with respect to the SM, while for the blue points the NP effect is in the range of 10-20\%. Only points allowed by $b\to s\nu\nu$ are shown. The black (green) solid line depicts the scenario where one generates the vector (scalar and tensor) operator only. We scanned over the couplings $\lambda_{23,33}$, $\kappa_{23,33}$ and $\hat{\lambda}_{23} \in [-1.5,1.5]$ and the LQ masses $M_1=M_3 \equiv M \in [1,2] \rm{TeV}$.}
	\label{fig:RD_RDs}
\end{figure}

This situation is illustrated in Fig.~\ref{fig:Bstautau_RDs}, where we show the correlations between $B_s\to\tau^+\tau^-$ and $R(D^{(*)})$. Note that for left-handed couplings \mbox{$R(D)/R(D)_{\rm SM}=R(D^*)/R(D^*)_{\rm SM}$} is predicted. The bound from $B_s-\bar B_s$ mixing limits the possible effect, both in $B_s\to\tau^+\tau^-$ and $R(D^{(*)})$, depending on the LQ mass. Heavier LQs lead to larger effects in $B_s-\bar B_s$ with respect to $B_s\to\tau^+\tau^-$ and $R(D^{(*)})$ than lighter LQs. For the same scenario, i.e. only left-handed couplings to tau leptons, we also show corrections to the $W\tau\nu$ coupling in Fig.~\ref{fig:bladibladibla}. Note that effect of $\Phi_1$ has opposite sign than the one of $\Phi_3$. Furthermore, if one aims at increasing $R(D^{(*)})$, the effect of $\Phi_1$ ($\Phi_3$) in $W \to \tau \nu$ is destructive (constructive) such that it increases (decreases) the slight tension in $\tau\to\mu\nu\bar\nu$ data.
\smallskip

\begin{figure}
	\centering
		\centering
		\includegraphics[width=0.85\textwidth]{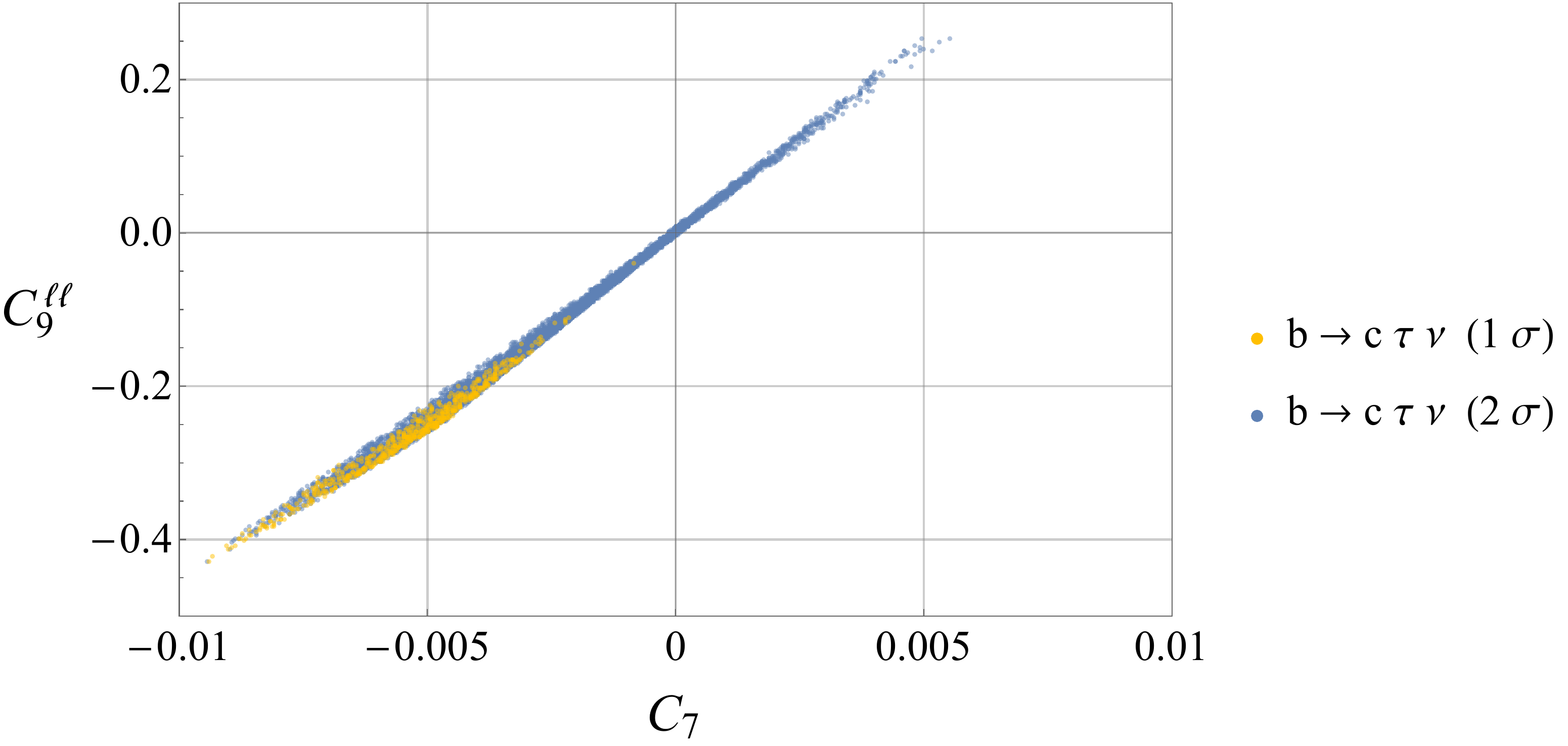}
		\caption{Correlations between $C_7$ and $C_{9}^{\ell\ell}$, both given at the $B$ meson scale. Here we imposed that the points satisfy $B_s-\bar B_s$ mixing (i.e. yield a maximal effect of 20\%) and lie within the $1\sigma$ (yellow) or $2\sigma$ (blue) region preferred by the global fit to  $b\to c\tau\nu$ data. Note that non-zero effects in $C_{7}(\mu_b)$ and $C_{9}^{\ell\ell}(\mu_b)$ are mandatory in order to explain $b\to c\tau\nu$ data at $1\sigma$ and that $C_{9}^{\ell\ell}(\mu_b)$ has the sign preferred by the fit if this is required. Both coefficients include $\mathcal{O}(\alpha_s)$ corrections. Again we scanned over the couplings $\lambda_{23,33}$, $\kappa_{23,33}$ and $\hat{\lambda}_{23} \in [-1.5,1.5]$ and the LQ masses $M_1=M_3 \equiv M \in [1,2] \rm{TeV}$.}
		\label{fig:C7-C9U}
\end{figure}

Next, let us allow for non-zero right-handed couplings $\hat\lambda_{23,33}$ of $\Phi_1$ to quarks and leptons. In this case the left-handed vector current encoded in $C_{VL}^{\tau\tau}$ (originating from $\Phi_1$ and $\Phi_3$ via $\lambda_{23,33}$ and $\kappa_{23,33}$ only) is now complemented by a $C_{SL}^{\tau\tau}=-4C_{TL}^{\tau\tau}$ effect from $\Phi_1$. This breaks the common rescaling of $R(D)/R(D)_{\rm SM}$ and $R(D^*)/R(D^*)_{\rm SM}$, depicted by the green line in Fig.~\ref{fig:RD_RDs}. The constraint from $B_s-\bar B_s$ only limits $C_{VL}$ but not $C_{SL}=-4C_{TL}$. The resulting correlations between $R(D)$ and $R(D^*)$ are shown in Fig. \ref{fig:RD_RDs}. One can see that for deviations of $R(D^{(*)})/R(D^{(*)})_{\rm SM}$ from unity of more than $\approx 10\%$, our model predicts $R(D)/R(D)_{\rm SM}>R(D^*)/R(D^*)_{\rm SM}$.\\
The size and correlation between $C_7$ and a LFU effect in $C_{9}^{\ell\ell}$, induced by the tau loop, is shown in Fig.~\ref{fig:C7-C9U}. Interestingly, to account for $b\to c\tau\nu$ data within $1\sigma$, we predict $-0.5<C_{9}^{\ell\ell}<-0.2$ (including right-handed couplings) which is in very good agreement with the global fit on $b\to s\ell^+\ell^-$ data, especially if it is complemented by a $C_{9}^{\mu\mu}=-C_{10}^{\mu\mu}$ LFUV effect~\cite{Alguero:2018nvb,Alguero:2019ptt}.
\smallskip

In the same way, $b\to d\tau\nu$ data can be addressed. Here, it was shown in Ref.~\cite{Crivellin:2019qnh} that already a 10\% effect with respect to the SM could lead to a neutron EDM observables in the near future.
\smallskip\\

\subsection{$b\to c\tau\nu$ and $b\to s\ell^+\ell^-$}
\begin{figure}
	\centering
		\centering
		\includegraphics[width=.45\linewidth]{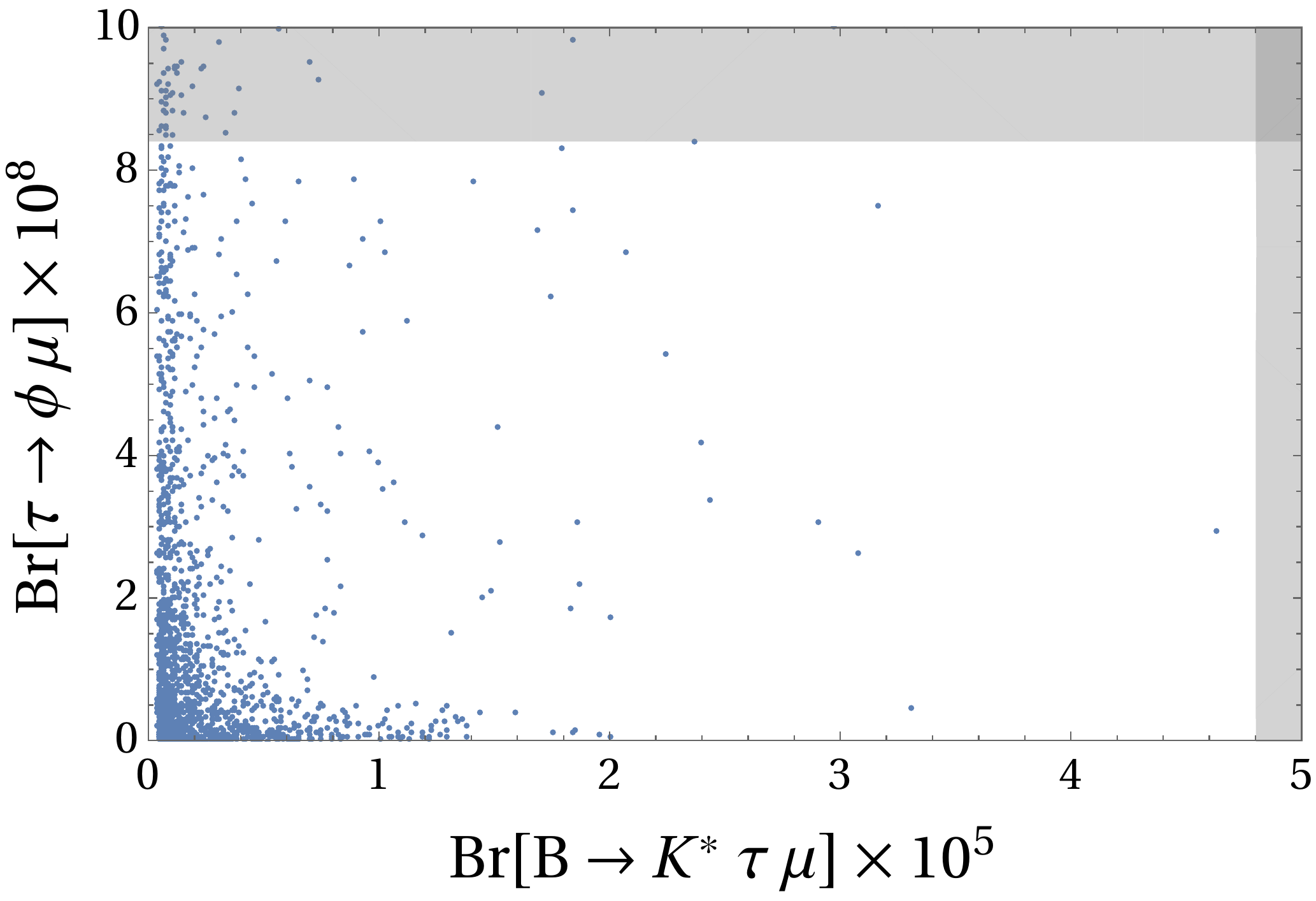}
		\includegraphics[width=.45\linewidth]{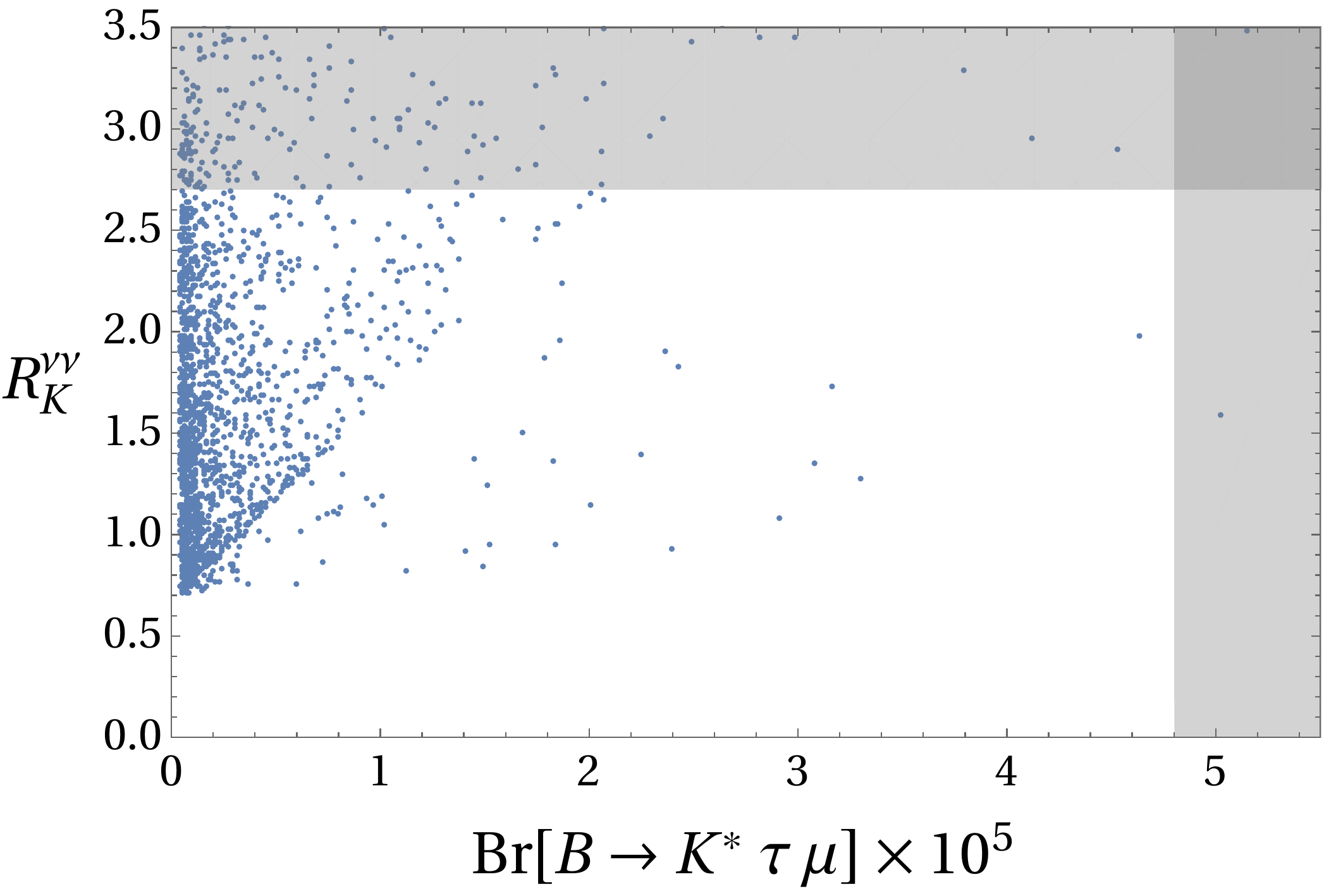}
		\caption{Correlations between $\text{Br}[B \to K^{*} \tau \mu]$ and $\text{Br}[\tau \to \phi \mu]$ (left) and between \mbox{$\text{Br}[B \to K^{*} \tau \mu]$} and $R_K^{\nu \bar\nu}$ (right). The blue points lie within the $1\sigma$ ranges of both the $b \to c \tau \nu$ and $b \to s \ell^{+} \ell^{-}$ fits, give an effect of less than 20\% to $B_s-\bar{B}_s$ mixing and do not violate any other constraints. We scanned over the couplings \mbox{$\{\lambda_{23,33},\kappa_{23,33},\hat{\lambda}_{23}\} \in [-1.5,1.5]$}, $\{\lambda_{22,32},\kappa_{22,32}\} \in [-0.3,0.3]$ and the LQ masses \mbox{$M_1=M_3 \equiv M \in [1,2] \, \rm{TeV}$.}}
		\label{fig:B_Ktaumu_tau_phimu}
\end{figure}

Let us now turn to the case where we allow for couplings to left-handed muons as well. Here, it is clear that, disregarding for the moment $R(D^{(*)})$ and thus tau couplings, one can explain $b\to s\ell^+\ell^-$ data with a tree-level $C_9^{\mu\mu}=-C_{10}^{\mu\mu}$ effect from $\Phi_3$ without running into the danger of violating bounds from other flavor observables. However, the situation gets more interesting if one aims at explaining $b\to s\ell^+\ell^-$ and $b\to c\tau\nu$ data simultaneously.  In this case LFV $\tau-\mu$ effects necessarily arise e.g. in $B\to K\tau\mu$, $\tau\to\phi\mu$, $Z\to\tau\mu$ and $\tau\to3\mu$. Note that our model does not possess scalar currents in the down sector, therefore $B_s\to\tau\mu$ does not receive a chiral enhancement. The correlations between $B\to K\tau\mu$ and $\tau\to\phi\mu$ are shown in Fig.~\ref{fig:B_Ktaumu_tau_phimu}, finding that they are in general anti-correlated despite fine-tuned points.
\smallskip

\subsection{$b\to c\tau\nu$, $b\to s\ell^+\ell^-$ and $a_\mu$}

\begin{figure}
	\centering
	\includegraphics[width=0.6\textwidth]{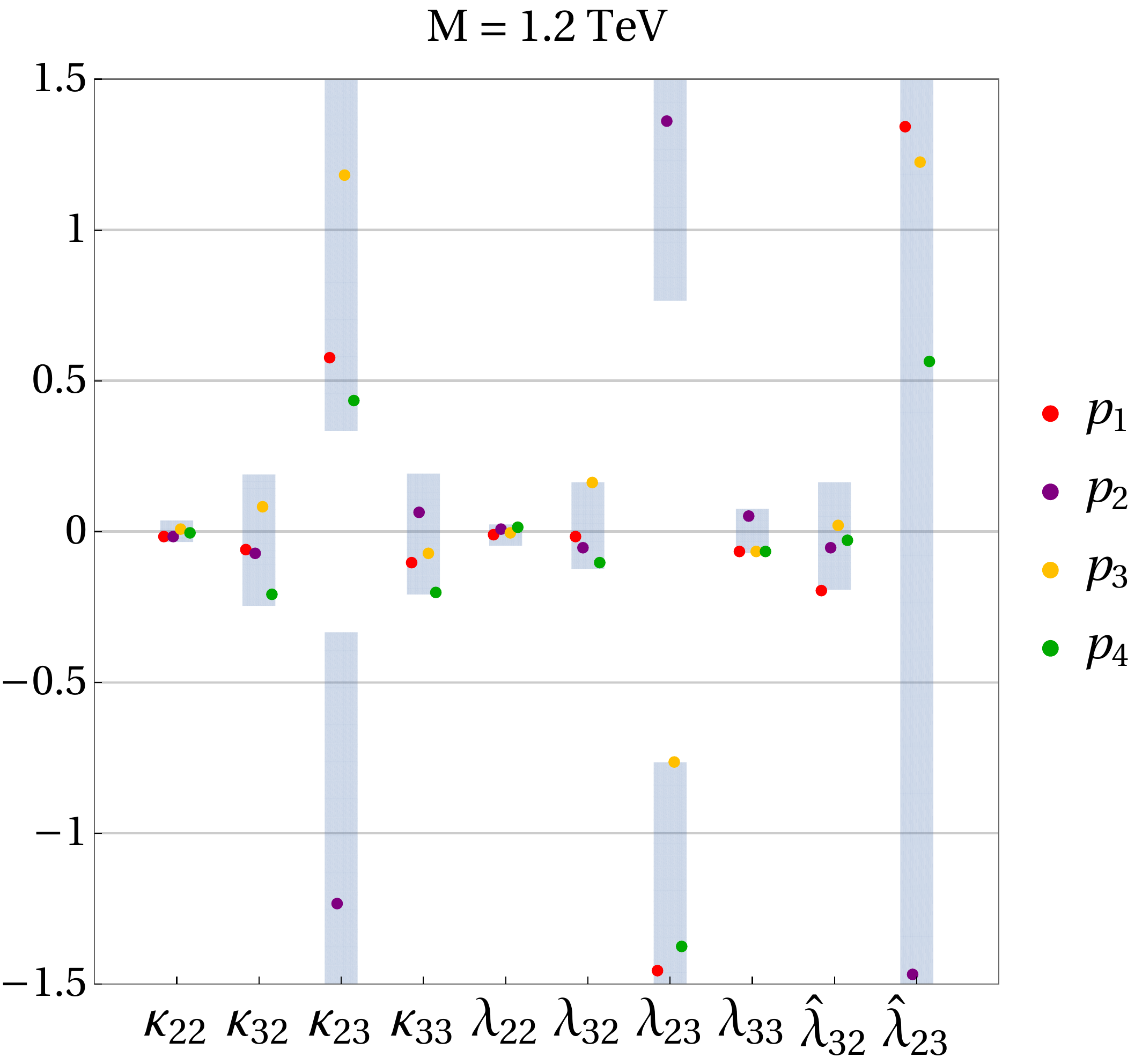}
	\caption{Possible ranges for the couplings of the points in parameter space which can explain all three anomalies at the $1\sigma$ level. We found these points by performing a parameter scan over the couplings $\{\lambda_{23,33},\kappa_{23,33},\hat{\lambda}_{23}\} \in [-1.5,1.5]$, $\{\lambda_{22,32},\kappa_{22,32},\hat{\lambda}_{32}\} \in [-0.3,0.3]$ and by setting the LQ masses $M_1=M_3 = 1.2 \, \rm{TeV}$. In color we depict the values of the four benchmark points given in Tab.~\ref{tab:benchpoints}. We found roughly 350 points that passed all constraints at the 95\% C.L. while allowing for an effect in $B_s - \bar{B}_s$ mixing of up to 30\%.}
	\label{fig:benchmarkplot}
\end{figure}

\begin{figure}
	\centering
	\includegraphics[width=0.6\textwidth]{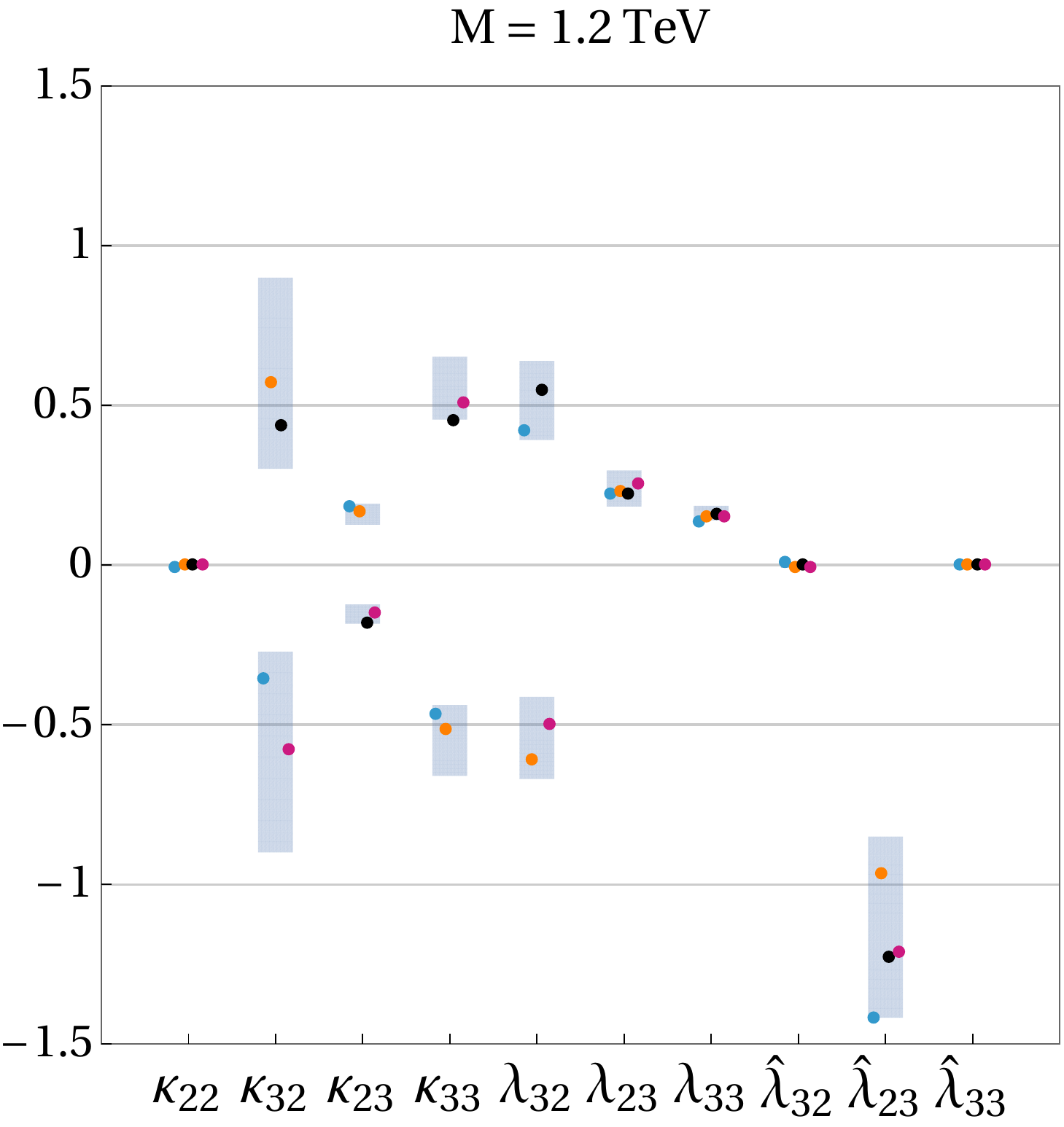}
	\caption{Possible ranges for the couplings of the points in parameter space which can explain all three anomalies at the $1\sigma$ level and are compatible with $D^{0}-\bar{D}^{0}$ mixing. We found these points by performing a parameter scan over the couplings $\{\lambda_{23,33},\kappa_{23,33},\hat{\lambda}_{23,33}\} \in [-1.5,1.5]$, $\{\lambda_{32},\kappa_{22,32},\hat{\lambda}_{32}\} \in [-0.3,0.3]$ and by setting the LQ masses $M_1=M_3 = 1.2 \, \rm{TeV}$. In color we depict the values of the four benchmark points given in Tab.~\ref{tab:benchpoints_new}.}
	\label{fig:benchmarkplot_new}
\end{figure}

Finally, we aim at explaining the anomaly in the AMM of the muon in addition to $b\to c\tau\nu$ and $b\to s\ell^+\ell^-$ data. Accounting for $\delta a_\mu$ alone is possible and the only unavoidable effect occurs in $Z\to\mu^+\mu^-$, which can however only be tested at the FCC-ee~\cite{ColuccioLeskow:2016dox}. Furthermore, explaining $\delta a_\mu$ together with $b\to s\ell^+\ell^-$ data does not pose a problem either since $\Phi_1$ can account for $\delta a_\mu$ while $\Phi_3$ can explain $b\to s\ell^+\ell^-$. However, once one wants to account for $b\to c\tau\nu$ data the situation becomes non-trivial. Scanning over 10 million points\footnote{First we individually scanned over two million points for couplings to muons only and over one million points for couplings to taus only. From each of both datasets roughly 3500 points passed all constraints while lying in the $1\sigma$ range of the global fits for $b \to s \ell^{+} \ell^{-}$ or $b \to c \tau \nu$, respectively. The combination of the two datasets was then used as seed for the final scan over all parameters.} we found approximately 350 points which can explain all three anomalies at the same time. The corresponding range for the couplings of these 350 points is shown in Fig.~\ref{fig:benchmarkplot}. Only allowing for an effect of 20\% in $B_s - \bar{B}_s$ mixing, the number of points is reduced to 40, where an effect as low as 10\% is possible. In addition, we choose (out of these 350 points) four benchmark points, shown in color in Fig.~\ref{fig:benchmarkplot}. The predictions for the various observables for these benchmark points are given in Tab.~\ref{tab:benchpoints}. Interestingly, even though in general $\tau \to \mu \gamma$ represents the most restrictive constraint on our model in case one aims at an explanation of all three anomalies, we still find points that give a relatively small contribution of roughly one order of magnitude below the current experimental bound. The branching ratio of $B_s \to \tau^{-} \tau^{+}$ is enhanced by a factor of roughly 100 with respect to the SM, which also is below the current experimental bound. While the effects in $\Lambda^W_{\tau \nu}$ are small, they are always positive, reducing the slight tension in the effective $W\tau\nu$ coupling. The effects in $B \to K \tau \mu$ and $\tau \to \phi \mu$ range from being negligible to close to the current experimental bounds while effects in $\tau \to \mu ee$ and $\tau \to 3\mu$ lie roughly two orders of magnitude below the current experimental limit. Furthermore, the effects in $Z\to\tau^{-}\tau^{+}$ would clearly be measurable at an FCC-ee~\cite{Abada:2019zxq}.
\smallskip

\begin{table}
\centering
	\resizebox{0.95\textwidth}{!}
	{
	\begin{tabular}{|p{0.1mm}p{3mm}|c|c|c|c|c|c|c|c|c|c|}
		\cline{3-12}
		\multicolumn{2}{c|}{} & $\kappa_{22}$ & $\kappa_{32}$ & $\kappa_{23}$ & $\kappa_{33}$ & $\lambda_{22}$ & $\lambda_{32}$ & $\lambda_{23}$ & $\lambda_{33}$ & $\hat{\lambda}_{32}$ & $\hat{\lambda}_{23}$\\
		\hline
		\tikz\draw[point1,fill=point1] (0,0) circle (.7ex); & $p_{1}$ & $-0.019$ & $-0.059$ & $0.58$ & $-0.11$ & $-0.0082$ & $-0.016$ & $-1.46$ & $-0.064$ & $-0.19$ & $1.34$ \\
		\tikz\draw[point2,fill=point2] (0,0) circle (.7ex); & $p_{2}$ & $-0.017$ & $-0.070$ & $-1.23$ & $0.066$ & $0.0078$ & $-0.055$ & $1.36$ & $0.052$ & $-0.053$ & $-1.47$ \\
		\tikz\draw[point3,fill=point3] (0,0) circle (.7ex); & $p_{3}$ & $0.0080$ & $0.081$ & $1.18$ & $-0.073$ & $-0.0017$ & $0.16$ & $-0.76$ & $-0.068$ & $0.023$ & $1.23$ \\
		\tikz\draw[point4,fill=point4] (0,0) circle (.7ex); & $p_{4}$ & $-0.0032$ & $-0.21$ & $0.44$ & $-0.20$ & $0.014$ & $-0.10$ & $-1.38$ & $-0.068$ & $-0.032$ & $0.57$ \\
		\hline
	\end{tabular}
	}
	\resizebox{0.95\textwidth}{!}
	{
	\begin{tabular}{|p{0.1mm}p{5.5mm}|c|c|c|c|c|c|c|c|c|}
		\cline{3-11}
		\multicolumn{2}{c|}{} & $C_{9}^{\mu\mu}=-C_{10}^{\mu\mu}$ & $C_{9}^{\ell\ell}$ & $\dfrac{R(D)}{R(D)_\text{SM}}$ & $\dfrac{R(D^{*})}{R(D^{*})_\text{SM}}$ & $\dfrac{B_s \to \tau \tau}{B_{s}\to\tau\tau\big|_\text{SM}}$ & \makecell{$\tau\to\mu\gamma$\\$\times 10^{8}$} & \makecell{$\delta a_{\mu}$ \\ $\times 10^{11}$} &   \makecell{$\tilde{V}_{cb}^{e}/\tilde{V}_{cb}^{\mu}-1$ \\ $\times 10^6$} & \makecell{$Z \rightarrow \tau \mu$\\$ \times 10^{10}$}\\
		\hline
		\tikz\draw[point1,fill=point1] (0,0) circle (.7ex); & $p_{1}$ & $-0.52$ & $-0.21$ & $1.15$ & $1.10$ & $59.88$ & $4.35$ & $207$ & $291$ & $0.117$  \\
		\tikz\draw[point2,fill=point2] (0,0) circle (.7ex); & $p_{2}$ & $-0.56$ & $-0.28$ & $1.14$ & $1.10$ & $99.76$ & $0.766$ & $199$ & $448$ & $2.38$  \\
		\tikz\draw[point3,fill=point3] (0,0) circle (.7ex); &$p_{3}$ & $-0.31$ & $-0.31$ & $1.14$ & $1.09$ & $112.5$ & $3.62$ & $255$ & $17$ & $0.129$ \\
		\tikz\draw[point4,fill=point4] (0,0) circle (.7ex); & $p_{4}$ & $-0.31$ & $-0.31$ & $1.13$ & $1.11$ & $112.5$ & $0.734$ & $230$ & $934$ & $45.6$ \\
		\hline
		\multicolumn{2}{c|}{} & $C_{SL}^{\tau\tau}=-4C_{TL}^{\tau\tau}$ & $C_{VL}^{\tau\tau}$ & $R^{K^{(*)}}_{\nu\bar{\nu}}$ & $\dfrac{\Delta m_{B_s}^{\text{NP}}}{\Delta m_{B_s}^{\text{SM}}}$ & \makecell{$B\to K \tau\mu$\\$\times 10^{5}$} & \makecell{$\tau\to\phi\mu$\\$\times 10^{8}$} & \makecell{$\tau \rightarrow \mu e e$ \\  $\times 10^{11}$} & \makecell{$|\Lambda_{33}^{\text{LQ}}(0)|$\\ $\times 10^5$} & \makecell{$\dfrac{\Delta_{33}^{L}(m_Z^2)}{\Lambda^{L\ell}_{\text{SM}}\times 10^{-5}}$ } \\
		\hline
		\tikz\draw[point1,fill=point1] (0,0) circle (.7ex); & $p_{1}$ & $0.023$ & $0.040$ & $2.33$ & $0.1$ & $0.512$ & $1.27$ & $44.94$ & $1.11$ &  $-3.64$\\
		\tikz\draw[point2,fill=point2] (0,0) circle (.7ex); & $p_{2}$ &  $0.020$ & $0.040$ & $0.87$ & $0.16$ & $3.32$ & $4.73$ & $7.783$ & $0.90$ &  $-3.02$\\
		\tikz\draw[point3,fill=point3] (0,0) circle (.7ex); &$p_{3}$ &  $0.023$ & $0.037$ & $1.08$ & $0.19$ & $4.07$ & $1.00$ & $37.89$ & $0.89$ & $-3.51$ \\
		\tikz\draw[point4,fill=point4] (0,0) circle (.7ex); & $p_{4}$ & $0.010$ & $0.047$ & $2.43$ & $0.18$ & $3.69$ & $0.0021$ & $18.60$ & $3.12$ & $-10.04$\\
		\hline
	\end{tabular}
	}
	\caption{$p_1$-$p_4$ are four benchmark points that can simultaneously explain all three flavor anomalies ($b\to s\ell^+\ell^-$, $b\to c\tau\nu$ and $\delta a_\mu$) at the $1\sigma$ level and pass all other constraints at the 95\%C.L.. Here we show the values for the fermion couplings, the results for $b\to s\ell^+\ell^-$, $b\to c\tau\nu$ and $\delta a_\mu$ as well as the predictions for several flavor observables which can be measured in the future. Note that the effect in $\tau \to 3 \mu$ (not depicted here) is of comparable size as the one in $\tau \to \mu e e$. The LQ masses were set to $M=M_1=M_3=1.2\,\text{TeV}$.}
\label{tab:benchpoints}
\end{table}

\subsubsection{Update}
In Ref.~\cite{Gherardi:2020qhc} it was pointed out that our benchmark points in Tab.~\ref{tab:benchpoints} are in conflict with $D^0-\bar D^0$ mixing. While this is true under the assumption of a vanishing SM contribution, we point out that with fine-tuning between the SM and the NP contribution our model is not excluded, as the SM effect can currently not be calculated. Furthermore, it was claimed that our points are in tension with $D_s \to \tau \nu$. While it is true that for points $p_1$ and $p_2$ in Tab.~\ref{tab:benchpoints} a very slight tension with the experiment  (below the $2 \sigma$ level) is observed, the points $p_3$ and $p_4$ do not suffer from any tension at all. Nevertheless we decided to extend our analysis and present 4 new benchmark points in Tab.~\ref{tab:benchpoints_new}. In this scenario we neglected the coupling $\lambda_{22}$ but used the coupling $\hat{\lambda}_{33}$ to tune $\tau \to \mu \gamma$ as suggested in Ref.~\cite{Gherardi:2020qhc}. In Fig.~\ref{fig:benchmarkplot_new} we show the allowed range for the new benchmark points. Note that these benchmark points are favored compared to the ones in Ref.~\cite{Gherardi:2020qhc} by $\tau$ pair searches (see e.g. Ref.~\cite{Buttazzo:2017ixm}) as our couplings to charm quarks are smaller and the LQ mass bigger.
\begin{table}
\centering
	\resizebox{0.95\textwidth}{!}
	{
	\begin{tabular}{|p{0.1mm}p{3mm}|c|c|c|c|c|c|c|c|c|c|}
		\cline{3-12}
		\multicolumn{2}{c|}{} & $\kappa_{22}$ & $\kappa_{32}$ & $\kappa_{23}$ & $\kappa_{33}$  & $\lambda_{32}$ & $\lambda_{23}$ & $\lambda_{33}$ & $\hat{\lambda}_{32}$ & $\hat{\lambda}_{23}$& $\hat{\lambda}_{33}$\\
		\hline
		\tikz\draw[point5,fill=point5] (0,0) circle (.7ex); & $p_{5}$  & $-0.0024$ & $-0.36$ & $0.18$ & $-0.47$ & $0.42$ & $0.23$ & $0.14$ & $0.0074$ & $-1.42$ & $0.0032$ \\
		\tikz\draw[point6,fill=point6] (0,0) circle (.7ex); & $p_{6}$ & $0.0020$ & $0.57$ &  $0.16$ &  $-0.52$ & $-0.61$ & $0.24$ & $0.16$ & $-0.0058$ & $-0.97$ & $0.003$ \\
		\tikz\draw[point7,fill=point7] (0,0) circle (.7ex); & $p_{7}$ & $0.0022$ & $0.44$ & $-0.18$ & $0.46$ & $0.55$ & $0.23$ & $0.16$ & $0.0055$ & $-1.23$ & $0.003$ \\
		\tikz\draw[point8,fill=point8] (0,0) circle (.7ex); & $p_{8}$ & $-0.0017$ & $-0.58$ & $-0.15$ & $0.51$ & $-0.50$ & $0.26$ & $0.15$ & $-0.0064$ & $-1.21$ & $0.003$ \\
		\hline
	\end{tabular}
	}
	\resizebox{0.95\textwidth}{!}
	{	
	\begin{tabular}{|p{0.1mm}p{5.5mm}|c|c|c|c|c|c|c|c|c|}
		\cline{3-11}
		\multicolumn{2}{c|}{} & $C_{9}^{\mu\mu}=-C_{10}^{\mu\mu}$ & $C_{9}^{\ell\ell}$ & $\dfrac{R(D)}{R(D)_\text{SM}}$ & $\dfrac{R(D^{*})}{R(D^{*})_\text{SM}}$ & $\dfrac{B_s \to \tau \tau}{B_{s}\to\tau\tau\big|_\text{SM}}$ & \makecell{$\tau\to\mu\gamma$\\$\times 10^{8}$} & \makecell{$\delta a_{\mu}$ \\ $\times 10^{11}$} &   \makecell{$\tilde{V}_{cb}^{e}/\tilde{V}_{cb}^{\mu}-1$ \\ $\times 10^6$} & \makecell{$Z \rightarrow \tau \mu$\\$ \times 10^{8}$}\\
		\hline
		\tikz\draw[point5,fill=point5] (0,0) circle (.7ex); & $p_{5}$ & $-0.41$ & $-0.30$ & $1.21$ & $1.11$ & $107$ & $4.23$ & $210$ & $-304$ & $8.20$  \\
		\tikz\draw[point6,fill=point6] (0,0) circle (.7ex); & $p_{6}$ & $-0.54$ & $-0.30$ & $1.17$ & $1.10$ & $108$ & $3.63$ & $238$ & $-224$ & $25.19$  \\
		\tikz\draw[point7,fill=point7] (0,0) circle (.7ex); &$p_{7}$ & $-0.45$ & $-0.29$ & $1.21$ & $1.11$ & $105$ & $3.07$ & $202$ & $-959$ & $12.07$ \\
		\tikz\draw[point8,fill=point8] (0,0) circle (.7ex); & $p_{8}$ & $-0.48$ & $-0.27$ & $1.20$ & $1.10$ & $93$ & $3.67$ & $217$ & $1194$ & $24.86$  \\
		\hline
		\multicolumn{2}{c|}{} & $C_{SL}^{\tau\tau}=-4C_{TL}^{\tau\tau}$ & $C_{VL}^{\tau\tau}$ & $R^{K^{(*)}}_{\nu\bar{\nu}}$ & $\dfrac{\Delta m_{B_s}^{\text{NP}}}{\Delta m_{B_s}^{\text{SM}}}$ & \makecell{$B\to K \tau\mu$\\$\times 10^{5}$} & \makecell{$\tau\to\phi\mu$\\$\times 10^{10}$} & \makecell{$\tau \rightarrow \mu e e$ \\  $\times 10^{9}$} & \makecell{$|\Lambda_{33}^{\text{LQ}}(0)|$\\ $\times 10^5$} & \makecell{$\dfrac{\Delta_{33}^{L}(m_Z^2)}{\Lambda^{L\ell}_{\text{SM}}\times 10^{-5}}$ } \\
		\hline
		\tikz\draw[point5,fill=point5] (0,0) circle (.7ex); & $p_{5}$ & $0.054$ & $0.028$ & $1.34$ & $0.19$ & $1.88$ & $0.21$ & $2.74$ & $13.76$ &  $-50.76$\\
		\tikz\draw[point6,fill=point6] (0,0) circle (.7ex); & $p_{6}$ &  $0.041$ & $0.029$ & $1.96$ & $0.19$ & $3.90$ & $0.12$ & $7.36$ & $16.78$ &  $-62.07$\\
		\tikz\draw[point7,fill=point7] (0,0) circle (.7ex); &$p_{7}$ &  $0.052$ & $0.029$ & $1.69$ & $0.19$ & $2.86$ & $0.18$ & $3.85$ & $12.80$ & $-49.77$ \\
		\tikz\draw[point8,fill=point8] (0,0) circle (.7ex); & $p_{8}$ & $0.050$ & $0.028$ & $1.46$ & $0.16$ & $3.44$ & $0.08$ & $7.19$ & $16.59$ & $-60.97$\\
		\hline
	\end{tabular}
	}
	\caption{Four benchmark points ($p_5$-$p_8$) that simultaneously explain $b\to s\ell^+\ell^-$, $b\to c\tau\nu$ and $\delta a_\mu$ at the $1\sigma$ level and pass all other constraints at the 95\%C.L.. Contrary to $p_1$-$p_4$, they do not yield a big effect in $D^{0}-\bar{D}^{0}$ mixing and therefore do not require fine-tuning with the SM contribution. Again the LQ masses were chosen to be $M=M_1=M_3=1.2\,\text{TeV}$.}
\label{tab:benchpoints_new}
\end{table}

\section{Conclusions}
\label{conclusions}

Motivated by the intriguing hints for LFU violating NP in $R(D^{(*)})$, $b\to s\ell^+\ell^-$ processes and $a_\mu$, we studied the flavor phenomenology of the LQ singlet-triplet model. We first defined the most general setup for the model, including an arbitrary number of LQ "generations" as well as mixing among them. With this at hand, we performed the matching of the model on the effective low energy theory and related the Wilson coefficients to flavor observables. Here, we included the potentially relevant loop effects, e.g. in $B_s-\bar B_s$ mixing, $b\to s\gamma$, LFU contributions to $C_9^{\ell\ell}$ and $a_\mu$, as well as in modified $Z$ and $W$ couplings.
\smallskip

Our phenomenological analysis proceeded in three steps: First, we disregarded the anomalies related to muons and considered the possibility of explaining $R(D^{(*)})$ and the resulting implication for other observables. We found that, including only couplings to left-handed fermions, the size of the possible effect depends crucially on the mass of the LQ: the larger (smaller) the mass (couplings) the bigger the relative effect in $B_s-\bar B_s$. Together with $b\to s\nu\bar\nu$, this is the limiting factor here. For $M=1\,$TeV and values of $\kappa_{33}$ up to $\pm 2$, a 20\% effect in $R(D^{(*)})$ is possible, while for $M=1.5\,$TeV and $|\kappa_{33}|<1$ only a 10\% effect with respect to the SM can be generated (see Fig.~\ref{fig:Bstautau_RDs}). At the same time, an enhancement of $B_s\to\tau^+\tau^-$ of the order of $10^2$ is predicted, which, via loop effects, leads to a LFU $C_9^{\ell\ell}\approx-0.3$. Once couplings to right-handed leptons are included, larger effects in $b\to c\tau\nu$ processes are possible and $R(D)/R(D)_{\rm SM}>R(D^*)/R(D)_{\rm SM}^*$ is predicted, see Figs.~\ref{fig:RD_RDs}~and~\ref{fig:C7-C9U}. 
\smallskip

In a second step, we aimed at a simultaneous explanation of $b\to s\ell^+\ell^-$ data together with $R(D^{(*)})$. In this case, effects in lepton flavor violating processes like $B\to K\tau\mu$ and $\tau\to\phi\mu$ are predicted as shown in Fig.~\ref{fig:B_Ktaumu_tau_phimu}. These effects are still compatible with current data but can be tested soon by LHCb and BELLE II.
\smallskip

Finally, including in addition the AMM of the muon in the analysis is challenging since then right-handed couplings to muons are required which, together with the couplings needed to explain $R(D^{(*)})$, lead to chirally enhanced effects in $\tau\to\mu\gamma$. It is still possible to find a common solution to all three anomalies but only a small region of the parameter space can do this. Nonetheless, we identified four benchmark points which can achieve such a simultaneous explanation to all three anomalies (see Fig.~\ref{fig:benchmarkplot}). 
\smallskip

In summary, the LQ singlet-triplet model is a prime candidate for explaining the flavor anomalies and we would like to emphasize that there is no renormalizable model on the market which is more minimal (only two new particles are needed here) and capable to address all three prominent flavor anomalies together.
\smallskip

{\it Acknowledgments} --- {We thank Christoph Greub for useful comments on the manuscript. The work of A.C. and D.M. is supported by a Professorship Grant (PP00P2\_176884) of the Swiss National Science Foundation. The work of F.S. is supported by the Swiss National Foundation grant 200020\_175449/1.}

\appendix
\section*{Appendix}
\addtocounter{section}{1}
\addcontentsline{toc}{section}{Appendix}

In this appendix we define the loop functions appearing in the calculation of the observables and give the most general expressions for the Wilson coefficients, including multiple LQ generations ($N$ singlets $\Phi_1$, $M$ triplets $\Phi_3$) and mixing among them. Let us recapitulate the definition of the masses:
\begin{itemize}
	\item The singlet and triplet representations with electromagnetic charge $Q_{em}=-1/3$ have the masses $m_{K}$ with $K=\{1,...,M+N\}$.
	\item The LQ with electromagnetic charge $Q_{em}=2/3$ and $Q_{em}=-4/3$, stemming from the triplet representations, have the same masses $\bar{m}_{J}$ with $J=\{1,...,M\}$.
\end{itemize}

\subsection{Loop Functions}

Throughout this article we used the loop functions $C_{0}$ and $D_{0,2}$, defined as
\begin{align}
\begin{aligned}
\frac{i}{16\pi^2}C_{0}(m_{0}^2,m_{1}^2,m_{2}^2)& =\mu^{2\epsilon}\int \frac{d^{D}\ell}{(2 \pi)^D} \frac{1}{\big(\ell^2-m_{0}^2\big)\big(\ell^2-m_{1}^2\big)\big(\ell^2-m_2^2\big)}\,,\\
\frac{i}{16\pi^2}D_{0}(m_{0}^2,m_{1}^2,m_{2}^2,m_{3}^2)&= \mu^{2\epsilon}\int \frac{d^{D}\ell}{(2 \pi)^D} \frac{1}{\big(\ell^2-m_{0}^2\big)\big(\ell^2-m_{1}^2\big)\big(\ell^2-m_2^2\big)\big(\ell^2-m_3^2\big)}\,,\\
\frac{i}{16\pi^2}D_{2}(m_{0}^2,m_{1}^2,m_{2}^2,m_{3}^2)&= \mu^{2\epsilon}\int \frac{d^{D}\ell}{(2 \pi)^D} \frac{\ell^2}{\big(\ell^2-m_{0}^2\big)\big(\ell^2-m_{1}^2\big)\big(\ell^2-m_2^2\big)\big(\ell^2-m_3^2\big)}\,,
\end{aligned}
\end{align}
with $D=4-2\epsilon$.
\smallskip

\begin{boldmath}
\subsection{$dd\ell\ell$}
\end{boldmath}
	
For $d_k\to d_j\ell_f^-\ell_i^+$ processes we match on the effective operators defined in \eq{eq:effHam}. The tree-level contribution gives
\begin{align}
C_{9,jk}^{fi}&=-C_{10,jk}^{fi}= \frac{{\sqrt 2}} {{4{G_F}{V_{td_k}}V_{td_j}^*}}\frac{\pi }{\alpha }\sum_{J=1}^{M}\frac{\Gamma^{J}_{d_{k}\ell_{i}}\Gamma_{d_{j}\ell_{f}}^{J*}}{{{\bar{m}_{J}^2}}}\,,
\end{align}
while the loop calculations yield
\begin{align}
\begin{split}
C_{7}^{jk}(\mu_{\text{LQ}})&=\frac{-\sqrt{2}}{4G_F V_{td_k}V_{td_j}^{*}}\Bigg[\frac{1}{72}\sum_{K=1}^{N+M}\frac{\Gamma_{d_{k}\nu_{i}}^{L,K}\Gamma_{d_{j}\nu_{i}}^{L,K*}}{m_{K}^2} +\frac{5}{36}\sum_{J=1}^{M}\frac{\Gamma_{d_{k}\ell_{i}}^{J}\Gamma_{d_{j}\ell_{i}}^{J*}}{\bar{m}_{J}^2}\Bigg]\,,\\
C_{8}^{jk}(\mu_{\text{LQ}})&=\frac{\sqrt{2}}{4G_F V_{td_k}V_{td_j}^{*}}\frac{1}{24}\Bigg[\sum_{K=1}^{N+M}\frac{\Gamma_{d_{k}\nu_{i}}^{L,K}\Gamma_{d_{j}\nu_{i}}^{L,K*}}{m_{K}^2} +\sum_{J=1}^{M}\frac{\Gamma_{d_{k}\ell_{i}}^{J}\Gamma_{d_{j}\ell_{i}}^{J*}}{\bar{m}_{J}^2}\Bigg]\,,\\
C_{9,jk}^{ii}(\mu_{\text{LQ}})&=\frac{\sqrt{2}}{216 G_{F}V_{td_{k}}V_{td_{j}}^{*}} \Bigg[ \sum_{K=1}^{N+M}\frac{\Gamma_{d_{k}\nu_{l}}^{L,K}\Gamma_{d_{j}\nu_{l}}^{L,K*}}{m_K^2} +2 \sum_{J=1}^{M}\frac{\Gamma_{d_{k}\ell_{l}}^{J}\Gamma_{d_{j}\ell_{l}}^{J*}}{\bar{m}_{J}^2}\! \left(\!14+9\log\!\left(\!\frac{\mu_{\text{LQ}}^2}{\bar{m}_{J}^2}\!\right)\!\!\right)\!\Bigg].
\end{split}
\end{align}
At the low scale of the processes, one has to include the effect of the diagram in the effective theory. This results in a so-called effective Wilson coefficient which also depends on the lepton mass in the loop and $q^2$
\begin{align}
{\mathcal{C}_{9,jk}^{ii \, \text{eff}}}(\mu)=\frac{\sqrt{2}}{216G_{F}V_{td_k}V_{td_j}^{*}}\Bigg[\sum_{K=1}^{N+M}\frac{\Gamma_{d_{j}\nu_{l}}^{L,K}\Gamma_{d_{k}\nu_{l}}^{L,K*}}{m_K^2}+2\sum_{J=1}^{M}\frac{\Gamma^{J}_{d_{j}\ell_{l}}\Gamma^{J*}_{d_{k}\ell_{l}}}{\bar{m}_{J}^2}\mathcal{F}\left(q^2,m_{\ell_{l}}^2,\bar{m}_{J}^{2},\mu^2\right)\Bigg]\,,
\end{align}
with
\begin{align}
\begin{split}
\mathcal{F}\left(q^2,m_{\ell}^2,M^2,\mu^2\right)&=\frac{1}{q^2}\left(9q^2\log\!\left(\!\frac{\mu^2}{M^2}\!\right)-q^2-36m_{\ell}^2\right)\\
&-\frac{18}{\left(q^2\right)^2\mathcal{X}(m_{\ell}^2,q^2)}\Big(\left(q^2\right)^2-2m_{\ell}^2q^2-8m_{\ell}^4\Big)\arctan\left(\frac{1}{\mathcal{X}(m_{\ell}^2,q^2)}\right)\,,
\end{split}
\end{align}
where we defined for convenience
\begin{align}
\mathcal{X}(a,b)=\sqrt{\frac{4a^2}{b^2}-1}\,.
\end{align}
\smallskip

\begin{boldmath}
	\subsection{$uu\gamma$ and EDM}
	\label{sec:app_EDM}
\end{boldmath}

We define the effective Hamiltonian as
\begin{equation}
{\cal H}_{{\rm{eff}}}^{{\rm{u \gamma}}} = C_\gamma^{jk}O_\gamma^{jk} + C_g^{jk}O_g^{jk} 
+ C_T^{jk\tau }O_T^{jk\tau }\,,
\end{equation}
with 
\begin{align}
\begin{aligned}
O_\gamma ^{jk} &= e\big[\bar u_j{\sigma^{\mu \nu }}{P_R}u_k\big]{F_{\mu \nu }}\,,\\
O_g^{jk} &= {g_s}\big[\bar u_j{\sigma^{\mu \nu }}{P_R}T^{a}u_k \, \big]{G_{\mu \nu }^a}\,,\\
O_T^{jk \tau } &= \big[\bar u_j{\sigma _{\mu \nu }}{P_R}u_k\big]\big[\bar \tau {\sigma ^{\mu \nu }}{P_R}\tau\big] \,,
\end{aligned}
\label{EDMoperators}
\end{align}
and obtain in the case of one generation of LQs and no mixing among them
\begin{align}
\begin{split}
{C}^{jk}_{\gamma}(\mu_{\text{LQ}})&=\frac{1}{1152 \pi^2}\Bigg[\!7\frac{m_{u_k}V_{kl}^{*}\lambda_{li}V_{jm}\lambda_{mi}^{*}+m_{u_j}{\hat{\lambda}}_{ki}\hat{\lambda}_{ji}^{*}}{M_{1}^2}-\frac{12m_{\ell_i}\hat{\lambda}_{ki}V_{jl}\lambda_{li}^{*}}{M_{1}^2}\bigg(\!4+3\log\!\bigg(\!\frac{\mu_{\text{LQ}}^2}{M_{1}^2}\!\bigg)\!\bigg)\\
&\qquad \qquad \phantom{12}+3\frac{m_{u_k}V_{kl}^{*}\kappa_{li}V_{jm}\kappa_{mi}^{*}}{M_{3}^2}\Bigg]\,,\\
{C}^{jk}_{g}(\mu_{\text{LQ}})&=-\frac{1}{384 \pi^2}\Bigg[\!\frac{m_{u_k}V_{kl}^{*}\lambda_{li}V_{jm}\lambda_{mi}^{*}+m_{u_j}\hat{\lambda}_{ki}\hat{\lambda}_{ji}^{*}}{M_{1}^{2}}+ \frac{6m_{\ell_i}\hat{\lambda}_{ki}V_{jl}\lambda_{li}^{*}}{M_{1}^2} +\frac{3m_{u_k}V_{kl}^{*}\kappa_{li}V_{jm}\kappa_{mi}^{*}}{M_{3}^2}\Bigg]\,, \\
C_T^{jk \tau}(\mu_{\text{LQ}}) &= \frac{V_{kl} \lambda_{l3}^* \hat{\lambda}_{j3}}{8 M_1^2}  \ .
\end{split}
\end{align}
The contributing diagram is depicted in Fig. \ref{fig:diagramm_bsll_EFT}.
For the neutron EDM we set $j=k=1$ and reproduce (setting $m_u=0$) our result from~\cite{Crivellin:2019qnh}, where also the relevant RGE can be found. In case of LQ mixing, we have
\begin{align}
\begin{aligned}
{C}_{7}^{jk}(\mu_{\text{LQ}})&=\frac{\sqrt{2}}{4G_{F}}\frac{1}{72}\Bigg[2\sum_{J=1}^{M}\frac{\Gamma_{u_{k}\nu_{i}}^{J}\Gamma_{u_{j}\nu_{i}}^{J*}}{\bar{m}_{J}^2}-\frac{7}{m_{u_k}}\sum_{K=1}^{M+N}\frac{m_{u_k}\Gamma^{L,K}_{u_{k}\ell_{i}}\Gamma_{u_{j}\ell_{i}}^{L,K*}+m_{u_j}\Gamma^{R,K}_{u_{k}\ell_{i}}\Gamma^{R,K*}_{u_{j}\ell_{i}}}{m_{K}^2}\\
&\hspace{20mm}+12\sum_{K=1}^{M+N}\frac{m_{\ell_{i}}}{m_{u_k}}\frac{\Gamma_{u_{k}\ell_{i}}^{R,K}\Gamma_{u_{j}\ell_{i}}^{L,K*}}{m_{K}^2}\bigg(4+3\log\!\bigg(\!\frac{\mu_{\text{LQ}}^2}{m_{K}^2}\!\bigg)\bigg)\Bigg]\,, \\
{C}_{8}^{jk}(\mu_{\text{LQ}})&=\frac{\sqrt{2}}{4G_{F}}\frac{1}{24}\Bigg[\sum_{J=1}^{M}\frac{\Gamma_{u_{k}\nu_{i}}^{J}\Gamma_{u_{j}\nu_{i}}^{J*}}{\bar{m}_{J}^2}+\frac{1}{m_{u_k}}\sum_{K=1}^{M+N}\frac{m_{u_k}\Gamma_{u_{k}\ell_{i}}^{L,K}\Gamma_{u_{j}\ell_{i}}^{L,K*}+m_{u_j}\Gamma_{u_{k}\ell_{i}}^{R,K}\Gamma_{u_{j}\ell_{i}}^{R,K*}}{m_{K}^2}\\
&\hspace{20mm}+6\sum_{K=1}^{M+N}\frac{m_{\ell_i}}{m_{u_k}}\frac{\Gamma_{u_{k}\ell_{i}}^{R,K}\Gamma_{u_{j}\ell_{i}}^{L,K*}}{m_{K}^2}\Bigg] \ , \\
{C}_T^{jk\tau} &= \frac{\Gamma^{L,K*}_{u_k \ell_{3}} \, \Gamma^{R,K}_{u_j \ell_{3}}}{8 m_K^2}\,.
\end{aligned}
\end{align}
\smallskip\\

\begin{boldmath}
\subsection{$du\ell\nu$}
\end{boldmath}
For the effective Hamiltonian defined in \eq{eq:Heff_dulnu} we find
\begin{align}
\begin{aligned}
C_{VL,jk}^{fi}&=\frac{-\sqrt{2}}{8G_{F}V_{u_{j}d_{k}}}\sum_{K=1}^{N+M}\frac{\Gamma^{K}_{d_{k}\nu_{i}}\Gamma^{L,K*}_{u_{j}\ell_{f}}}{m_{K}^{2}}\,, \\
C_{SL,jk}^{fi}=-4C_{TL,jk}^{fi}&=\frac{\sqrt{2}}{8G_{F}V_{u_jd_k}}\sum_{K=1}^{M+N}\frac{\Gamma^{K}_{d_{k}\nu_{i}}\Gamma^{R,K*}_{u_{j}\ell_{f}}}{m_{K}^2}\,.
\end{aligned}
\end{align}
\smallskip\\

\begin{boldmath}
\subsection{$dd\nu\nu$ and $B_{s}-\bar{B}_{s}$ Mixing}
\end{boldmath}
The effective Hamiltonians for $dd\nu\nu$ and $B_{s}-\bar{B}_{s}$ mixing are given by \eq{eq:Heff_ddnunu} and \eq{eq:Heff_Bs-mixing}, respectively. We find for $b\to s\nu\bar\nu$
\begin{align}
C_{L,jk}^{fi}&=\frac{\sqrt{2}}{4G_{F}V_{td_{k}}V_{td_{j}}^{*}}\frac{\pi}{\alpha}\sum_{K=1}^{N+M}\frac{\Gamma_{d_{k}\nu_{i}}^{K}\Gamma_{d_{j}\nu_{f}}^{K*}}{m_{K}^{2}}\,,
\end{align}
and for $B_{s}-\bar{B}_{s}$ mixing
\begin{align}
\begin{aligned}
{C_1} &= \frac{{ - 1}}{{128{\pi ^2}}}\Bigg(\, {\sum\limits_{\{K,P\} = 1}^{N + M} {\Gamma _{{d_2}{\nu _i}}^{K*}\Gamma _{{d_3}{\nu _j}}^{K}\Gamma _{{d_2}{\nu _j}}^{P*}\Gamma _{{d_3}{\nu _i}}^{P}{C_0}\left( {0,m_K^2,m_P^2} \right)}}  \\&\hspace{20mm}{+\sum\limits_{\{J,Q\} = 1}^M {\Gamma _{{d_2}{\ell _i}}^{Q*}\Gamma _{{d_3}{\ell _j}}^{Q}\Gamma _{{d_2}{\ell _j}}^{J*}\Gamma _{{d_3}{\ell _i}}^{J}{C_0}\left( {0,\bar{m}_{Q}^2,\bar{m}_{J}^2} \right)} } \Bigg)\,.
\end{aligned}
\end{align}
\smallskip\\

\begin{boldmath}
\subsection{$\ell\ell\gamma$, $Z\ell\ell$ and $Z\nu\nu$}
\end{boldmath}
In case of $\ell_{i}\to\ell_{f}\gamma$ transitions and the effective Hamiltonian given by \eq{eq:Heff_llgamma} we have
\begin{align}
\begin{split}
C^{L}_{\ell_{f}\ell_{i}}=-\!\!&\sum_{K=1}^{N+M}\!\Bigg[\frac{m_{\ell_{f}} \Gamma^{L,K}_{u_{j}\ell_{i}}\Gamma^{L,K*}_{u_{j}\ell_{f}}+m_{\ell_{i}} \Gamma^{R,K}_{u_{j}\ell_{i}}\Gamma^{R,K*}_{u_{j}\ell_{f}}}{28 m_{K}^2}-\frac{m_{u_{j}}\Gamma^{L,K}_{u_{j}\ell_{i} }\Gamma^{R,K*}_{u_{j}\ell_{f}}}{4m_{K}^2}\!\left(\!7+4\log\!\left(\!\frac{m^2_{u_{j}}}{m_{K}^2}\!\right)\!\right)\!\Bigg]\\
+&\sum_{J=1}^{M}\frac{m_{\ell_{f}}\Gamma^{J}_{d_{j}\ell_{i}}\Gamma^{J*}_{d_{j}\ell_{f}}}{4\bar{m}_{J}^{2}}\,,
\end{split}
\end{align}
with $N_{c}=3$ already included. For the off-shell photon, as given by the amplitude in \eq{eq:offshell_photon_amp}, we obtain
\begin{align}
\begin{split}
\tilde{\Xi}^{L}_{\ell_{f}\ell_{i}}&=\frac{-N_c}{576\pi^2}\bigg[ \delta_{fi}+\sum_{K=1}^{N+M}\frac{\Gamma_{u_{j}\ell_{f}}^{L,K*}\Gamma_{u_{j}\ell_{i}}^{L,K}}{m_K^2}F\Big(\frac{m_{u_j}^2}{m_K^2}\Big)+\sum_{J=1}^{M}\frac{\Gamma_{d_{j}\ell_{i}}^{J*}\Gamma_{d_{j}\ell_{f}}^{J}}{\bar{m}_{J}^{2}}G\Big(\frac{m_{d_j}^2}{\bar{m}_{J}^2}\Big)\bigg]\,,\\
\tilde{\Xi}^{R}_{\ell_{f}\ell_{i}}&=\frac{-N_c}{576\pi^2}\bigg[\delta_{fi}+\sum_{K=1}^{M+N}\frac{\Gamma_{u_{j}\ell_{f}}^{R,K*}\Gamma_{u_{j}\ell_{i}}^{R,K}}{m_K^2}F\Big(\frac{m_{u_j}^2}{m_K^2}\Big)\bigg] \ ,
\end{split}
\label{eq:app_photon_offshell}
\end{align}
where the loop functions $F(y)$ and $G(y)$ are defined in \eq{eq:loop_functions_photon}.\smallskip

For $Z$ decays, where the amplitude is given by \eq{eq:def_Zll_and_Zvv} and the $\Delta_{fi}^{L(R)}$ are introduced in \eq{eq:effectice_Z_couplings}, we find
\begin{align}
\begin{split}
{\Delta}^{L}_{fi}(q^2)&=\sum_{K=1}^{N+M}\Gamma_{u_{j}\ell_{f}}^{L,K*}\Gamma_{u_{j}\ell_{i}}^{L,K} \mathcal{F}_{L}\left(m_{u_j}^2,q^2,m_{K}^2\right) +\sum_{J=1}^{M}\Gamma_{d_{j}\ell_{f}}^{J*}\Gamma_{d_j\ell_{i}}^{J}\mathcal{G}_{L}\left(q^2,\bar{m}_{J}^2\right)\,,\\
{\Delta}^{R}_{fi}(q^2)&=\sum_{K=1}^{N+M}\Gamma_{u_{j}\ell_{f}}^{R,K*}\Gamma_{u_{j}\ell_{i}}^{R,K} \mathcal{F}_{R}\left(m_{u_j}^2,q^2,m_{K}^2\right)\,,
\end{split}
\label{eq:Zlleff_full}
\end{align}
with
\begin{align}
\begin{split}
\mathcal{F}_{L}&\left(m_{u}^2,q^2,M^2\right)=\frac{N_{c}}{864\pi^2 M^2} \bigg(\!\!\Big(3q^2(4s_w^2-3)+27m_{u}^2\Big)\log\!\left(\!\frac{m_{u}^2}{M^2}\!\right)-s_w^2(5q^2+48m_{u}^2)\\
&+3(q^2+3m_{u}^2)+6\mathcal{X}(m_u^2,q^2)\Big(4s_w^2(q^2+2m_{u}^2)-3q^2+3m_{u}^2\Big)\arctan\!\left(\!\frac{1}{\mathcal{X}(m_u^2,q^2)}\right)\!\!\bigg)\,,\\
\mathcal{G}_{L}&\left(q^2,M^2\right)=-\frac{N_{c}\,q^2}{864\pi^2M^2}\bigg(\!(6s_w^2-9)\log\!\left(\!\frac{q^2}{M^2}\!\right)+2s_w^2(1-3i\pi)+9i\pi\bigg)\,,\\
\mathcal{F}_{R}&\left(m_{u}^2,q^2,M^2\right)=\frac{N_{c}}{864\pi^2 M^2 }\bigg(\!\Big(12 s_w^2 q^2-27m_{u}^2\Big)\log\!\left(\!\frac{m_{u}^2}{M^2}\!\right) -s_w^2\left(5q^2+48m_{u}^2\right)+27m_{u}^2\\
&+6\mathcal{X}(m_u^2,q^2)\Big(4s_w^2(q^2+2m_{u}^2)-9m_{u}^2\Big)\arctan\!\left(\!\frac{1}{\mathcal{X}(m_u^2,q^2)}\!\right)\!\!\bigg)\,,
\end{split}
\label{eq:Zll_loop_function}
\end{align}
again using
\begin{align}
\mathcal{X}(a^2,b^2)=\sqrt{\frac{4a^2}{b^2}-1}\,.
\end{align}
At the level of the effective couplings ($q^2=0$) we have
\begin{align}
\begin{split}
\Delta^{L}_{fi}(0)&=\sum_{K=1}^{N+M}\Gamma_{u_{3}\ell_{f}}^{L,K*}\Gamma_{u_{3}\ell_{i}}^{L,K} \mathcal{F}_{L}\left(m_{t}^2,0,m_{K}^2\right)\,,\\
\Delta^{R}_{fi}(0)&=\sum_{K=1}^{N+M}\Gamma_{u_{3}\ell_{f}}^{R,K*}\Gamma_{u_{3}\ell_{i}}^{R,K} \mathcal{F}_{R}\left(m_{t}^2,0,m_{K}^2\right)\,.
\end{split}
\end{align}
The functions $\mathcal{F}_{L/R}$ then become
\begin{align}
\begin{split}
\mathcal{F}_{L}(m_{t}^2,0,M^2)&=\frac{m_{t}^2 N_c}{32\pi^2 M^2}\left(1+\log\!\left(\!\frac{m_{t}^2}{M^2}\!\right)\right)=-\mathcal{F}_{R}(m_{t}^2,0,M^2)\,.
\end{split}
\end{align}
The amplitude for $Z\to\nu\bar{\nu}$ is again given by \eq{eq:def_Zll_and_Zvv}. For the $\Sigma_{fi}^{\text{LQ}}\big(q^2\big)$, introduced in \eq{eq:effectice_Z_couplings}, we obtain
\begin{align}
{\Sigma}_{fi}^{\text{LQ}}(q^2)=\sum_{K=1}^{N+M}\Gamma_{d_{j}\nu_{f}}^{L,K*}\Gamma_{d_{j}\nu_{i}}^{L,K}\mathcal{H}_{1}(q^2,m_K^2)+\sum_{J=1}^{M}\Gamma_{u_{j}\nu_{f}}^{J*}\Gamma_{u_{j}\nu_{i}}^{J}\mathcal{H}_{2}(m_{u_{j}}^2,q^2,\bar{m}_{J}^2)\,,
\label{eq:Znunu_full}
\end{align}
with
\begin{align}
\begin{split}
\mathcal{H}_{1}&(q^2,M^2)=\frac{N_{c}\,q^2}{864\pi^2 M^2}\bigg(3(3-2s_w^2)\log\!\left(\!\frac{q^2}{M^2}\!\right)-3i\pi(3-2s_w^2)-3+s_w^2\bigg)\,,\\
\mathcal{H}_{2}&(m_{u}^2,q^2,M^2)=\frac{N_c}{864\pi^2M^2} \bigg(3\Big((4s_w^2-3)q^2+9m_{u}^2\Big)\log\!\left(\!\frac{m_{u}^2}{M^2}\!\right)-2s_w^2(q^2+24m_{u}^2)\\&
+9m_{u}^2+6\mathcal{X}(m_u^2,q^2) \Big(4s_w^2(q^2+2m_{u}^2)-3q^2+3m_{u}^2\Big)\arctan\!\bigg(\!\frac{1}{\mathcal{X}(m_u^2,q^2)}\!\bigg)\!\bigg)\,,
\end{split}
\label{eq:Zvv_loop_functions}
\end{align}
where we again neglected to down-type quark masses, but kept the dependencies on the up-type ones due to the heavy top quark. If we work with effective couplings instead of full amplitudes, the results are
\begin{align}
\Sigma_{fi}^{\text{LQ}}(0)=\sum_{J=1}^{M}\Gamma_{u_{3}\nu_{f}}^{J*}\Gamma_{u_{3}\nu_{i}}^{J}\mathcal{H}_{2}(m_{t}^2,0,\bar{m}_{J}^2)\,,
\end{align}
with
\begin{align}
\mathcal{H}_{2}(m_{t}^2,0,M^2)=&\frac{N_c m_{t}^2}{32\pi^2 M^2}\bigg(1+\log\!\left(\!\frac{m_{t}^2}{M^2}\!\right)\!\bigg)\,.
\end{align}
\smallskip\\

\begin{boldmath}
\subsection{$W\ell\nu$}
\end{boldmath}
For the $\Lambda^{\text{LQ}}_{fi}\big(q^2\big)$, defined in \eq{eq:ampl_wlnu} and \eq{eq:wlnu_eff}, we obtain
\begin{align}
\begin{split}
{\Lambda}^{\text{LQ}}_{fi}&(q^2) =
\frac{N_c}{64 \pi^2}\Bigg\{\sum_{K=1}^{N+M}\bigg[V^*_{u_j d_k}\Gamma^{L,K*}_{u_j \ell_f} \Gamma^{L,K}_{d_k \nu_i} \mathcal{F}_{W}\left(m_{u_j}^2,q^2,m_K^2\right)
+\Gamma^{L,K*}_{u_3\ell_f}\Gamma^{L,K}_{u_3\ell_i}\frac{m_{t}^2}{m_K^2}\bigg] \\
&+\sum_{J=1}^M\Gamma^{J*}_{u_3\nu_f} \Gamma^{J}_{u_3\nu_i}\frac{m_t^2}{\bar{m}_{J}^2} +2\sqrt{2}\sum_{K=1}^{N+M} \sum_{J=1}^M W_{J+N,K} \Gamma^{L,K*}_{u_3 \ell_j} \Gamma^J_{u_3 \nu_i} \frac{m_{t}^2}{m_K^2-\bar{m}_{J}^2}\log\! \left(\!\frac{m_K^2}{\bar{m}_{J}^2}\!\right)\\
&-\frac{2\sqrt{2}}{3}\sum_{K=1}^{N+M} \sum_{J=1}^M q^2\Big(W_{J+N,K} \Gamma^{L,K*}_{u_j \ell_f} \Gamma^J_{u_j \nu_i}- W_{J+N,K}^{*} \Gamma^{L,K*}_{d_k \ell_f} \Gamma^J_{d_k \nu_i}\Big)\mathcal{H}_{W}\left(m_K^2,\bar{m}_{J}^2\right)\Bigg\}
\end{split}
\label{eq:weff_full}
\end{align}
with
\begin{align}
\begin{split}
\mathcal{F}_{W}\left(m_{u}^2,q^2,M^2\right)=&\frac{1}{9 M^2}\bigg[6(2q^2-3m_{u}^2) \log\!\left(\!\frac{m_{u}^2}{M^2}\!\right) -\Big(4q^2-3m_{u}^2-\frac{6m_u^4}{q^2}\Big)\\
&+6\Big(2q^2-3m_{u}^2+\frac{m_u^6}{(q^2)^2}\Big) \log\!\Big(1-\frac{q^2}{m_u^2}\Big)\bigg]\\
\mathcal{H}_{W}(x^2,y^2)=&\frac{x^2+y^2}{(x^2-y^2)^2}-\frac{2\,x^2 y^2}{(x^2-y^2)^3}\log\!\left(\frac{x^2}{y^2}\right) \ .
\end{split}
\end{align}
Additionally, there are terms that do not trivially decouple, however, they vanish in the decoupling limit. They read
\begin{align}
\begin{split}
&\overline{\Lambda}_{fi}^{\text{LQ}}(\mu^2) = \frac{N_c}{64 \pi^2} \Bigg\{\sum_{K=1}^{N+M} \bigg[-V^*_{u_j d_k}\Gamma^{L,K*}_{u_j \ell_f} \Gamma^{L,K}_{d_k \nu_i}\bigg(2 \log\!\bigg(\! \frac{\mu^2}{m_K^2}\!\bigg)+1 \bigg) \\ 
&\hspace{28mm}- \Big(\Gamma^{L,K*}_{u_j\ell_f}\Gamma^{L,K}_{u_j\ell_i} + \Gamma^{L,K*}_{d_j\nu_f}\Gamma^{L,K}_{d_j\nu_i}\Big)\bigg( \log\! \bigg(\!\frac{\mu^2}{m_K^2}\!\bigg) + \frac{1}{2} \bigg)\bigg] \\ &\hspace{2mm}-\sum_{J=1}^M \Big(\Gamma^{J*}_{u_j\nu_f} \Gamma^{J}_{u_j\nu_i}+\Gamma^{J*}_{d_j\ell_f} \Gamma^{J}_{d_j\ell_i} \Big)\bigg( \log\! \bigg(\!\frac{\mu^2}{\bar{m}_{J}^2}\!\bigg) + \frac{1}{2} \bigg) \\ 
&\hspace{2mm}-\sqrt{2}\sum_{J=1}^{M} \sum_{K=1}^{N+M}
\bigg[ W_{J+N,K} \Gamma^{L,K*}_{u_j \ell_f} \Gamma^J_{u_j \nu_i}\bigg(2 \log\!\bigg(\!\frac{\mu^2}{\bar{m}_{J}^2}\!\bigg) - \frac{2 m_K^2}{m_K^2-\bar{m}_{J}^2}\log\! \bigg(\!\frac{m_K^2}{\bar{m}_{J}^2}\!\bigg) + 3 \bigg) \\ &\hspace{23mm} - W_{J+N,K}^{*} \Gamma^{J*}_{d_k \ell_f} \Gamma^{L,K}_{d_k \nu_i}\bigg(2 \log\!\bigg(\! \frac{\mu^2}{m_K^2}\!\bigg) - \frac{2 \bar{m}_{J}^2}{m_K^2-\bar{m}_{J}^2} \log\! \bigg(\!\frac{m_K^2}{\bar{m}_{J}^2}\!\bigg) +3 \bigg)  \bigg] \Bigg\}\,.
\end{split}
\end{align}
Note that the scale dependence $\mu$ drops out exactly. If we work at the level of effective couplings, we have
\begin{align}
\begin{split}
\Lambda_{fi}^{\text{LQ}}(0) =& \frac{N_c m_t^2}{64\pi^2} \Bigg[\sum_{K=1}^{N+M}\Bigg(\frac{\Gamma^{L,K*}_{u_3\ell_f} \Gamma^{L,K}_{u_3\ell_i}}{m_K^2}-\frac{2 V^*_{u_3 d_k} \Gamma^{L,K}_{d_k \nu_i} \Gamma^{L,K*}_{u_3 \ell_f}}{m_K^2} \log\! \bigg(\!\frac{m_{t}^2}{m_K^2}\!\bigg)\Bigg)\\ 
&+\sum_{J=1}^M\frac{\Gamma^{J*}_{u_3\nu_f} \Gamma^{J}_{u_3\nu_i}}{\bar{m}_{J}^2} +2\sqrt{2}\sum_{K=1}^{N+M} \sum_{J=1}^M W_{J+N,K}  \frac{\Gamma^{L,K*}_{u_3 \ell_f} \Gamma^J_{u_3 \nu_i}}{m_K^2-\bar{m}_{J}^2}\log\! \bigg(\!\frac{m_K^2}{\bar{m}_{J}^2}\!\bigg)\Bigg] \, .
\end{split}
\end{align}
In the limit of no LQ mixing, the loop functions used in \eq{eq:wlnu} become
\begin{align}
\begin{split}
\mathcal{F}_{1}\left(m_{u}^2,q^2,M^2\right)&=\mathcal{F}_{W}\left(m_{u}^2,q^2,M^2\right)+\frac{m_{u}^2}{M^2}\\
\mathcal{F}_2\left(m_{u}^2,q^2,M^2\right)&=\mathcal{F}_{W}\left(m_{u}^2,q^2,M^2\right)-\frac{m_{u}^2}{M^2}\,.
\end{split}
\label{eq:Wlv_loop_functions}
\end{align}
\smallskip\\

\begin{boldmath}
\subsection{$\tau \to 3\mu$, $\tau \to \mu e^{+}e^{-}$ and $\mu\to3e$}
\end{boldmath}
The relevant effective Hamiltonian is given in \eq{eq:Heff_tau3mu}. The contributions of the photon and $Z$ penguin diagrams are given by \eq{eq:tau3mu_photon} and \eq{eq:tau3mu_Z}, respectively. Now we use the effective couplings as defined in \eq{eq:app_photon_offshell} (photon) and \eq{eq:Zlleff_full} ($Z$ boson).\smallskip

Finally, we have the box diagrams. Contrary to the vector current operators, the scalar operators $O_{\ell\ell\ell\ell}^{S}$ are always proportional to  $m_{q}^2/M_{LQ}^2$. Therefore, we only consider contributions from the top quark. The box contributions read
\begin{align}
\begin{split}
C_{abfi}^{V\,LL}=&\frac{-N_c}{256\pi^2}\!\!\!\sum_{\{K,P\}=1}^{N+M}\!\!\!\left(\Gamma_{u_{k}\ell_{a}}^{L,P*}\Gamma_{u_{k}\ell_{b}}^{L,K}\Gamma_{u_{j}\ell_{f}}^{L,K*}\Gamma_{u_{j}\ell_{i}}^{L,P}+\Gamma_{u_{j}\ell_{a}}^{L,K*}\Gamma_{u_{k}\ell_{b}}^{L,K}\Gamma_{u_{k}\ell_{f}}^{L,P*}\Gamma_{u_{j}\ell_{i}}^{L,P}\right)D_{2}\big(m_{u_k}^2,m_{u_j}^2,m_K^2,m_P^2\big)\\
&-\frac{N_c}{256\pi^2}\!\!\!\sum_{\{J,Q\}=1}^{M}\!\!\!\left(\Gamma_{d_{k}\ell_{a}}^{Q*}\Gamma_{d_{k}\ell_{b}}^{J}\Gamma_{d_{j}\ell_{f}}^{J*}\Gamma_{d_{j}\ell_{i}}^{Q}+\Gamma_{d_{j}\ell_{a}}^{J*}\Gamma_{d_{k}\ell_{b}}^{J}\Gamma_{d_{k}\ell_{f}}^{Q*}\Gamma_{d_{j}\ell_{i}}^{Q}\right)C_{0}\big(0,\bar{m}_{J}^2,\bar{m}_{Q}^2\big)\,,\\
C_{abfi}^{V\,LR}=&\frac{-N_c}{128\pi^2}\!\!\!\sum_{\{K,P\}=1}^{N+M}\!\!\!\Big[\Gamma_{u_{k}\ell_{a}}^{L,P*}\Gamma_{u_{k}\ell_{b}}^{L,K}\Gamma_{u_{j}\ell_{f}}^{R,K*}\Gamma_{u_{j}\ell_{i}}^{R,P}D_{2}\big(m_{u_k}^2,m_{u_j}^2,m_K^2,m_P^2\big)\\
&\phantom{12345678901234}-2\Gamma_{u_{3}\ell_{a}}^{L,K*}\Gamma_{u_{3}\ell_{b}}^{L,K}\Gamma_{u_{3}\ell_{f}}^{R,P*}\Gamma_{u_{3}\ell_{i}}^{R,P}m_t^2 D_{0}\big(m_t^2,m_t^2,m_{K}^2,m_P^2\big)\Big]\,,\\
C_{abfi}^{S\,LL}=&\frac{-m_t^2 N_c}{64\pi^2}\!\!\!\sum_{\{K,P\}=1}^{N+M}\!\!\!\left(2\Gamma_{u_{3}\ell_{a}}^{R,P*}\Gamma_{u_{3}\ell_{b}}^{L,K}\Gamma_{u_{3}\ell_{f}}^{R,K*}\Gamma_{u_{3}\ell_{i}}^{L,P}-\Gamma_{u_{3}\ell_{a}}^{R,K*}\Gamma_{u_{3}\ell_{b}}^{L,K}\Gamma_{u_{3}\ell_{f}}^{R,P*}\Gamma_{u_{3}\ell_{i}}^{L,P}\right) D_{0}\big(m_t^2,m_t^2,m_K^2,m_P^2\big)\,.
\end{split}
\label{eq:l_lll_boxes}
\end{align}
Again, $C^{V/S \, RL(RR)}_{abfi}$ are obtained from $C^{V/S \, LR(LL)}_{abfi}$ by interchanging $L$ and $R$.
\smallskip\\

\begin{boldmath}
\subsection{$\tau \to \ell\nu\bar{\nu}$ and $\mu \to e\nu\bar{\nu}$}
\end{boldmath}

As it was the case for the previous results, we consider the top as the only non-zero quark mass and in cases where the result is proportional to the quark mass (squared), we directly write the result in terms of the top. The effective Hamiltonian for the process is given in \eq{eq:Heff_taumununu}. The box diagrams read
\begin{align}
\begin{split}
D^{L,fi}_{\ell_{a}\ell_{b}}=-\frac{N_c}{64\pi^2}\Bigg\{&\sum_{\{K,P\}=1}^{N+M}\Gamma_{u_{k}\ell_{a}}^{L,P*}\Gamma_{u_{k}\ell_{b}}^{L,K}\Gamma_{d_{j}\nu_{f}}^{L,K*}\Gamma_{d_{j}\nu_{i}}^{L,P}C_{0}\big(m_{u_{k}}^2,m_K^2,m_P^2\big)\\
&+\sum_{K=1}^{N+M}\sum_{J=1}^{M}\Big[\Gamma_{u_{j}\ell_{a}}^{L,K*}\Gamma_{u_{k}\ell_{b}}^{L,K}\Gamma_{u_{k}\nu_{f}}^{J*}\Gamma_{u_{j}\nu_{i}}^{J}D_{2}\big(m_{u_k}^2,m_{u_j}^2,m_K^2,\bar{m}_{J}^2\big)\\
&\phantom{12345678910}+\Gamma_{d_{j}\ell_{a}}^{J*}\Gamma_{d_{k}\ell_{b}}^{J}\Gamma_{d_{k}\nu_{f}}^{L,K*}\Gamma_{d_{j}\nu_{i}}^{L,K}C_{0}\big(0,\bar{m}_{J}^2,m_K^2\big)\Big] \Bigg\}\,,\\
D^{R,fi}_{\ell_{a}\ell_{b}}=-\frac{N_c}{64\pi^2}\Bigg\{&\sum_{\{K,P\}=1}^{N+M}\Gamma_{u_{k}\ell_{a}}^{R,P*}\Gamma_{u_{k}\ell_{b}}^{R,K}\Gamma_{d_{j}\nu_{f}}^{L,K*}\Gamma_{d_{j}\nu_{i}}^{L,P}C_{0}\big(m_{u_{k}}^2,m_{K}^2,m_P^2\big)\\
&-2\sum_{K=1}^{N+M}\sum_{J=1}^{M}\Gamma_{u_{3}\ell_{a}}^{R,K*}\Gamma_{u_{3}\ell_{b}}^{R,K}\Gamma_{u_{3}\nu_{f}}^{J*}\Gamma_{u_{3}\nu_{i}}^{J}m_{t}^2D_{0}\big(m_{t}^2,m_{t}^2,m_{K}^2,\bar{m}_{J}^2\big)\Bigg\}\,.
\label{eq:l_lvv_boxes}
\end{split}
\end{align}
The contributions of the $W$ and $Z$ penguins are given by \eq{eq:taumununu_W} and \eq{eq:taumununu_Z}, respectively. Now the effective couplings from \eq{eq:weff_full}, \eq{eq:Zlleff_full} and \eq{eq:Znunu_full} have to be used.

\newpage

\bibliographystyle{JHEP}
\bibliography{BIB}

\end{document}